\documentclass[a4paper,11pt]{article}
\usepackage[skins]{tcolorbox}
\usepackage{mathtools,slashed}
\usepackage[T1]{fontenc}
\usepackage{slashed,verbatim,subfigure}
\usepackage[numbers,sort&compress]{natbib}
\usepackage{amsmath}
\usepackage{mathrsfs}
\usepackage{amsbsy}
\usepackage{amssymb}
\usepackage{appendix}
\usepackage{graphicx}
\usepackage[final]{pdfpages}
\usepackage{dcolumn}
\usepackage{bm}
\usepackage{cancel}
\usepackage[colorlinks=true,linktocpage=true,
linkcolor=blue,citecolor=blue]{hyperref}
\usepackage{url}
\definecolor{db5}{cmyk}{0.5,0.5,0,0.5}
\definecolor{mauve}{cmyk}{0.3,0.7,0.1,0.3}
\definecolor{palemauve}{cmyk}{0.3,0.7,0.1,0.0}
\definecolor{pb}{cmyk}{0.4,0.1,0,0.1}
\definecolor{pgreen}{cmyk}{0.4,0.0,0.3,0.0}
\definecolor{pink}{cmyk}{0.0,0.5,0.3,0.0}
\newcommand{\bs}{\begin{slide}}
\newcommand{\es}{\end{slide}}
\newcommand{\tcb}{\textcolor {blue}}
\usepackage[a4paper]{geometry}
\catcode`@11
\def\seceqaa{\@addtoreset{equation}{section}
	\def\theequation{A\arabic{equation}}}
\def\seceqbb{\@addtoreset{equation}{section}
	\def\theequation{B\arabic{equation}}}
\def\seceqcc{\@addtoreset{equation}{section}
	\def\theequation{C\arabic{equation}}}
\def\seceqdd{\@addtoreset{equation}{section}
	\def\theequation{D\arabic{equation}}}
\def\seceqee{\@addtoreset{equation}{section}
	\def\theequation{E\arabic{equation}}}
\def\seceqff{\@addtoreset{equation}{section}
	\def\theequation{F\arabic{equation}}}	
\def\seceqgg{\@addtoreset{equation}{section}
	\def\theequation{G\arabic{equation}}}
\def\seceqhh{\@addtoreset{equation}{section}
	\def\theequation{H\arabic{equation}}}
\def\seceqii{\@addtoreset{equation}{section}
	\def\theequation{H\arabic{equation}}}	
\def\seceqjj{\@addtoreset{equation}{section}
	\def\theequation{H\arabic{equation}}}
\catcode`@11
\usepackage{authblk}
\newcommand{\be}{\begin{eqnarray}}
\newcommand{\ee}{\end{eqnarray}}
\begin{document}
\large
\title{Pole-Skipping and Chaos in Hot ${\cal M}$QCD}
\author[,1]{Gopal Yadav\footnote{email- gopalyadav@cmi.ac.in}}  \author[,2] {Shivam Singh Kushwah\footnote{email- shivams\_kushwah@ph.iitr.ac.in}} \author[,2] {Aalok Misra\footnote{email- aalok.misra@ph.iitr.ac.in}} 
\affil[1]{Chennai Mathematical Institute,
SIPCOT IT Park, Siruseri 603103, India
}
\affil[2]{Department of Physics, 
Indian Institute of Technology Roorkee, Roorkee 247667, Uttarakhand, India}
\date{}
\maketitle
\begin{abstract}
We address the question of whether thermal QCD at high temperature is chaotic from the ${\cal M}$ theory dual of QCD-like theories at intermediate coupling as constructed in \cite{OR4}. The equations of motion of the gauge-invariant combination $Z_s(r)$ of scalar metric perturbations is shown to possess an irregular singular point at the horizon radius $r_h$. Very interestingly, at a specific value of the imaginary frequency and momentum used to read off the analogs of the ``Lyapunov exponent'' $\lambda_L$ and ``butterfly velocity'' $v_b$ not only does $r_h$ become a regular singular point, but truncating the incoming mode solution of $Z_s(r)$ as a power series around $r_h$,  yields a ``missing pole'', i.e., $C_{n, n+1}=0,\ {\rm det}\ M^{(n)}=0, n\in\mathbb{Z}^+$ is satisfied for {\it a single} $n\geq3$ depending on the values of the string coupling $g_s$, number of (fractional) $D3$ branes $(M)N$ and flavor $D7$-branes $N_f$ in the parent type IIB set \cite{metrics}, e.g., for the QCD(EW-scale)-inspired $N=100, M=N_f=3, g_s=0.1$, one finds a missing pole at $n=3$. For integral $n>3$, truncating $Z_s(r)$ at ${\cal O}((r-r_h)^n)$, yields $C_{n, n+1}=0$ at order $n,\ \forall n\geq3$. Incredibly, (assuming preservation of isotropy in $\mathbb{R}^3$ even with the inclusion of higher derivative corrections) the aforementioned gauge-invariant combination of scalar metric perturbations receives no ${\cal O}(R^4)$ corrections. Hence, (the aforementioned analogs of) $\lambda_L, v_b$ are unrenormalized up to ${\cal O}(R^4)$ in ${\cal M}$ theory.
\end{abstract}

\tableofcontents

\section{Introduction}
\label{Introduction}
Quantum chromodynamics (QCD) is a theory of strong interaction that behaves differently in the low-energy and high-energy regimes, and there is a first-order phase transition from the confined phase ($T<T_c$) to the deconfined phase ($T>T_c$) at $T=T_c$ where $T_c$ is the deconfinement temperature. In the heavy/chiral quark-mass limit,  the first-order phase transition at vanishing chemical potential turns into a cross-over for non-vanishing chemical potential \cite{He:2013qq}. This is consistent with the UV-complete top-down (type IIB) holographic dual of thermal QCD-like theories of \cite{metrics} wherein it was shown in \cite{QCD-trace-anomaly-Misra+Gale} that one obtained a vanishing baryon chemical potential in the heavy/chiral quark-mass limit; the Hawking-Page phase transition between the confined to the deconfined phase to obtain a lattice-compatible deconfinement temperature, was also worked out in \cite{QCD-trace-anomaly-Misra+Gale}. 

The degrees of freedom in the confined phase are Mesons, Baryons, etc., whereas in the deconfined phase, degrees of freedom are those of quark-gluon plasma. Quark-gluon plasma can be treated as a many-body system as the number of degrees of freedom are very large in the deconfined phase of QCD, and one can ask the question about the chaotic nature of quark-gluon plasma. This paper addresses this question from the perspective of top-down holography \cite{Maldacena}. As far as we know, two models describe QCD from a top-down approach: Sakai-Sukimoto model \cite{SS} and the UV complete type IIB string dual constructed in \cite{metrics}. The Sakai-Sugimoto model is not UV complete with mesinos (superpartners of mesons) isospectral with mesons, and with non-vanishing mesino-mesino-meson interaction \cite{mesino-SS} unlike real QCD. Recently, two of us (AM and GY) found that mesinos are superheavy and mesino-mesino-meson interaction is vanishing \cite{Aalok+Gopal-Mesino} in ${\cal M}$-theory dual inclusive of the ${\cal O}(R^4)$ corrections as constructed in \cite{OR4} (based on \cite{MQGP}) which describes QCD at intermediate coupling. The aforementioned ${\cal M}$-theory dual was constructed by constructing the type IIA SYZ mirror of type IIB setup \cite{metrics} and then uplifting to ${\cal M}$-theory.

{Classical chaos is measured by the stability of trajectory under small change in initial condition in phase space. Suppose we have a reference trajectory $X(t)$ with initial condition $X(0)=X_0$. Now we change the initial condition as: $X_0 \rightarrow X_0+\delta X_0$, this will lead to a new trajectory $X(t)+\delta X(t)$. If the system is chaotic then distance between the new trajectory and reference trajectory increases exponentially with time, i.e., $\delta X(t) \sim \delta X_0 e^{\lambda_L t}$ where $\lambda_L$ is the Lyapunov exponent. If $\lambda_L=0$ then  $\delta X(t) \sim \delta X_0$ and the system is a non-chaotic system.} {Quantum chaos is characterized by the Lyapunov exponent ($\lambda_L$) and butterfly velocity ($v_b$). For chaotic systems, quantum chaos is related to OTOC(out-of-time-ordered correlator). 
OTOC of the CFTs with large $N$ degrees of freedom which have holographic dual has the following form \cite{Maldacena:2015waa,Shenker:2013pqa,Shenker:2013yza,Kitaev,Shenker:2014cwa,Roberts:2014isa}:
\begin{eqnarray}
\label{OTOC-i}
& & F(t) = {\rm tr}[yVyW(t)yVyW(t)] = f_0-\frac{f_1}{N^2} e^{\lambda_L t}+{\cal O}(N^{-4}),
\end{eqnarray}
where $y^4=\frac{e^{-\beta_0 H}}{Z}, \beta_0=\frac{1}{T_0}$, $H$ is the Hamiltonian of the system, $Z={\rm tr}\left(e^{-\beta_0 H}\right)$ and $V$, $W$ are generic operators; $f_0$ and $f_1$ are order one positive constants which depend on the operators $V$ and $W$. Near the scrambling time $t_*=\frac{\beta_0}{2 \pi}\log N^2$, $F(t)$ starts decreasing. There is another time scale known as dissipation time ($t_d$), for large $N$ CFTs, $t_d \sim \beta_0$. Hence dissipation time and scrambling time are separated.
The speed of propagation of perturbation in space is described by the butterfly velocity ($v_b$). In the language of AdS/CFT duality, the causal structure of the bulk geometry is determined by the butterfly velocity \cite{Qi:2017ttv} and for AdS-BH spacetime, $v_b=\sqrt{\frac{d+1}{2 d}}$ \cite{Mezei:2016wfz} where $d$ is the number of spatial dimensions of boundary theory. The Lyapunov exponent is bounded as, $\lambda_L \leq \frac{2 \pi}{\beta}, \beta=\frac{1}{T}$ and for maximally chaotic systems, $\lambda_L=\frac{2 \pi}{\beta}$ \cite{Maldacena:2015waa}. The Lyapunov exponent and butterfly velocity  are calculated using holography from a shock wave analysis \cite{Shenker:2013pqa,Shenker:2013yza,Roberts:2014isa,Shenker:2014cwa} or from the pole-skipping phenomenon \cite{Blake:2018leo,Blake:2019otz,Natsuume:2019xcy,Natsuume:2019sfp,Natsuume:2020snz,Ning:2023ggs}\footnote{See also \cite{Rabinovici:2022beu,Altland:2022xqx,Grozdanov:2023tag,deBoer:2017xdk,Fischler:2018kwt,Das:2022jrr,Das:2021qsd,David:2019bmi,Karlsson:2021duj,Ramirez:2020qer,Amano:2022mlu,Abbasi:2023myj,Blake:2021hjj,Jensen:2016pah}.} . The classification of pole-skipping points and pole-skipping in the context of hyperbolic black holes are discussed in \cite{Ahn:2019rnq,Ahn:2020baf}\footnote{We thank H.~S. Jeong for bringing his works to our attention.}. }

The chaotic nature of holographic QCD has been studied in \cite{Ageev:2021poy} where the author found the relation between drag force($F$), jet quenching parameter($\hat{q_y}$), and butterfly velocity ($v_b$):
\begin{eqnarray}
\label{hQCD-chaos}
& & F=\frac{d p_\sigma}{dt}=-v-\frac{v^3}{(d-1)v_b^2}+...., \ \ \ \hat{q_y}={\cal A}\left(\frac{v_b^{(x)}}{v_b^{(y)}}\right)^2 T \sigma_x,
\end{eqnarray}
where $p_\sigma=\frac{p}{\sigma}$, $\sigma$, $d$, $v_b$, $T$, and $v$ are the renormalized momentum, renormalization factor, spatial dimension of boundary theory, butterfly velocity, and temperature and the velocity of charge carrier moving through the strongly interacting plasma at temperature $T$ respectively. {Further, $\sigma_x$ being the coefficient associated with leading order drag force coefficient acting on the projectiles along $x$ direction and ${\cal A}$ is a constant and it depends upon $a$ and infrared exponents $\nu$ where $a$ and $\nu$ appear in the $d$-dimensional metric of the form:
\begin{eqnarray}
ds^2=-\frac{f(z)}{z^{2\nu}}dt^2+\frac{dz^2}{z^{2\nu_z}f(z)}+\frac{dx^2}{z^{2\nu}}+\frac{dy^2}{z^{2\nu_y}}+\sum_{\alpha}\frac{{d x_\alpha}^2}{z^{2\nu_\alpha}},
\end{eqnarray}
with $f(z)=1-\left(\frac{z}{z_h}\right)^a$. The precise form of ${\cal A}$ is given in the supplimental file of \cite{Ageev:2021poy}.} It should be mentioned that results of \cite{Ageev:2021poy} are obtained for a radial-coordinate ($u=1/r$)-dependent diagonal metric where the form of the bulk gravitational action is not explicitly mentioned. The Lyapunov exponent and butterfly velocity for the aforementioned general metric are given as \cite{Ageev:2021poy}:
\begin{eqnarray}
\label{Ageev-results}
& & \lambda_L=2 \pi T, \ \ \ \ \ v_b^2=\frac{g_{tt}'(u_h)}{g_{x^1 x^1}'(u_h)(d-1)},
\end{eqnarray}
where $u_h$ is the black hole horizon. The author in \cite{Losacco:2022xen} studied the chaotic dynamics of {quark-antiquark ($q\overline{q}$) pair} in the presence of chemical potential and magnetic field by studying the Poincaré section and Lyapunov exponent and found that in the absence of the magnetic field, increasing the chemical potential stabilizes the trajectory of open strings (hanging in the bulk) in Poincaré section and for non-zero magnetic field one obtains stable trajectories for certain range of the magnetic field. Chaos in ${\cal N}=2$ supersymmetric QCD has been studied in \cite{Hashimoto:2016wme} using AdS/CFT duality, and authors found that for small $N_c$ {(where $N_c$ is the number of color degrees of freedom in QCD)}, ${\cal N}=2$ supersymmetric QCD is more chaotic. {This was shown by calculating the Lyapunov exponent of the time evolution of a homogeneous quark condensate using AdS/CFT. Authors found that Lyupunov exponent decreases as $N_c$ increases and vice-versa. Hence for small $N_c$, Lyapunov exponent is large and for large $N_c$, Lyapunov exponent is small. This justifies the aforementioned statement that small $N_c$, ${\cal N}=2$ supersymmetric QCD is more chaotic than the large $N_c$, ${\cal N}=2$ supersymmetric QCD. Authors also discussed that there is a phase transtion from non-chaotic to chaotic phase at an energies $E_{\rm chaos}\stackrel{>}{\sim} (6 \times 10^2)\times m_q^4 \frac{N_c}{\lambda^2}$ where $m_q$ is the quark mass and $\lambda$ is the 't Hooft coupling. From $E_{\rm chaos}$ it is clear that ${\cal N}=2$ supersymmetric QCD will be more (easily) chaotic when $N_c$ is small or $\lambda$ is large as the same would imply that the threshold above which chaos sets in, gets lowered and therefore is more easily surpassable/achievable.}
 Authors in \cite{Akutagawa:2019awh} studied the chaos of QCD strings at large $N_c$ and strong coupling where QCD string is a pair of quark and antiquark of separation length $L_q$. The gravity dual used in \cite{Akutagawa:2019awh} is solitonic D4-geometry \cite{Witten:1998zw}. To study chaos, the authors applied pulse force at the endpoints of the QCD string and found the appearance of chaos when the applied pulse force is larger than a critical value. For a large separation between quark-antiquark pair, the QCD string is less chaotic, and the origin of chaos is the endpoints of the QCD string. 

The above results in the context of chaos in QCD are obtained either from a bottom-up approach or a top-down approach, but the studies have been done in the confined phase of QCD (i.e. $T<T_c$), and the results of \cite{Hashimoto:2016wme} are valid for ${\cal N}=2$ supersymmetric QCD. The approach of \cite{Losacco:2022xen,Hashimoto:2016wme,Akutagawa:2019awh} is to plot the Poincaré section of string dynamics and see whether QCD is chaotic. Although the results of \cite{Ageev:2021poy} are valid for the high-temperature phase ($T>T_c$) of QCD-like theory, again, this is a bottom-up approach, and the precise form of the gravitational action is not given. The gravity dual constructed in \cite{MQGP,OR4} corresponds to real QCD, and top-down construction is more fundamental where one starts from the original type IIB/IIA string theory or ${\cal M}$-theory. {\it The motivation of the paper is to see from the ${\cal M}$-theory dual inclusive of ${\cal O}(R^4)$ corrections \cite{OR4},  based on what is known as ``pole skipping'', whether the deconfined phase ($T>T_c$) of holographic thermal QCD at intermediate coupling is chaotic. We also want to see the effect of ${\cal O}(R^4)$ terms appearing in the eleven-dimensional supergravity action of \cite{OR4} on the Lyapunov exponent ($\lambda_L$) and butterfly velocity ($v_b$)}.

The remaining paper is organized in the following way. We start with a review of ${\cal M}$-theory uplift of type IIB string dual \cite{metrics} in the presence of ${\cal O}(R^4)$ terms [based on \citep{metrics,MQGP,NPB,OR4}] in section \ref{review}. In section \ref{LP-BV}, we compute the Lyapunov exponent and butterfly velocity in ${\cal M}$-theory dual in the presence of ${\cal O}(R^4)$ terms using the pole-skipping phenomenon via subsections \ref{beta0-LP-BV} and \ref{beta-LP-BV} respectively. Summary of this paper is given in \ref{conclusion}. There are three appendices. Linearized equations of motion at ${\cal O}(\beta^0)$ and ${\cal O}(\beta)$ are given in \ref{EOMs-beta0} and \ref{EOMs-beta-ap} of appendix \ref{EOMs}. The coefficients appearing in linearized equations of motion at ${\cal O}(\beta^0)$ and ${\cal O}(\beta)$ are given in appendices \ref{yi-si} and \ref{pi-zi-vi} respectively.

\section{Review}
\label{review}

In this section, via two sub-sections, we briefly review the ${\cal M}$-theory uplift inclusive of ${\cal O}(R^4)$ terms of the type IIB string theory dual of thermal QCD-like theories (subsection \ref{MQGP}), and the pole-skipping phenomenon leading up to the Lyapunov exponent and butterfly velocity (subsection \ref{lambdaL_vb}).

\subsection{${\cal M}$ Theory Uplift in the Presence of ${\cal O}(R^4)$ Terms}
\label{MQGP}
\begin{table}[h]
\begin{center}
\begin{tabular}{|c|c|c|}\hline
&&\\
S. No. & Branes & World Volume \\
&&\\ \hline
&&\\
1. & $N\ D3$ & $\mathbb{R}^{1,3}(t,x^{1,2,3}) \times \{r=0\}$ \\
&&\\  \hline
&&\\
2. & $M\ D5$ & $\mathbb{R}^{1,3}(t,x^{1,2,3}) \times \{r=0\} \times S^2(\theta_1,\phi_1) \times {\rm NP}_{S^2_a(\theta_2,\phi_2)}$ \\
&&\\  \hline
&&\\
3. & $M\ \overline{D5}$ & $\mathbb{R}^{1,3}(t,x^{1,2,3}) \times \{r=0\}  \times S^2(\theta_1,\phi_1) \times {\rm SP}_{S^2_a(\theta_2,\phi_2)}$ \\
&&\\  \hline
&&\\
4. & $N_f\ D7$ & $\mathbb{R}^{1,3}(t,x^{1,2,3}) \times \mathbb{R}_+(r\in[|\mu_{\rm Ouyang}|^{\frac{2}{3}},r_{\rm UV}])  \times S^3(\theta_1,\phi_1,\psi) \times {\rm NP}_{S^2_a(\theta_2,\phi_2)}$ \\
&&\\  \hline
&&\\
5. & $N_f\ \overline{D7}$ & $\mathbb{R}^{1,3}(t,x^{1,2,3}) \times \mathbb{R}_+(r\in[{\cal R}_{D5/\overline{D5}}-\epsilon,r_{\rm UV}]) \times S^3(\theta_1,\phi_1,\psi) \times {\rm SP}_{S^2_a(\theta_2,\phi_2)}$ \\
&&\\  \hline
\end{tabular}
\end{center}
\caption{Branes configuration of type IIB string dual \cite{metrics}.}
\label{brane-setup}
\end{table}
The equivalence class of theories that are UV conformal, IR confining, and have the ``quarks'' transforming in the fundamental representation of the color and flavor groups are referred to as thermal QCD-like theories. The authors in \cite{metrics} constructed the UV-complete type IIB string dual associated with these large-$N$ thermal QCD-like theories and ${\cal M}$-theory uplift of \cite{metrics} was constructed in the absence and presence of ${\cal O}(R^4)$ terms in \cite{MQGP} and \cite{OR4} respectively. Let us discuss them briefely here.

\begin{itemize}

\item {\bf Brane picture of \cite{metrics}}: {The table \ref{brane-setup} provides a summary of the brane construction of the type IIB string dual; branes are embedded in type IIB string dual of large $N$ thermal QCD at infinite 't Hooft coupling \cite{metrics}}. The brane picture is comprised of $N$ space-time filling $D3$-branes which are located {at} the tip of a {six dimensional} warped resolved conifold, $M$ space-time filling $D5$ branes at the same tip of the above conifold encircling the vanishing crushed $S^2$ as well as on the North Pole (NP) belonging to the resolved squashed $S^2$ of the radius $a$ (resolution parameter), and space-time filling $\overline{D5}$-branes on the tip of the conifold wrapping the previously mentioned vanishing crushed $S^2(\theta_1,\phi_1)$ and located at the South Pole (SP) associated with the resolved squashed $S^2(\theta_2,\phi_2)$. Moreover, $N_f$ space-time filling flavor $D7$-branes encircle the vanishing sqaushed $S^3(\theta_1,\phi_1,\psi)$ and are located at the North Pole associated with the squashed resolved $S^2(\theta_2,\phi_2)$, diving into the IR to the extent $|\mu_{\rm Ouyang}|^{\frac{2}{3}}$, in which $|\mu_{\rm Ouyang}|$ {corresponds to the Ouyang embedding parameter's modulus  associated with the flavor $D7$-branes}: 
\begin{equation}
\label{Ouyang-definition}
\left(9 a^2 r^4 + r^6\right)^{1/4}e^{\frac{i}{2}(\psi - \phi_1-\phi_2)}\sin\frac{\theta_1}{2} \sin\frac{\theta_2}{2}=\mu_{\rm Ouyang}.
\end{equation}
There are also the same numbers of $\overline{D7}$ on the South Pole associated with the blown-up squashed $S^2(\theta_2,\phi_2)$ and are wrapping a vanishing squashed $S^3(\theta_1,\phi_1,\psi)$. UV conformality is ensured by the same number of $D5/D7$-branes and $\overline{D5}/\overline{D7}$-branes in the UV. A flavor gauge group $SU(N_f) \times SU(N_f)$ in the UV is implied by the existence of $N_f$ flavor $D7$ and $\overline{D7}$-branes. This gauge group breaks down to $SU(N_f)$ because there are no $\overline{D7}$-branes in the IR.  This brane configuration is an equivalent to chiral symmetry breaking. 

\item {\bf Bulk picture of \cite{metrics}}: Deforming the vanishing squashed $S^3$ in the conifold causes IR confinement in the gravity dual. Given our interest in QCD at finite temperatures, this can be achieved through thermal ($T<T_c$) and black hole ($T>T_c$) backgrounds on the gravity dual side. The conifold additionally needs to have a $S^2$-blow-up/resolution (with radius/resolution parameter $a$) as a result of the finite temperature as well as finite separation between $D5$ and $\overline{D5}$-branes on the brane side. Moreover, the back-reaction effect is included in the ten-dimensional warp factor and fluxes. Consequently, we deduce that in the large-$N$ limit, the string dual associated with thermal QCD-like theories includes a warped resolved deformed conifold. The type IIB string dual belonging to \cite{metrics} has an additional benefit - within the intermediate-$N$ MQGP limit\footnote{The MQGP limit originally proposed in \cite{MQGP} was in fact the 
very-large-$N$ MQGP limit with $N\gg1, \frac{\left(g_s M^2\right)^{m_1}\left(g_s N_f\right)^{m_2}}{N}\ll1,\ m_{1,2}\in\mathbb{Z}^+\cup\{0\}$.} \cite{MQGP}, \cite{ACMS},
\begin{eqnarray}
\label{MQGP_limit}
& & g_s\equiv\frac{1}{{\cal O}(1)},\nonumber\\
& &  M, N_f \equiv {\cal O}(1), \nonumber
\end{eqnarray}
\begin{eqnarray}
& & N>1:\  \frac{\left(g_s M^2\right)^{m_1}\left(g_s N_f\right)^{m_2}}{N}<1,\ m_{1,2}\in\mathbb{Z}^+\cup\{0\},
\end{eqnarray}
 the number of the colors $N_c$ becomes equal to $M$ in the IR towards the completion of a Seiberg-like duality cascade. 
It was observed in \cite{ACMS} that for the values of $g_s, M, N_f$ in Table \tcb{2}, 
\begin{table}[h]
\label{Parameters-real-QCD}
\begin{center}
\begin{tabular}{|c|c|c|c|} \hline
S. No. & Parameterc & Value chosen consistent with (\ref{MQGP_limit}) & Physics reason \\ \hline
1. & $g_s$ & 0.1 & QCD fine structure constant \\ 
&&& at EW scale \\ \hline
2. & $M$ & 3 & Number of colors after a \\ 
&&& Seiberg-like duality cascade \\
&&& to match real QCD \\ \hline
3. & $N_f$ & 2 or 3 & Number of light quarks in real QCD \\ \hline
\end{tabular}
\end{center}
\caption{QCD-Motivated values of $g_s, M, N_f$}
\end{table}
  $N=100\pm{\cal O}(1)$ is picked out to obtain explicitly  Contact 3-Structures and their associated transverse $SU(3)$ structures; a different set of values of $(g_s, M, N_f)$ will pick out another intermediate $N$.  When dealing with the embedding of the flavor $D7$-branes in the vanishing-Ouyang-modulus limit ($|\mu_{\rm Ouyang}|\ll1$ in (\ref{Ouyang-definition})), $N_f =$ 2 or 3, correspond to the lightest quark flavors \cite{Vikas+Gopal+Aalok}.

Ten dimensional type IIB supergravity solution of \cite{metrics} has the following form:
\begin{equation}
\label{metric}
ds^2 = \frac{1}{\sqrt{h}}
\left(-g_1 dt^2+dx_1^2+dx_2^2+dx_3^2\right)+\sqrt{h}\biggl[g_2^{-1}dr^2+r^2 d{\cal M}_5^2\biggr],
\end{equation}
 where $h$ is the 10D warp factor defined later, $g_i$'s are the black hole functions and are given as:
$ g_{1,2}(r,\theta_1,\theta_2)= 1-\frac{r_h^4}{r^4} + {\cal O}\left(\frac{g_sM^2}{N}\right)$
with $r_h$ being the black hole horizon, and the origin of ($\theta_1, \theta_2$) is due to
 ${\cal O}\left(\frac{g_sM^2}{N}\right)$ corrections. 
Five dimensional compact metric appearing in (\ref{metric}) is given by:
\begin{eqnarray}
\label{RWDC}
& & d{\cal M}_5^2 =  h_1 (d\psi + {\rm cos}~\theta_1~d\phi_1 + {\rm cos}~\theta_2~d\phi_2)^2 +
h_2 (d\theta_1^2 + {\rm sin}^2 \theta_1 ~d\phi_1^2) +   \nonumber\\
&&  + h_4 (h_3 d\theta_2^2 + {\rm sin}^2 \theta_2 ~d\phi_2^2) + h_5~{\rm cos}~\psi \left(d\theta_1 d\theta_2 -
{\rm sin}~\theta_1 {\rm sin}~\theta_2 d\phi_1 d\phi_2\right) + \nonumber\\
&&  + h_5 ~{\rm sin}~\psi \left({\rm sin}~\theta_1~d\theta_2 d\phi_1 +
{\rm sin}~\theta_2~d\theta_1 d\phi_2\right),
\end{eqnarray}
$r> a, h_5\sim\frac{({\rm deformation\ parameter})^2}{r^3}\ll \frac{a^2}{r^2} \forall r \gg({\rm deformation\ parameter})^{\frac{2}{3}}$.  The $h_i$'s appearing in the compact five dimensional metric (\ref{RWDC}) are given as:
\begin{eqnarray}
\label{h_i}
& & \hskip -0.45in h_1 = \frac{1}{9} + {\cal O}\left(\frac{g_sM^2}{N}\right),\  h_2 = \frac{1}{6} + {\cal O}\left(\frac{g_sM^2}{N}\right),\ h_4 = h_2 + \frac{a^2}{r^2},\nonumber\\
& & h_3 = 1 + {\cal O}\left(\frac{g_sM^2}{N}\right),\ h_5\neq0,\
\end{eqnarray}
Equations (\ref{RWDC}) and (\ref{h_i}) are implying that we have a non-extremal resolved warped deformed conifold which invlove an $S^2$-blowup (as $h_4 - h_2 = \frac{a^2}{r^2}$), an $S^3$-blowup (as $h_5\neq0$) and squashing of an $S^2$ (as $h_3$ is not strictly unity). The geometry of the horizon is warped squashed $S^2\times S^3$ and in the deep IR, we get a warped squashed $S^2(a)\times S^3(\epsilon)$ where $\epsilon$ is the deformation parameter. In the IR, ten dimensional warp factor which includes backreaction is given by: 
\begin{eqnarray}
\label{h-def}
&& \hskip -0.45in h(r, \theta_{1,2}) =\frac{L^4}{r^4}\Bigg[1+\frac{3g_sM_{\rm eff}^2}{2\pi N}{\rm log}r\left\{1+\frac{3g_sN^{\rm eff}_f}{2\pi}\left({\rm
log}r+\frac{1}{2}\right)+\frac{g_sN^{\rm eff}_f}{4\pi}{\rm log}\left({\rm sin}\frac{\theta_1}{2}
{\rm sin}\frac{\theta_2}{2}\right)\right\}\Biggr],
\end{eqnarray}
whereas $M_{\rm eff}/N_f^{\rm eff}$ are not always equal to $M/N_f$; up to ${\cal O}\left(\frac{g_sM^2}{N}\right)$ we assume that they are equal.


  In the infrared(IR),  up to ${\cal O}(g_s N_f)$ \cite{metrics}:
\begin{eqnarray}
\label{three-form-fluxes}
& & \hskip -0.4in (a)\ {\widetilde F}_3  =  2M { A_1} \left(1 + \frac{3g_sN_f}{2\pi}~{\rm log}~r\right) ~e_\psi \wedge
\frac{1}{2}\left({\rm sin}~\theta_1~ d\theta_1 \wedge d\phi_1-{ B_1}~{\rm sin}~\theta_2~ d\theta_2 \wedge
d\phi_2\right)\nonumber\\
&& \hskip -0.3in -\frac{3g_s MN_f}{4\pi} { A_2}~\frac{dr}{r}\wedge e_\psi \wedge \left({\rm cot}~\frac{\theta_2}{2}~{\rm sin}~\theta_2 ~d\phi_2
- { B_2}~ {\rm cot}~\frac{\theta_1}{2}~{\rm sin}~\theta_1 ~d\phi_1\right)\nonumber \\
&& \hskip -0.3in -\frac{3g_s MN_f}{8\pi}{ A_3} ~{\rm sin}~\theta_1 ~{\rm sin}~\theta_2 \left(
{\rm cot}~\frac{\theta_2}{2}~d\theta_1 +
{ B_3}~ {\rm cot}~\frac{\theta_1}{2}~d\theta_2\right)\wedge d\phi_1 \wedge d\phi_2, \nonumber\\
& & \hskip -0.4in (b)\ H_3 =  {6g_s { A_4} M}\Biggl(1+\frac{9g_s N_f}{4\pi}~{\rm log}~r+\frac{g_s N_f}{2\pi}
~{\rm log}~{\rm sin}\frac{\theta_1}{2}~
{\rm sin}\frac{\theta_2}{2}\Biggr)\frac{dr}{r}\nonumber \\
&& \hskip -0.3in \wedge \frac{1}{2}\Biggl({\rm sin}~\theta_1~ d\theta_1 \wedge d\phi_1
- { B_4}~{\rm sin}~\theta_2~ d\theta_2 \wedge d\phi_2\Biggr)
+ \frac{3g^2_s M N_f}{8\pi} { A_5} \Biggl(\frac{dr}{r}\wedge e_\psi -\frac{1}{2}de_\psi \Biggr)\nonumber  \\
&& \hskip -0.4in  \wedge \Biggl({\rm cot}~\frac{\theta_2}{2}~d\theta_2
-{ B_5}~{\rm cot}~\frac{\theta_1}{2} ~d\theta_1\Biggr), \nonumber\\
& &  \hskip -0.4in (c)\  F_5 = \frac{1}{g_s}\left[d^4x\wedge d h^{-1}+*\left(d^4x\wedge d h^{-1}\right)\right],\nonumber\\
\end{eqnarray}
where $e_\psi=d\psi +\cos \theta_1 d\phi_1+\cos \theta_2 d\phi_2$; $A_i,B_i$ are the asymmetry factors and are given as:\\ $ A_i=1 +{\cal O}\left(\frac{a^2}{r^2}\ {\rm or}\ \frac{a^2\log r}{r}\ {\rm or}\ \frac{a^2\log r}{r^2}\right) + {\cal O}\left(\frac{{\rm deformation\ parameter }^2}{r^3}\right),$\\ $  B_i = 1 + {\cal O}\left(\frac{a^2\log r}{r}\ {\rm or}\ \frac{a^2\log r}{r^2}\ {\rm or}\ \frac{a^2\log r}{r^3}\right)+{\cal O}\left(\frac{({\rm deformation\ parameter})^2}{r^3}\right)$.

\item {\bf Color-Flavor Length Scale's Enhancement}: The incorporation of terms higher order in $g_s N_f$ in the RR and NS-NS three-form fluxes (\ref{three-form-fluxes}), as well as the next-to-leading order terms in $N$ within the metric (\ref{h_i}), results in an Infra-Red (IR) color-flavor enhancement associated with the length scale in the MQGP limit (\ref{MQGP_limit}) with respect to a Planckian length scale in the Klebanov-Strassler(KS)'s model {\cite{Klebanov:2000hb} even} for ${\cal O}(1)$ $M$. This ensures the suppression of stringy and quantum corrections. This was partly addressed in \cite{NPB}. {We summarize the argument below, and also extend the same to argue suppression of loop/quantum corrections. 

\begin{itemize}
\item {\bf Suppression of stringy/$\alpha^\prime$ corrections}:
We can rewrite the ten-dimensional warp factor (\ref{h-def}) near (\ref{alpha_theta_12})/(\ref{alpha_theta_12_prime}), wherein using (\ref{rh-estimate}) we can ignore the angular part (as near (\ref{alpha_theta_12})/(\ref{alpha_theta_12_prime}), $\left|\log\left(\sin\left(\frac{\theta_1}{2}\right)\sin\left(\frac{\theta_2}{2}\right)\right)\right|\sim\log N<|\log r|\sim N^{1/3}$):
\begin{equation}
\label{h10d}
h = \frac{4\pi g_s}{r^4}\Biggl[N_{\rm eff}(r) + \frac{9 g_s M^2_{\rm eff} g_s N_f^{\rm eff}}{2\left(2\pi\right)^2}\log r \Biggr],
\end{equation}
where \cite{metrics}
\begin{eqnarray}
\label{NeffMeffNfeff}
& & N_{\rm eff}(r) = \frac{F_5 + B_2\wedge F_3}{\rm Vol(resolved\ warped\ deformed\ conifold\ base)}\nonumber\\
& &  = N\left[ 1 + \frac{3 g_s M_{\rm eff}^2}{2\pi N}\left(\log r + \frac{3 g_s N_f^{\rm eff}}{2\pi}\left(\log r\right)^2\right)\right],\nonumber\\
& & M_{\rm eff}(r) = \int_{S^3\ {\rm dual\ to}\ e_\psi\wedge(\Omega_{11} - B_1\Omega_{22})}\tilde{F}_3 \nonumber\\
& & = M + \frac{3g_s N_f M}{2\pi}\log r + \sum_{m\geq1}\sum_{n\geq1} N_f^m M^n f_{mn}(r),\nonumber\\
& & N^{\rm eff}_f(r) = \frac{4\pi C_0}{\left(\psi - \phi_1 - \phi_2\right)} = N_f + \sum_{m\geq1}\sum_{n\geq0} N_f^m M^n g_{mn}(r),
\end{eqnarray}
where $e_\psi = d\psi + \cos\theta_1 d\phi_1 + \cos\theta_2 d\phi_2; \Omega_{ii} = \sin\theta_i d\theta_i\wedge d\phi_i, i=1, 2$. Typically, $f_{mn}(r)$ and $g_{mn}(r)$ are proportional to positive integral powers of $\log r$ in the IR. Higher order in $g_s N_f$ in (\ref{three-form-fluxes}) and the next-to-leading order terms in (\ref{h_i}) are responsible for the origin of terms appearing in double summation in $M_{\rm eff}$ in (\ref{NeffMeffNfeff}). Under Seiberg duality which is effected via $r\rightarrow r e^{-\frac{2\pi}{3g_s M_{\rm eff}}}$ \cite{Ouyang}, $N_{\rm eff}\rightarrow N_{\rm eff} - M_{\rm eff} + N_f^{\rm eff}$.  Hence, after a Seiberg-like duality cascade, one expects an $r=r_0: N_{\rm eff}(r_0)=0$. Therefore, the length scale of (\ref{metric}) in the IR is given as\footnote{Using $L=\left(4\pi g_s N_{\rm eff}\right)^{\frac{1}{4}}$ and $h \sim \frac{L^4}{r^4}$.}:
\begin{eqnarray}
\label{length-IR}
& & \hskip -0.4in L_{(\ref{metric})}\sim N_f^{\frac{3}{4}}\sqrt{\left(\sum_{m\geq0}\sum_{n\geq0}N_f^mM^nf_{mn}(r_0)\right)}\left(\sum_{l\geq0}\sum_{p\geq0}N_f^lM^p g_{lp}(r_0)\right)^{\frac{1}{4}}\left(\log r_0\right)^{\frac{1}{4}} L_{\rm KS},
\end{eqnarray}
\noindent where $L_{\rm KS}=\left({g_s M}\right)^{\frac{1}{4}}\sqrt{\alpha^\prime}$. The relationship (\ref{length-IR}) along with (\ref{Neff}) - (\ref{rh-estimate}), suggests that the length scale for the MQGP limit (\ref{MQGP_limit}) wherein $M$ and $N_f$ are of ${\cal O}(1)$, has a color-flavor enhancement  in the IR({arising from the logarithmic IR enhancement in the double series summation in (\ref{length-IR})}) as compared to KS. Therefore, within the IR, regardless of $N_c^{\rm IR}=M=3$ and $N_f=2(u/d)+1(s)$, $L\gg L_{\rm KS}(\sim L_{\rm Planck})$ within the MQGP limit (\ref{MQGP_limit}) implies that {the stringy ($\alpha^\prime$) corrections have been suppressed so we can rely on supergravity calculations with sub-dominant contributions arising from the higher derivative (type II $l_s^6/{\cal M}-{\rm theory}\ l_p^6$) corrections. In this paper, this is mirrored by the fact that the analogs of ``Lyapunov exponent'' and ``butterfly velocity'' in our setup, receive no ${\cal O}(R^4)$ corrections.} 

\item {\bf Absence of perturbative quantum/loop corrections beyond one loop at ${\cal O}(R^4)$}: We will now summarize the argument of \cite{Green and Gutperle} that an $SL(2,\mathbb{Z})$ completion of the effective $R^4$ action,  forbids perturbative corrections to this term beyond one loop in the zero-instanton sector. One can show that  the complete  effective $R^4$ action in the Einstein's frame, can be expressed in the form,
\begin{equation}
\label{modular-completion-i}
S_{R^4} = (\alpha')^{-1} \left[ {\cal A} \zeta(3)\tau_2^{3/2}   + {\cal B}
\tau_2^{-1/2} + {\cal C}
e^{2\pi i  \tau} +\cdots \right] R^4 \equiv  (\alpha')^{-1}
f(\tau,\bar{\tau})R^4,
\end{equation}
where ${\cal A}, {\cal B}$ are known numerical constants, $R^4$ denotes the contractions that figure in (\ref{D=11_O(l_p^6)}) - (\ref{J0+E8-definitions}), and $\cdots$ indicates possible perturbative and nonperturbative corrections to the coefficient of $R^4$; ${\cal C}$ can depend on $\tau = C_{\rm RR} + i e^{-\phi} \equiv \tau_1 + i \tau_2$ (type IIB) and $\bar \tau$. The first term in (\ref{modular-completion-i}) represents the $(\alpha^\prime)^3$ tree-level contribution and the second the one-loop relative to the former.

However, (\ref{modular-completion-i}) must be invariant under
$SL(2,\mathbb{Z})$ transformations:  $\tau\to (a\tau+b)(c\tau+d)^{-1}$ ($a,b,c,d\in\mathbb{Z}: ad-bc=1$), which provides very strong constraints on its structure. $R^4$ factor being already $SL(2,\mathbb{Z})$-invariant, $f(\tau,\bar \tau)$ must also be separately invariant, which hence implies a sum over  all instantons and anti-instantons.  These strong constraints are satisfied by the following simple function proposed by \cite{Green and Gutperle},
\begin{eqnarray}
\label{modular-completion-ii}
& & \hskip -0.5in  f(\tau,\bar{\tau}) = \sum_{\mathbb{Z}\ni(p,n )\neq
(0,0)}{\tau_2^{3/2}\over |p+n\tau|^3}\nonumber\\
& & \hskip -0.5in = 2\zeta(3)\tau_2^{3/2} +
  {2\pi^2\over
3} \tau_2^{-1/2}  +8 \pi  \tau_2^{ 1/2}  \sum_{m \ne 0 n\ge 1   }
\left|{m\over n}\right|
e^{2\pi i mn\tau_1} K_1 (2\pi |mn|\tau_2)\nonumber\\
&&  \hskip -0.5in = 2\zeta(3)\tau_2^{3/2} + {2\pi^2\over  3} \tau_2^{-1/2}   \nonumber\\
&&  \hskip -0.5in +4\pi^{3/2}  \sum_{m,n \ge 1} \left({m\over n^3}\right)^{1/2}
(e^{2\pi i mn
\tau} + e^{-2\pi i mn \bar  \tau} ) \left(1 + \sum_{k=1}^\infty  (4\pi mn
\tau_2)^{-k} {\Gamma(  k -1/2)\over \Gamma(- k -1/2) } \right) ,
\end{eqnarray}
where performing a perturbative expansion in $\frac{1}{\tau_2}$ of the non-perturbative instanton contribution of charge $mn$, one uses the asymptotic expansion for $K_1(z)$ for
large $z$  in the last line of (\ref{modular-completion-ii}).

The perturbative terms in (\ref{modular-completion-ii}) terminate after the one-loop term.  The non-perturbative terms have the form of a sum over
single multiply-charged instantons and anti-instantons with action proportional
to $|mn|$. The terms in parenthesis in the last line of (\ref{modular-completion-ii}) represent the infinite sequence of  perturbative corrections around the
instantons of  charge $mn$.   It is hence evident that in the zero-instanton sector, there are no perturbative corrections figuring in the action up to ${\cal O}(R^4)$, i.e., $S_{R^4}$, beyond one loop. Of course, using the aforementioned argument of suppression of stringy/$\alpha^\prime$-corrections, even the one-loop contribution (as it multiplies the $R^4$-contractions) will be suppressed. This will be important in demonstrating the "non-renormalization" of the analogs of the Lyapunov exponent and butterfly velocity later in the paper.

\end{itemize}
}

\item{{\bf Planckian and large-$N$ suppression vs IR enhancement}:  It was shown in \cite{OR4} that in the $\psi=2n\pi,n=0, 1, 2$-coordinate patch  that in the IR: $r = \chi r_h, \chi\equiv {\cal O}(1)$, and up to ${\cal O}(\beta)$:
\begin{equation}
\label{IR-beta-N-suppressed-logrh-rh-neg-exp-enhanced}
f_{MN} \sim \beta\frac{\left(\log {\cal R}_h\right)^{m}}{{\cal R}_h^n N^{\beta_N}},\ m\in\left\{0,1,3\right\},\ n\in\left\{0,2,5,7\right\},\
\beta_N>0,
\end{equation}
where ${\cal R}_h\equiv\frac{r_h}{{\cal R}_{D5/\overline{D5}}}$ and $\beta$ in (\ref{IR-beta-N-suppressed-logrh-rh-neg-exp-enhanced}) refers to Planck length corrections defined in (\ref{beta-def}). Now, $|{\cal R}_h|\ll1$. As estimated in \cite{Bulk-Viscosity-McGill-IIT-Roorkee}, $|\log {\cal R}_h|\sim N^{\frac{1}{3}}$, implying there is a competition between Planckian and large-$N$ suppression and infra-red enhancement arising from $m,n\neq0$ in (\ref{IR-beta-N-suppressed-logrh-rh-neg-exp-enhanced}). Choosing a hierarchy: $\beta\sim e^{-\gamma_\beta N^{\gamma_N}}, \gamma_\beta,\gamma_N>0: \gamma_\beta N^{\gamma_N}>7N^{\frac{1}{3}} + \left(\frac{m}{3} - \beta_N\right)\log N$ (ensuring that the IR-enhancement does not overpower Planckian suppression - we took the ${\cal O}(\beta)$ correction to $G^{\cal M}_{yz}$, which had the largest IR enhancement, to set a lower bound on $\gamma_{\beta,N}$/Planckian suppression). If $\gamma_\beta N^{\gamma_N}\sim7N^{\frac{1}{3}}$, then one will be required to go to a higher order in $\beta$. This hence answers the question, when one can truncate at ${\cal O}(\beta)$.
}

\item {\bf ${\cal M}$-Theory Uplift of \cite{metrics}}: In \cite{OR4}, the ${\cal O}(R^4)$ terms in eleven-dimensional supergravity action have been incorporated in order to look into the intermediate coupling regime of thermal QCD. The type IIA Strominger-Yau-Zaslow (SYZ) mirror of the type IIB setup was initially constructed, and then uplifted to ${\cal M}$-theory.  In order to arrive at the type IIA SYZ mirror associated with type IIB setup, a triple T-duality has been carried out along a local special Lagrangian (sLag) $T^3(x,y,z)$, where $(x,y,z)$ are actually the toroidal equivalents of $(\phi_1,\phi_2,\psi)$. This can be recognized with some the $T^2$-invariant sLag of \cite{M.Ionel and M.Min-OO (2008)} in large-complex structure limit where the large-complex structure limit has been achieved by ensuring the base ${\cal B}(r,\theta_1,\theta_2)$ (of a $T^3(\phi_1,\phi_2,\psi)$-fibration over ${\cal B}(r,\theta_1,\theta_2)$)to be  large \cite{MQGP}, \cite{NPB}. T-dualization of all the color and flavor $D$-branes of type IIB setup results in color and flavor $D6$-branes. Taking the ${\cal M}$-theory uplift to be valid for all angles $\theta_{1,2}, \psi$, is analogous to \cite{SYZ-free-delocalization} (which includes $D5$-branes wrapping a resolved squashed $S^2$). This uplift maps to a genuine $G_2$ structure that satisfy the {equations of motion} regardless of whether the delocalization\footnote{{This is explained in \cite{MQGP}, \cite{SYZ-free-delocalization}. The basic idea is given that a resolved warped {\it deformed} conifold that figures in the type IIB gravity dual (See Fig. \tcb{1}) does not possess an isometry along $\psi$, to construct the type IIA SYZ mirror and its ${\cal M}$-theory uplift, to begin with, one works in the delocalized limit $\psi = \langle\psi\rangle$ as well as replace $S^2(\theta_{1,2},\phi_{1,2})$ by $T^2(\theta_{1,2},x/y)$ via (\ref{xyz-defs}). Then, analogous to \cite{SYZ-free-delocalization}, one can show that the ${\cal M}$-theory uplift and hence its type IIA descendant, are free of this delocalization.}} is removed.  Furthermore, it is clear that one will need to work near tiny values of $\theta_{1,2}$ while working in the previously described vanishing-Ouyang-embedding's-modulus limit (basically restricting to the first-generation quarks[+s quark]) from (\ref{Ouyang-definition}). For instance, we do work in the vicinity of 
   \begin{eqnarray}
\label{alpha_theta_12}
& & (\theta_1, \theta_2) = \left(\frac{\alpha_{\theta_1}}{N^{1/5}}, \frac{\alpha_{\theta_2}}{N^{3/10}}\right),\ \ \ \ \ \ \ \alpha_{\theta_{1,2}}\equiv{\cal O}(1).
\end{eqnarray}
Additionally, the somewhat differing powers of $N$ for the delocalized $\theta_{1,2}$ serve as a reminder that the resolved $S^2(\theta_2,\phi_2)$ and vanishing $S^2(\theta_1,\phi_1)$ in the pair of squashed $S^2$s do not sit at the ``same ground''.  From the on-shell action's perspective, by substituting $N^{1/5}\sin\theta_1$ or $N^{3/10}\sin\theta_2$ for the ${\cal O}(1)$ delocalization parameters $\alpha_{\theta_{1,2}}$, respectively, the outcomes up to ${\cal O}(\frac{1}{N})$ become independent of the delocalization (as shown in \cite{OR4}). Afterwards, one can select an alternative delocalization by substituting $\sin\theta_{1,2}$ with
\begin{eqnarray}
\label{alpha_theta_12_prime} 
& & \left(\frac{\tilde{\alpha}_{\theta_1}}{N^{\gamma_{\theta_1}}}, \frac{\tilde{\alpha}_{\theta_2}}{N^{\gamma_{\theta_2}}}\right), \gamma_{\theta_1}\neq\frac{1}{5}, \gamma_{\theta_2}\neq\frac{3}{10},
\ \ \ \ \ \ \ \ \tilde{\alpha}_{\theta_{1,2}}\equiv{\cal O}(1).
\end{eqnarray}
We define the ${\cal M}$-theory uplift (finite-but-large-$N$/intermediate coupling) metric of \cite{metrics} as follows \cite{MQGP}, \cite{OR4}, 
\begin{eqnarray}
\label{TypeIIA-from-M-theory-Witten-prescription-T>Tc}
\hskip -0.1in ds_{11}^2 & = & e^{-\frac{2\phi^{\rm IIA}}{3}}\Biggl[\frac{1}{\sqrt{h(r,\theta_{1,2})}}\left(-g(r) dt^2 + \left(dx^1\right)^2 +  \left(dx^2\right)^2 +\left(dx^3\right)^2 \right)
\nonumber\\
& & \hskip -0.1in+ \sqrt{h(r,\theta_{1,2})}\left(\frac{dr^2}{g(r)} + ds^2_{\rm IIA}(r,\theta_{1,2},\phi_{1,2},\psi)\right)
\Biggr] + e^{\frac{4\phi^{\rm IIA}}{3}}\left(dx^{11} + A_{\rm IIA}^{F_1^{\rm IIB} + F_3^{\rm IIB} + F_5^{\rm IIB}}\right)^2,
\end{eqnarray} 
where $A_{\rm IIA}^{F^{\rm IIB}_{i=1,3,5}}$ (type IIA RR 1-forms) are generated from type IIB  fluxes ($F_{1,3,5}^{\rm IIB}$) by the application of SYZ mirror to type IIB string dual \cite{metrics}, type IIA dilaton profile is $\phi^{\rm IIA}$, and $g(r) = 1 - \frac{r_h^4}{r^4}$. The thermal gravitational dual for low temperatures QCD, denoted as $T<T_c$, is defined as follows: 
\begin{eqnarray}
\label{TypeIIA-from-M-theory-Witten-prescription-T<Tc}
\hskip -0.1in ds_{11}^2 & = & e^{-\frac{2\phi^{\rm IIA}}{3}}\Biggl[\frac{1}{\sqrt{h(r,\theta_{1,2})}}\left(-dt^2 + \left(dx^1\right)^2 +  \left(dx^2\right)^2 + \tilde{g}(r)\left(dx^3\right)^2 \right)
\nonumber\\
& & \hskip -0.1in+ \sqrt{h(r,\theta_{1,2})}\left(\frac{dr^2}{\tilde{g}(r)} + ds^2_{\rm IIA}(r,\theta_{1,2},\phi_{1,2},\psi)\right)
\Biggr] + e^{\frac{4\phi^{\rm IIA}}{3}}\left(dx^{11} + A_{\rm IIA}^{F_1^{\rm IIB} + F_3^{\rm IIB} + F_5^{\rm IIB}}\right)^2,\nonumber\\
& & 
\end{eqnarray}
where $\tilde{g}(r) = 1 - \frac{r_0^4}{r^4}$. Observing that $t\rightarrow x^3,\ x^3\rightarrow t$ in (\ref{TypeIIA-from-M-theory-Witten-prescription-T>Tc}) followed by a Double Wick rotation in the new $x^3, t$ coordinates yields (\ref{TypeIIA-from-M-theory-Witten-prescription-T<Tc}); the ten-dimensional warp factor is denoted by $h(r,\theta_{1,2})$ \cite{metrics, MQGP}. This may also be expressed as follows: $-g_{tt}^{\rm BH}(r_h\rightarrow r_0) = g_{x^3x^3}\ ^{\rm Thermal}(r_0),$ $ g_{x^3x^3}^{\rm BH}(r_h\rightarrow r_0) = -g_{tt}\ ^{\rm Themal}(r_0)$ in the outcomes of \cite{VA-Glueball-decay}, \cite{OR4} (Refer to \cite{Kruczenski et al-2003} regarding Euclidean/black $D4$-branes in type IIA). We are going to take the spatial component of the solitonic $M3$ brane [it, locally, might be seen as a homologous sum of $S^2_{\rm squashed}$ wrapping across a solitonic $M5$-brane \cite{DM-transport-2014}] in (\ref{TypeIIA-from-M-theory-Witten-prescription-T<Tc}), in which the extremely small $M_{\rm KK}$ is provided through $\frac{2r_0}{ L^2}\left[1 + {\cal O}\left(\frac{g_sM^2}{N}\right)\right]$ and their world volume provided by $\mathbb{R}^2(x^{1,2})\times S^1(x^3)$. The period of $S^1(x^3)$ is provided by an extremely big: $\frac{2\pi}{M_{\rm KK}}$, where $L = \left( 4\pi g_s N\right)^{\frac{1}{4}}$ and $r_0$ denote the extremely tiny IR cut-off that defines the thermal background (see also \cite{Armoni et al-2020}). Thus, 4D Physics is recovered by $\lim_{M_{\rm KK}\rightarrow0}\mathbb{R}^2(x^{1,2})\times S^1(x^3) = \mathbb{R}^3(x^{1,2,3})$. As for the thermal background corresponding to $T<T_c$, in (\ref{TypeIIA-from-M-theory-Witten-prescription-T<Tc}), $\tilde{g}(r)$ will be set to unity as the working metric.\par

{
The eleven-dimensional supergravity action used in \cite{OR4} that includes ${\cal O}(R^4)$ terms is the following:
\begin{eqnarray}
\label{D=11_O(l_p^6)}
& & \hskip -0.5in S = \frac{1}{2\kappa_{11}^2}\Biggl[\int_{M_{11}}\sqrt{g}R + \int_{\partial M_{11}}\sqrt{h}K -\frac{1}{2}\int_{M_{11}}\sqrt{g}G_4^2
-\frac{1}{6}\int_{M_{11}}C_3\wedge G_4\wedge G_4\nonumber\\
& & \hskip -0.5in  + \frac{\left(4\pi\kappa_{11}^2\right)^{\frac{2}{3}}}{{(2\pi)}^4 3^2.2^{13}}\Biggl(\int_{\cal{M}} d^{11}\!x \sqrt{g}\left(J_0-\frac{1}{2}E_8\right) + 3^2.2^{13}\int C_3 \wedge X_8 + \int t_8 t_8 G^2 R^3 + \cdot \cdot \Biggr)\Biggr] - {\cal S}^{\rm ct},\nonumber\\
\end{eqnarray}
where:
\begin{eqnarray}
\label{J0+E8-definitions}
& & \hskip -0.8inJ_0  =3\cdot 2^8 (R^{HMNK}R_{PMNQ}{R_H}^{RSP}{R^Q}_{RSK}+
{1\over 2} R^{HKMN}R_{PQMN}{R_H}^{RSP}{R^Q}_{RSK}),\nonumber\\
& & \hskip -0.8inE_8  ={ 1\over 3!} \epsilon^{ABCM_1 N_1 \dots M_4 N_4}
\epsilon_{ABCM_1' N_1' \dots M_4' N_4' }{R^{M_1'N_1'}}_{M_1 N_1} \dots
{R^{M_4' N_4'}}_{M_4 N_4},,\nonumber\\
& & \hskip -0.8in t_8t_8G^2R^3 = t_8^{M_1...M_8}t^8_{N_1....N_8}G_{M_1}\ ^{N_1 PQ}G_{M_2}\ ^{N_2}_{\ \ PQ}R_{M_3M_4}^{\ \ \ \ N_3N_4}R_{M_5M_6}^{\ \ \ \ N_5N_6}R_{M_7M_8}^{\ \ \ \ N_7N_8},
\nonumber\\
& & \hskip -0.8in X_8 = {1 \over 192} \left( {\rm tr}\ R^4 -
{1\over 4} ({\rm tr}\ R^2)^2\right),\nonumber\\
& & \hskip -0.8in\kappa_{11}^2 = \frac{(2\pi)^8 l_p^{9}}{2}.
\end{eqnarray}
$t_8$ symbol has the following structure \cite{OR4}:
{\footnotesize
\begin{eqnarray}
\label{t_8}
t_8^{N_1\dots N_8}   &=& \frac{1}{16} \big( -  2 \left(   g^{ N_1 N_3  }g^{  N_2  N_4  }g^{ N_5   N_7  }g^{ N_6 N_8  }
 + g^{ N_1 N_5  }g^{ N_2 N_6  }g^{ N_3   N_7  }g^{  N_4   N_8   }
 +  g^{ N_1 N_7  }g^{ N_2 N_8  }g^{ N_3   N_5  }g^{  N_4 N_6   }  \right) \nonumber \\
 & &  +
 8 \left(  g^{  N_2     N_3   }g^{ N_4    N_5  }g^{ N_6    N_7  }g^{ N_8   N_1   }
  +g^{  N_2     N_5   }g^{ N_6    N_3  }g^{ N_4    N_7  }g^{ N_8   N_1   }
  +   g^{  N_2     N_5   }g^{ N_6    N_7  }g^{ N_8    N_3  }g^{ N_4  N_1   }
\right) \nonumber \\
& &  - (N_1 \leftrightarrow  N_2) -( N_3 \leftrightarrow  N_4) - (N_5 \leftrightarrow  N_6) - (N_7 \leftrightarrow  N_8) \big),
\end{eqnarray}
}
where $g^{ N_i N_j}$ with ($i,j=1,2,..,8$) are in the inverse metric components. ${\cal S}^{\rm ct}$ denotes the counter terms required for holographic renormalization of (\ref{D=11_O(l_p^6)}). We found the bulk on-shell eleven dimensional supergravity has the following form \cite{Gopal+Vikas+Aalok}:
{\footnotesize
\begin{equation}
\label{on-shell-D=11-action-up-to-beta}
 S^{\rm on-shell} = -\frac{1}{2}\Biggl[-2S_{\rm EH}^{(0)} + 2 S_{\rm GHY}^{(0)}+ \beta\left(\frac{20}{11}S_{\rm EH} - 2\int_{M_{11}}\sqrt{-g^{(1)}}R^{(0)}
+ 2 S_{\rm GHY} - \frac{2}{11}\int_{M_{11}}\sqrt{-g^{(0)}}g_{(0)}^{MN}\frac{\delta J_0}{\delta g_{(0)}^{MN}}\right)\Biggr],
\end{equation}
}
where superscripts ``$(0)$'' and ``$(1)$'' are denoting ${\cal O}(\beta^0)$ and ${\cal O}(\beta)$ contributions to the various terms appearing in (\ref{on-shell-D=11-action-up-to-beta}). The terms appearing in (\ref{on-shell-D=11-action-up-to-beta}) have the UV divergences of the following form:
\begin{eqnarray}
\label{UV_divergences}
& &\left. \int_{M_{11}}\sqrt{-g}R\right|_{\rm UV-divergent},\ \left.\int_{\partial M_{11}}\sqrt{-h}K\right|_{\rm UV-divergent} \sim r_{\rm UV}^4 \log r_{\rm UV},\nonumber\\
& & \left.\int_{M_{11}} \sqrt{-g}g^{MN}\frac{\delta J_0}{\delta g^{MN}}\right|_{\rm UV-divergent} \sim
\frac{r_{\rm UV}^4}{\log r_{\rm UV}}.
\end{eqnarray}
In \cite{Gopal+Vikas+Aalok}, we showed that above UV divergnces (\ref{UV_divergences}) can be canceled by an appropriate linear combination of the boundary  terms: $\left.\int_{\partial M_{11}}\sqrt{-h}K\right|_{r=r_{\rm UV}}$ and \\$\left.\int_{\partial M_{11}}\sqrt{-h}h^{mn}\frac{\partial J_0}{\partial h^{mn}}\right|_{r=r_{\rm UV}}$}.
\par
The following are the equations of motion associated with three form potential $C$ and metric:
\begin{eqnarray}
\label{eoms}
& & {\rm EOM}_{\rm MN}:\ R_{MN} - \frac{1}{2}g_{MN}{\cal R} - \frac{1}{12}\left(G_{MPQR}G_N^{\ PQR} - \frac{g_{MN}}{8}G_{PQRS}G^{PQRS} \right)\nonumber\\
 & &  = - \beta\left[\frac{g_{MN}}{2}\left( J_0 - \frac{1}{2}E_8\right) + \frac{\delta}{\delta g^{MN}}\left( J_0 - \frac{1}{2}E_8\right)\right],\nonumber\\
& & d*G = \frac{1}{2} G\wedge G +3^22^{13} \left(2\pi\right)^{4}\beta X_8,\nonumber\\
& &
\end{eqnarray}
where \cite{Becker-sisters-O(R^4)}:
\begin{equation}
\label{beta-def}
\beta \equiv \frac{\left(2\pi^2\right)^{\frac{1}{3}}\left(\kappa_{11}^2\right)^{\frac{2}{3}}}{\left(2\pi\right)^43^22^{12}} \sim l_p^6.
\end{equation}
In (\ref{D=11_O(l_p^6)})/(\ref{eoms}), the elven-dimensional Riemann curvature tensor, Ricci tensor, and the Ricci scalar are denoted by the symbols $R_{MNPQ}, R_{MN}, {\cal R}$. Following was the ansatz constructed in order to solve (\ref{eoms}):
\begin{eqnarray}
\label{ansaetze}
& & \hskip -0.8ing_{MN} = g_{MN}^{\beta^0} +\beta g_{MN}^{\beta},\nonumber\\
& & \hskip -0.8inC_{MNP} = C^{(0)}_{MNP} + \beta C_{MNP}^\beta.
\end{eqnarray}
equations of motion corresponding to $C_{MNP}$ may be expressed symbolically as follows\footnote{{In (\ref{deltaC=0consistent}), $\epsilon_{11}\partial C^{\beta^0} \partial C^{\beta}$ is a schematic way of writing $\epsilon^{M_1...M_{11}}\partial_{[M_4}C_{M_5M_6M_7]}^{\beta^0}\partial_{[M_8}C_{M_9M_{10}M_{11}]}^{\beta}$ with $\epsilon_{11}$ denoting $\epsilon^{M_1...M_{11}}$.}}:
\begin{eqnarray}
\label{deltaC=0consistent}
& & \beta \partial\left(\sqrt{-g}\partial C^{\beta}\right) + \beta \partial\left[\left(\sqrt{-g}\right)^{\beta}\partial C^{\beta^0}\right] + \beta\epsilon_{11}\partial C^{\beta^0} \partial C^{\beta} = {\cal O}(\beta^2) \sim 0 [{\rm up\ to}\ {\cal O}(\beta)].
\nonumber\\
& & \end{eqnarray}
$C^{\beta}_{MNP}=0$ has been proven in \cite{OR4} to be a consistent truncation of ${\cal O}(\beta)$ corrections the ${\cal M}$-theory uplift of \cite{MQGP}, \cite{NPB} provided ${\cal C}_{zz} - 2 {\cal C}_{\theta_1z} = 0, |{\cal C}_{\theta_1x}|\ll1$ where $C_{MN}$ are the constants of integration appearing in the solutions to the equations of motion of $h_{MN}$ and the delocalized toroidal coordinates $T^3(x, y, z)$ corresponding to some $(r, \theta_1, \theta_2) = (\langle r\rangle, \langle\theta_1\rangle, \langle\theta_2\rangle)$ are defined as \cite{MQGP}: 
\begin{eqnarray}
\label{xyz-defs}
& & dx = \sqrt{\frac{1}{6} + {\cal O}\left(\frac{g_sM^2}{N}\right)}h^{\frac{1}{4}}\Bigl(\langle r\rangle, \langle \theta_{1,2}\rangle\Bigr)\langle r\rangle
\sin\langle\theta_{1}\rangle d\phi_1,\nonumber\\ 
& & dy = \sqrt{\frac{1}{6} + \frac{a^2}{r^2} + {\cal O}\left(\frac{g_sM^2}{N}\right)}h^{\frac{1}{4}}\Bigl(\langle r\rangle, \langle \theta_{1,2}\rangle\Bigr)
\langle r\rangle\sin\langle\theta_{2}\rangle d\phi_2,\nonumber\\  
& & dz =  \sqrt{\frac{1}{6} + {\cal O}\left(\frac{g_sM^2}{N}\right)}h^{\frac{1}{4}}\Bigl(\langle r\rangle, \langle \theta_{1,2}\rangle\Bigr)\langle r\rangle d\psi,
\end{eqnarray}
where the 10-D warp factor is given in (\ref{h-def}).
 In light of this, ${\cal O}(R^4)$ corrections are applied exclusively to the metric and are defined as:
\begin{eqnarray}
\label{fMN-definitions}
\delta g_{MN} =g^{\beta}_{MN} = g_{MN}^{\beta^0} f_{MN}(r).
\end{eqnarray}
With ${\cal O}(R^4)$ corrections included, the ${\cal M}$ theory metric typically takes a particular form:
\begin{equation}
\label{fMN-def}
g_{MN} = g_{MN}^{\beta^0}\left(1+\beta f_{MN}(r)\right).
\end{equation}
where $f_{MN}(r)$ are given in \cite{OR4}. The metric components (\ref{fMN-def}) are worked out near the $\psi=2n\pi, \ n=0, 1, 2$-coordinate patches wherein $g_{rM}=0, M\neq r$ and the $M_5=(S^1 \times_w R^3) \times R_{>0}$ and the unwarped $\tilde{M}_6(S^1_{\cal M} \times_w T^{\rm NE})$ of $SU(3)$ structure wherein $S^1_{\cal M}$ is the ${\cal M}$-theory circle and $T^{\rm NE}$ is the non-Einsteinian deformation of $T^{1,1}$, decouple.

\item {\bf Conceptual Physics issues miscellanea}:
\begin{itemize}

\item {\it Regime of validity of the top-down holographic model and $\Lambda_{\rm QCD}$}: Naturally, we desires to know what range of scales the novel top-down holographic model is predicted to match the QCD.  The main principle that provides a solution to this query lies in the fact that the range of variation of the radial coordinate on the supergravity dual which corresponds to the energy scale in QCD-like theories can be provided by: $\left\{\left.r\right| N_{\rm eff}(r) = \int_{M_5(\theta_{1,2},\phi_{1,2},\psi)}\left(F_5 + B_2\wedge C_3\right)\approx0\right\}\cap\left\{\left.r\right|M_{\rm eff}\equiv{\cal O}(1)\right\} $ where $M_5(\theta_{1,2},\phi_{1,2},\psi)$ is the base of the non-K\"{a}hler resolved warped deformed conifold. It is possible to prove that $N_{\rm eff}(r)\sim N\left[1 + g_s\frac{g_s M^2(g_s N_f)}{N}\log^3r\right]$ \cite{Bulk-Viscosity} in the IR (wherein $|\log r|\gg1$) in the MQGP limit. From (\ref{NeffMeffNfeff}), estimate: $M\left[1 + \frac{3g_s N_f M \log r}{2\pi}\right]\equiv {\cal O}(1)$, or $\log r =  - \frac{2\pi\left(-{\cal O}(1) + M\right)}{3g_s M N_f}$], and putting in $N_{\rm eff}$ for $N\sim10^2, N_f=3$, results in, $M\sim{\cal O}(1)$ similar to the MQGP limit.
\\
Within the context of the geometrical data associated with our top-down holographic model, we must comprehend the non-perturbative QCD scale, $\Lambda_{\rm QCD}$, in order to grasp the upper bound in energy on the QCD side or $r$ on the gravitational dual side. Recall that the {${SU(M+N)}$} and $SU(N)$ gauge couplings, $g_{SU(M+N)}$ and $g_{SU(N)}$, satisfy \cite{metrics}
$4\pi^2\left(\frac{1}{g_{SU(M+N)}^2} + \frac{1}{g_{SU(N)}^2}\right) e^{\phi^{\rm IIB}(r,\theta_{10},\theta_{20})}= \pi ;\ 4\pi^2\left(\frac{1}{g_{SU(M+N)}^2} - \frac{1}{g_{SU(N)}^2}\right) e^{\phi^{\rm IIB}(r,\theta_{10},\theta_{20})}\\
\sim\frac{1}{2\pi\alpha^\prime}\int_{S^2(\theta_1,\phi_1)}B^{\rm IIB}\sim \left.g_s M_{\rm eff}(r)N_f^{\rm eff}(r)\log r\right|_{r\in\rm IR-UV\ interpolating\ region}$. Keep in mind that $\frac{r}{{\cal R}_{D5/\overline{D5}}}, \frac{a}{{\cal R}_{D5/\overline{D5}}}$ are actually present anywhere as $r, a$ where $\sqrt{3}a$ has been taken on to be ${\cal R}_{D5/\overline{D5}}$, where $a$ represents the resolution parameter corresponding to the blown-up $S^2$ \cite{Bulk-Viscosity}. Therefore, ``$\Lambda_{\rm QCD} = {\cal R}_{D5/\overline{D5}}\sim\sqrt{3}a$'' since the previously mentioned gauge couplings become extremely high near $r={\cal R}_{D5/\overline{D5}}$, signaling the beginning of non-perturbative QCD.

\item
{\it Hierarchy of scales in the gravitational dual}: The boundary of the far infrared is indicated by an infrared cut-off, $r_0$\footnote{There exists an additional scale which is potentially present in the type IIB holographic dual. This is given by the $D7$-branes being embedded via the Ouyang's embedding (\ref{Ouyang-definition}), where the radial separation between the color $D3$-branes and the ``deepest'' embedded flavor $D7$-branes into the IR is measured by the modulus corresponding to the Ouyang embedding parameter $|\mu|^{\frac{2}{3}}$. This quantity corresponds to the mass of the lightest quark. Consequently, the hierarchy might be refined by assuming that the IR is $r\in[r_0, |\mu|^{\frac{2}{3}}]$, and that the IR-UV interpolating region is $r\in[|\mu|^{\frac{2}{3}}, {\cal R}_{D5/\overline{D5}}]$, and $r\in[{\cal R}_{D5/\overline{D5}}, r_{\rm UV}]$ is the UV region.}.  The separation between $D5-\overline{D5}$ branes ${\cal R}_{D5/\overline{D5}}$ serves as an additional scale, such that at energies greater than ${\cal R}_{D5/\overline{D5}}$, the $D5-\overline{D5}$ strings ends up massive, and the resulting gauge group turns into the Klebanov-Witten-like UV-conformal $SU(M+N)\times SU(M+N)$. At energies below ${\cal R}_{D5/\overline{D5}}$, the gauge group is reduced to the Klebanov-Strassler-like $SU(M+N)\times SU(N)$; there's is additionally a $U(1)^M$ to support massless strings beginning and ending upon the same $\overline{D5}$-brane. Since one is usually carrying out calculations in the ``near-horizon'' limit, where a ten-dimensional warp factor $h(r,\theta_{1,2})\sim \frac{L^4}{r^4}\left(1 + f\left(r; N, M, N_f\right)\right)$, as well in the UV (i.e., one throws the 1 that figures, for example, in the warp factor appearing in the Klebanov-Witten supergravity dual: $h = 1 + \frac{L^4}{r^4}$), where $L\equiv\left(4\pi g_s N\alpha^\prime\right)^{\frac{1}{4}}$, therefore the UV cut-off $r_{\rm UV}\stackrel{<}{\sim}L$. Therefore, the gravitational equivalent of $\Lambda_{\rm QCD}$ and $r_{\rm UV}$ remain separate within our top-down holographic dual.

\end{itemize}
\item {\bf Hierarchy $t_8^2G^2R^3<E_8<J_0$}: The leading-order-in-$N$ contribution to $J_0$ could be confirmed as follows \cite{OR4}:
\begin{eqnarray}
\label{J0-1}
J_0 = \frac{1}{2} R^{\phi_2r \theta_1r}  R_{r \psi\theta_1r}  R_{\phi_2}^{\ \ r \phi_1r}  R^{\psi}_{r \phi_1r} -R^{\phi_2r \theta_1r}
   R_{\phi_1r\theta_1r}  R_{\phi_2}^{\ \ r \phi_1r}  R^{\theta_1}_{\ \ r\theta_1r} ,
\end{eqnarray}
whereas, for example, close to (\ref{alpha_theta_12}),
\begin{eqnarray}
\label{J0-2}
& & \hskip -0.5in J_0 \sim \frac{ a^{10} \left(\frac{1}{{\log N}}\right)^{8/3} M \left(\frac{1}{N}\right)^{7/4} \left(9 a^2+r^2\right) \left(r^4-{r_h}^4\right)
   \log (r) \Sigma_1\sum_{n_1,n_2,n_3:n_1+n_2+n_3=6}\kappa_{n_1, n_2, n_3}a^{2n_1}r^{2n_2}r_h^{2n_3}
  }{ \sqrt{{g_s}} {N_f}^{5/3} r^{16} \left(3 a^2-r^2\right)^8 \left(6 a^2+r^2\right)^7 \alpha _{\theta _2}^3}\nonumber\\
& &  \sim \left(\frac{1}{N}\right)^{7/4},
\end{eqnarray}
where $\Sigma_1$ is defined in (\ref{Sigma_1-3-def}), and $ \sum_{n_1,n_2,n_3:n_1+n_2+n_3=6}\kappa_{n_1, n_2, n_3} a^{2n_1}r^{2n_2}r_h^{2n_3} =  -81 a^8
   r^4+243 a^8 {r_h}^4+27 a^6 r^6-36 a^6 r^2 {r_h}^4+15 a^4 r^8-27 a^4 r^4 {r_h}^4+a^2 r^{10}-2 a^2 r^6 {r_h}^4$. For an arbitrarily small $\theta_{1,2}$, it can be shown that:
\begin{eqnarray}
\label{J0-arb-small-theta12}
& & \hskip -0.5in  J_0 \sim  \frac{ a^{10} \left(\frac{1}{{\log N}}\right)^{8/3} M \left(\frac{1}{N}\right)^{29/20} \left(9 a^2+r^2\right) \left(r^4-{r_h}^4\right)
   \log (r)
  }{ \sqrt{{g_s}} {N_f}^{5/3} r^{16} \left(3 a^2-r^2\right)^8 \left(6 a^2+r^2\right)^7 \sin^3{\theta _2}}\nonumber\\
  & & \hskip -0.5in \times \Biggl(\left(19683
   \sqrt{6} \sin^6\theta_1+6642 \sin^2{\theta _2} \sin^3{\theta _1}-40 \sqrt{6} \sin^4{\theta _2}\right)\sum_{n_1,n_2,n_3:n_1+n_2+n_3=6}\kappa_{n_1, n_2, n_3} a^{2n_1}r^{2n_2}r_h^{2n_3}\Biggr).\nonumber\\
   & &
\end{eqnarray}
Further, the variation of $E_8$:
{\footnotesize
\begin{eqnarray}
\label{deltaE8}
& & \delta E_8 \sim -\frac{2}{3}\delta g_{\tilde{M}\tilde{N}} g^{N_1^\prime\tilde{N}} \epsilon^{ABCM_1N_1,,,M_4N_4}\epsilon_{ABCM_1^\prime N_1^\prime...M_4^\prime N_4^\prime} R^{M_1^\prime\tilde{M}}_{\ \ \ \ \ M_1N_1}R^{M_2^\prime N_2^\prime}_{\ \ \ \ \ M_2N_2}R^{M_3^\prime N_3^\prime}_{\ \ \ \ \ M_2N_2}R^{M_4^\prime N_4^\prime}_{\ \ \ \ \ M_4N_4}\nonumber\\
& & + \frac{\delta g_{\tilde{M}\tilde{N}}}{3}\Biggl[ 2 \epsilon^{ABCM_1\tilde{N},,,M_4N_4}\epsilon_{ABCM_1^\prime N_1^\prime...M_4^\prime N_4^\prime}
g^{N_1^\prime\tilde{N}_1}g^{M_1^\prime\tilde{M}}D_{\tilde{N}_1}D_{M_1}\left(R^{M_2^\prime N_2^\prime}_{\ \ \ \ \ M_2N_2}R^{M_3^\prime N_3^\prime}_{\ \ \ \ \ M_2N_2}R^{M_4^\prime N_4^\prime}_{\ \ \ \ \ M_4N_4}\right)\nonumber\\
& & + \epsilon^{ABCM_1N_1,,,M_4N_4}\epsilon_{ABCM_1^\prime N_1^\prime...M_4^\prime N_4^\prime}
g^{N_1^\prime\tilde{N}_1}g^{M_1^\prime\tilde{M}}[D_{\tilde{N}_1},D_{M_1}]\left(R^{M_2^\prime N_2^\prime}_{\ \ \ \ \ M_2N_2}R^{M_3^\prime N_3^\prime}_{\ \ \ \ \ M_2N_2}R^{M_4^\prime N_4^\prime}_{\ \ \ \ \ M_4N_4}\right)\nonumber\\
& & - 2 \epsilon^{ABCM_1\tilde{M},,,M_4N_4}\epsilon_{ABCM_1^\prime N_1^\prime...M_4^\prime N_4^\prime}
g^{N_1^\prime\tilde{N}}g^{M_1^\prime\tilde{L}}D_{\tilde{L}_1}D_{M_1}\left(R^{M_2^\prime N_2^\prime}_{\ \ \ \ \ M_2N_2}R^{M_3^\prime N_3^\prime}_{\ \ \ \ \ M_2N_2}R^{M_4^\prime N_4^\prime}_{\ \ \ \ \ M_4N_4}\right)\Biggr],
\end{eqnarray}
}
where \cite{Tseytlin-epsilonD^2R^4-kroneckerdeltaR^4}
{\footnotesize
\begin{eqnarray}
\label{epsilonD^2R^4}
& &  \epsilon^{ABCM_1M_2...M_8}\epsilon_{ABCM_1^\prime M_2^\prime...M_8^\prime}R^{M_1^\prime M_2^\prime}_{\ \ \ \ \ M_1M_2}R^{M_3^\prime M_4^\prime}_{\ \ \ \ \ M_3M_4}R^{M_5^\prime M_6^\prime}_{\ \ \ \ \  M_5M_6}R^{M_7^\prime M_8^\prime}_{\ \ \ \ \ M_7M_8}\nonumber\\
& &  = -3!8!
\delta^{M_1}_{[M_1^\prime}...\delta^{M_8}_{M_8^\prime]}R^{M_1^\prime M_2^\prime}_{\ \ \ \ \ M_1M_2}R^{M_3^\prime M_4^\prime}_{\ \ \ \ \ M_3M_4}R^{M_5^\prime M_6^\prime}_{\ \ \ \ \ M_5M_6}R^{M_7^\prime M_8^\prime}_{\ \ \ \ \ M_7M_8}.
\end{eqnarray}
}
Eight of the eleven space-time indices must be chosen, together with the proper anti-symmetrization, in order to assess the contribution of $E_8$, (\ref{epsilonD^2R^4}). Consider the following: \\ $R^{M_1N_1}_{\ \ \ \ \ \ M_1N_1} R^{M_2N_2}_{\ \ \ \ \ \ M_2N_2}R^{M_3N_3}_{\ \ \ \ \ \ M_3N_3}R^{M_4N_4}_{\ \ \ \ \ M_4N_4}$, one of the types of terms that one will obtain using (\ref{epsilonD^2R^4}). After a laborious and thorough calculations, it is then possible to see that the above generates a ``$\frac{1}{N^2}$'' scaling for arbitrarily tiny $\theta_{1,2}$, not only limited to (\ref{Ouyang-definition}), via the most prominent term in the MQGP limit, which is as follows:
\begin{eqnarray}
\label{E8-dominant-large-N}
E_8 \ni R^{tx^1}_{\ \ \ tx^1}R^{x^2x^3}_{\ \ \ x^2x^3}R^{r\theta_1}_{\ \ \ r\theta_1}\left(R^{\psi x^{10}}_{\ \ \ \psi x^{10}}
+ R^{\phi_1\psi}_{\ \ \ \phi_1\psi} +  R^{yz}_{\ \ \ \phi_2\psi}\right) \sim {\cal O}\left(\frac{1}{N^2}\right).
\end{eqnarray}
A comparable $N$ dependency is obtained by an analogous analysis for the remaining kinds of summands in (\ref{epsilonD^2R^4}). Thus, for any $\alpha>0$, $\frac{E_8}{J_0}\sim\frac{1}{N^\alpha}$. The authors in \cite{OR4} really made the argument that one obtains the hierarchy $t_8^2G^2R^3<E_8<J_0$. Hence, we will consider only ``$J_0$'' term for the calculation purpose in this paper similar to \cite{Vikas+Gopal+Aalok,Gopal+Vikas+Aalok,Gopal-Tc-Vorticity,Gopal+Aalok,ACMS,Aalok+Gopal-Mesino}\footnote{See also \cite{Yadav:2023glu} where some of the results in the context of holographic thermal QCD at intermediate coupling are discussed in the thesis of one of us (GY).}.
\end{itemize}

\subsection{Pole Skipping, the Lyapunov Exponent and Butterfly Velocity}
\label{lambdaL_vb}

Let us discuss briefly how we can obtain pole-skipping points using \cite{Natsuume:2023lzy}\footnote{Here, we have discussed the prescription of \cite{Natsuume:2023lzy}; in our work, we have used the metric in ingoing Eddington-Finkelstein coordinate and radial cordianate as $r$.}. Consider Schwarzschild-AdS$_5$ black hole:
\begin{eqnarray}
\label{SAdS5}
& & ds^2=\frac{{r_h}^2}{u}\left(-f dt^2+dx_1^2+dx_2^2+dx_3^2\right)+\frac{du^2}{4 u^2 f},
\end{eqnarray}
{where $r_h$ is the black hole horizon, $f=1-u^2=1-\left(\frac{r_h}{r}\right)^4$ with $u=\frac{r_h^2}{r^2}$}. {Consider the metric perturbation of the form: $g_{\mu \nu}=g_{\mu \nu}^{(0)}+ \eta h_{\mu \nu}$. 
In $h_{r \mu}=0$ gauge, in general, there can be three different modes of the metric perturbation $h_{ab}$ based on the transformation property under the $SO(2)$ rotational symmetry in the $x_2$-$x_3$ plane \cite{Policastro:2002se}:

\begin{itemize}
\item Scalar modes: $h_{tt}, h_{t x_1}, h_{x_1,x_1}, h_{x_2 x_2}=h_{x_3 x_3}$.
\item Vector mode:  $h_{t x_2}, h_{x_1 x_2}$ or $h_{t x_3}, h_{x_1 x_3}$.
 \item Tensor mode: $h_{x_2 x_3}$. 
 \end{itemize}
In Eddington-Finkelstein coordinate above metric perturbations are given as:
\begin{itemize}
\item Scalar modes: $h_{vv}, h_{v x_1}, h_{x_1,x_1}, h_{x_2 x_2}=h_{x_3 x_3}, h_{rr}$.
\item Vector modes: $h_{v x_2}, h_{x_1 x_2}$.
 \item Tensor mode: $h_{x_2 x_3}$. 
 \end{itemize}
 We consider only the scalar modes perturbation in this paper. See \cite{Sil:2016jmc} where all three modes have been studied in the context of thermal QCD. If we compute the linearzied Einstein equations of motion with the metric perturbations defined above then we will get coupled differential equations which are not easy to solve. Using \cite{Kovtun:2005ev}, one can construct the gauge invariant combination of metric perturbations and all the coupled differential equations will be converted into a single differential equation. For example, for the scalar mode, gauge invariant combination for our case is defined in equation  (\ref{Zs}). Since here we are discussing the prescription of \cite{Natsuume:2023lzy} therefore we will consider the scalar mode perturbation as $Z_s(u)$ without going into the details and proceed further.}
\par
For the scalar perturbation, $Z_s(u) e^{-{i} w t +i q x^1}$, field equation has the following form near the horizon:
\begin{eqnarray}
\label{FE-NH}
& & Z_s''(u)+\frac{1}{u-1}Z_s'(u)+\frac{{\cal W}}{4 (u-1)^2} Z_s(u)=0,
\end{eqnarray}
where ${\cal W}=\frac{w}{2 \pi T}$ and the solutions of (\ref{FE-NH}) are:
\begin{eqnarray}
\label{FE-NH-solns}
& & Z_s(u) \propto (u-1)^{\pm \iota \frac{{\cal W}}{2}}.
\end{eqnarray}
Imposing incoming wave boundary condition at the horizon, {consider the following ansatz:
\begin{eqnarray}
\label{ansatz-Zs(u)}
& & Z_s(u)=(u-1)^{- \iota \frac{{\cal W}}{2}}z_s(u).
\end{eqnarray}
Substituting (\ref{ansatz-Zs(u)}) into (\ref{FE-NH}) yields:
\begin{eqnarray}
\label{Ex-PS}
& &z_s''(u)+P(u)z_s'(u)+Q(u)z_s(u)=0.
\end{eqnarray}
with $P(u)$ and $Q(u)$ given by the following:
\begin{eqnarray}
\label{l-m}
& & P(u)=\frac{P_{-1}}{u-1},\nonumber\\
& & Q(u)=\frac{Q_{-1}}{(u-1)^2}.....
\end{eqnarray}
with $P_{-1}=1- \iota {\cal W}, Q_{-1} = w(1-w)/4$. One can write the solution in terms of power series as below:
\begin{eqnarray}
\label{soln-power-series}
& & z_s(u)=\sum_{n=0} A_n (u-1)^{n+\lambda}.
\end{eqnarray}
Subsitution of (\ref{soln-power-series}) into (\ref{Ex-PS}) results in the indicial equation:
\begin{equation}
\label{indicial}
\lambda^2 + (P_{-1}-1)\lambda + Q_{-1} = 0,
\end{equation}
whose roots are $\lambda=0, 1-P_{-1}$.} We keep the incoming mode ($\lambda=0$) and disregard the outgoing mode ($\lambda=1-P_{-1}=\iota {\cal W}$). We can determine the coefficients $A_n$ from the recursion relation. Writing the field equation in the form of the matrix \cite{Blake:2019otz}:
\begin{eqnarray}
\label{Matrix-eq}
& & C z_s(u)=0,\nonumber\\
& & \begin{pmatrix}
C_{11} & C_{12} & 0 & 0 & 0 & 0 & . & . & .  \\
C_{21} & C_{22} & C_{23}  & 0 & 0 & 0 & . & . & . \\
. & . & .  & . & . & . & . & . & . \\
. & . & .  & . & . & . & . & . & . \\
\end{pmatrix}
 \begin{pmatrix}
A_{0}   \\
A_{1} \\
.  \\
. \\
\end{pmatrix}=0,
\end{eqnarray}
where $C_{n, n+1}=n\left(n-1+P_{-1}\right)$. One can obtain the matrix $M^{(n)}$ by keeping the first $n$ rows and the first $n$ column of matrix $C$, and the pole-skipping points are determined by solving the following equation:
\begin{eqnarray}
\label{PS-Eqs}
& & C_{n,n+1}=0, \ \ \ \ {\rm det} M^{(n)}=0.
\end{eqnarray}
{
Pole-skipping points $(w_n,q_n)$ [$n=0,1,2,...$ represents the order of near horizon expansion of the bulk equations of motion] are those points in the complex $(w,q)$ plane at which poles are skipped in the retarded Green's function because, at these special points, numerator and denominator both vanish simultaneously in retarded Green's function. The Lyapunov exponent ($\lambda_L$) and butterfly velocity ($v_b$) can be obtained using \cite{Sil:2020jhr,Blake:2017ris}:
}
\begin{eqnarray}
\label{w-q-lambdaL-vb-relation}
& & w=\iota \lambda_L, \ \ \ \ q=\frac{\iota \lambda_L}{v_b}.
\end{eqnarray}
{ We will determine the pole skipping points for only scalar modes perturbation in this paper. See \cite{Sil:2020jhr} where the author has determined pole-skipping points for all three channels (scalar, vector and tensor).}

\section{Non-Conformal ``Lyapunov Exponent'' and ``Butterfly velocity'' From ${\cal M}$ Theory}
\label{LP-BV}
In this section, we compute the Lyapunov exponent and butterfly velocity in ${\cal M}$-theory dual (inclusive of ${\cal O}(R^4)$ corrections \cite{OR4}) of thermal QCD-like theories at intermediate coupling from the pole-skipping analysis. Before going into the actual computation, first we setup the formalism which involve some results of section \ref{review}.\par
As we discussed in section \ref{review} that contribution from the $E_8$ term is subdominant in comparison to $J_0$ in the large-$N$ limit and hence after dropping $E_8$ in (\ref{eoms}), equations of motion for metric (\ref{eoms}) is rewritten as:
\begin{eqnarray}
\label{EOM-D=11}
R_{MN}-\frac{1}{2}g_{MN}R-\frac{1}{12}\left(G_{MPQR}G_N^{\ PQR}-\frac{g_{MN}}{8}G_4^2\right)=-\beta \Biggl[\frac{g_{MN}}{2}J_0+\frac{\delta J_0}{\delta g^{MN}}\Biggr].
\end{eqnarray}
Taking the trace of the above equation for the general space-time with dimensionality $D$, 
\begin{eqnarray}
\label{trace-EOM-D=11}
& &
R\left(1-\frac{D}{2}\right)-\frac{G_4^2}{12}\left(1-\frac{D}{8}\right)=\beta \Biggl[\frac{D}{2}J_0+g^{MN}\frac{\delta J_0}{\delta g^{MN}}\Biggr].
\end{eqnarray}
Since, $R=R^{\beta^0}+\beta R^{\beta}$, $G_4^2=\left(G_4^2\right)^{\beta^0}$, $J_0=J_0^{\beta^0}$, and $g^{MN}\frac{\delta J_0}{\delta g^{MN}}=\left(g^{MN}\frac{\delta J_0}{\delta g^{MN}}\right)^{\beta^0}$, therefore for $D=11$, equation (\ref{trace-EOM-D=11}) implies the following relation from the ${\cal O}(\beta^0)$ and ${\cal O}(\beta)$ terms:
\begin{eqnarray}
\label{trace-EOM-D=11-i}
& &
R^{\beta^0}=\frac{G_4^2}{144},\nonumber\\
& & J_0=-\frac{2}{11}\left(R^\beta+g^{MN}\frac{\delta J_0}{\delta g^{MN}}\right).
\end{eqnarray}
 Now, we can write equations of motion at ${\cal O}(\beta^0)$ and ${\cal O}(\beta)$ from the substitution of (\ref{trace-EOM-D=11-i}) into the equation (\ref{EOM-D=11}) as follows:
\begin{eqnarray}
\label{EOMs-beta0-beta}
& & {\bf {EOM_{MN}^{\beta^0}}}: \ \ \ 
R_{MN}^{\beta^0}-\frac{1}{2}g_{MN}^{\beta^0}R^{\beta^0}-\frac{1}{12}\left(G_{MPQR}G_N^{\ PQR}\right)^{\beta^0}+\left(\frac{144}{196}\right)g_{MN}^{\beta^0}R^{\beta^0}=0, \nonumber\\
& & {\bf {EOM_{MN}^{\beta}}}: \ \ \ R_{MN}^\beta + g_{MN}^\beta R^{\beta^0}-\frac{13}{22}g_{MN}^{\beta^0} R^{\beta}-\frac{1}{4}\left(G_{MPQR}G_{N\tilde{P}\tilde{Q}\tilde{R}}g^{P\tilde{P}, \beta}g^{Q\tilde{Q},\beta^0}g^{R\tilde{R},\beta^0}\right)\nonumber\\
& & =- \Biggl[-\frac{g_{MN}^{\beta^0}}{11}\left(g^{PQ}\frac{\delta J_0}{\delta g^{PQ}}\right)^{\beta^0}+\left(\frac{\delta J_0}{\delta g^{MN}}\right)^{\beta^0} \Biggr].
\end{eqnarray}
Use will be made of the following to compute the ${\cal O}(R^4)$ contribution to the Linearized Einstein's equation in ${\rm EOM_{MN}^{\beta}}$:
{\footnotesize
\begin{eqnarray}
\label{delta J_0}
 \delta J_0 & \stackrel{\rm MQGP,\ IR}{\xrightarrow{\hspace*{1.5cm}}} & 3\times 2^8 \delta R^{HMNK} R_H^{\ RSP}\Biggl(R_{PQNK}R^Q_{\ RSM} + R_{PSQK}R^Q_{\ MNR}  \nonumber\\
 & & + 2\left[R_{PMNQ}R^Q_{\ RSK} + R_{PNMQ}R^Q_{\ SRK}\right]\Biggr) \nonumber\\
& &  \equiv 3\times 2^8 \delta R^{HMNK}\chi_{HMNK}\nonumber\\
& & = -\delta g_{\tilde{M}\tilde{N}}\Biggl[ g^{M\tilde{N}} R^{H\tilde{N}NK}\chi_{HMNK}
+ g^{N\tilde{N}} R^{HM\tilde{M}K}\chi_{HMNK} + g^{K\tilde{M}}R^{HMN\tilde{N}}\chi_{HMNK}
\nonumber\\
& & + \frac{1}{2}\Biggl(g^{H\tilde{N}}[D_{K_1},D_{N_1}]\chi_H^{\tilde{M}N_1K_1} +
g^{H\tilde{N}}D_{M_1}D_{N_1} \chi_H^{M_1[{N}_1\tilde{M}]}  - g^{H\tilde{H}} D_{\tilde{H}}D_{N_1}\chi_H^{\tilde{N}[N_1\tilde{M}]}\Biggr)\Biggr],\nonumber\\
& &
\end{eqnarray}
}
where:
{\footnotesize
\begin{eqnarray}
\label{chi-def}
& & \chi_{HMNK} \equiv R_H^{\ \ RSP}\left[R_{PQNK} R^Q_{\ \ RSM}
+ R_{PSQK} R^Q_{\ \ MNR} + 2\left(R_{PMNQ} R^Q_{\ \ RSK} + R_{PNMQ}R^Q_{\ \ SRK}\right)\right].
\end{eqnarray}
}
Now, we compute the linearized equations of motion of (\ref{EOMs-beta0-beta}) for the perturbed metric defined as below:
\begin{eqnarray}
\label{metric-perturbations}
& & 
g_{MN}^{\beta^0}=\left(g_{MN}^{\beta^0}\right)^{(0)}+\eta h_{MN}^{\beta^0}; \ \ \ g_{MN}^{\beta}=\left(g_{MN}^{\beta}\right)^{(0)}+ \eta h_{MN}^{\beta}, \nonumber\\
& &
R_{MN}^{\beta^0}=\left(R_{MN}^{\beta^0}\right)^{(0)}+\eta \delta R_{MN}^{\beta^0}; \ \ \ R_{MN}^{\beta}=\left(R_{MN}^{\beta}\right)^{(0)}+\eta \delta R_{MN}^{\beta}
\end{eqnarray}
where $h_{MN}^{\beta^0} \neq h_{MN}^{\beta}$, $\delta R_{MN}^{\beta^0} \neq \delta R_{MN}^{\beta}$. {In (\ref{metric-perturbations}); we have introduced a parameter $\eta$  to indicate that we are working up to linear order perturbation in the metric, Ricci tensors etc. First terms in $g_{MN}^{\beta^0}$, $g_{MN}^{\beta}$, $R_{MN}^{\beta^0}$, $R_{MN}^{\beta}$ are the unperturbed part of the metric and Ricci tensors. This parameter helps us to identify the unperturbed and perturbed contributions of the Einsetin's equations of motion in the following way. To compute the linearized equations of motion of ${EOM_{MN}^{\beta^0}}$ appearing in (\ref{EOMs-beta0-beta}), we have to compute $R_{MN}^{\beta^0}$, $R^{\beta^0}$ using the metric $g_{MN}^{\beta^0}=\left(g_{MN}^{\beta^0}\right)^{(0)}+\eta h_{MN}^{\beta^0}$ and then substitue these in ${EOM_{MN}^{\beta^0}}$ of (\ref{EOMs-beta0-beta}) and extract the coefficient of $\eta$. Similarly, we can extract the linearized equations of motion from ${EOM_{MN}^{\beta}}$ appearing in (\ref{EOMs-beta0-beta}) using (\ref{metric-perturbations}) and then extract the coefficient of $\eta$}.
 The form of $h_{MN}^{\beta^0/\beta}$ are similar to \cite{Sil:2020jhr}:
\begin{eqnarray}
& & h_{vv}^{\beta^0/\beta}(r)=e^{-i w v+iq x^1} g_{vv}(r) H_{vv}^{\beta^0/\beta}(r), \ \ \ \ \  h_{v x^1}^{\beta^0/\beta}(r)=e^{-i w v+iq x^1} g_{x^1 x^1}(r) H_{v x^1}^{\beta^0/\beta}(r), \nonumber\\
& & h_{x^1 x^1}^{\beta^0/\beta}(r)=e^{-i w v+iq x^1} g_{x^1 x^1}(r) H_{x^1 x^1}^{\beta^0/\beta}(r), \ \ \ h_{x^2 x^2}^{\beta^0/\beta}(r)=e^{-i w v+iq x^1} g_{x^2 x^2}(r) H_{x^2 x^2}^{\beta^0/\beta}(r),
\end{eqnarray}
where $g_{vv}(r), \ g_{v r}(r), g_{x_1 x_1}(r)=g_{x_2 x_2}(r)=g_{x_3 x_3}(r)$ are the metric components in terms of the ingoing Eddington-Finkelstein coordinates which are obtained using:
\begin{eqnarray}
\label{v-rstar}
& & v=t+r_{*}, \ \ \ \ \ \ r_{*}=\int \sqrt{\frac{g_{rr}(r)}{g_{tt}(r)}} dr.
\end{eqnarray} 
As the eleven-dimensional metric is worked out near the $\psi=2n\pi, \ n=0, 1, 2$-coordinate patches wherein $g_{rM}=0, M\neq r$ and the $M_5=(S^1 \times_w \mathbb{R}^3) \times \mathbb{R}_{>0}$ and the unwarped $\tilde{M}_6(S^1_{\cal M} \times_w T^{\rm NE})$ of $SU(3)$ structure wherein $S^1_{\cal M}$ is the ${\cal M}$-theory circle and $T^{\rm NE}$ is the non-Einsteinian deformation of $T^{1,1}$, decouple,  the five-dimensional part of the metric (\ref{fMN-def}) using (\ref{v-rstar})  is written as:
\begin{eqnarray}
\label{5D-metric-EF}
ds^2=g_{vv} dv^2+g_{vr} dv dr+g_{x_1 x_1} dx_1^2+g_{x_2 x_2} dx_2^2+g_{x_3 x_3} dx_3^2,
\end{eqnarray}
where,  near the type IIA $D6$ flavor branes \cite{MQGP-meson-spectroscopy}, (\ref{alpha_theta_12}) and $(\tilde{x},\tilde{z}) = (0, {\rm constant})$  -  $(\tilde{x}, \tilde{y}, \tilde{z})$ diagonalizing the $T^3$ sLag used for constructing the type IIA SYZ mirror, up to leading order in $N$:
\begin{eqnarray}
\label{gvv-etc}
& & g_{vv}(r)=-\frac{3^{2/3} \sqrt{\frac{1}{N}} N_f^{2/3} r^2 \left(1-\frac{r_h^4}{r^4}\right) (2 \log(N)-6
   \log (r))^{2/3}}{8 \pi ^{7/6} \sqrt{g_s}} -\beta \Biggl(\frac{b^8 \left(9 b^2+1\right)^3 \Sigma _1 \left(6 a^2+r_h^2\right) }{72 \sqrt[3]{3} \pi ^{13/6} \left(18 b^4-3
   b^2-1\right)^5 \sqrt{g_s} }\nonumber\\
   & & \times \frac{\left(4374 b^6+1035 b^4+9 b^2-4\right) M \left(\frac{1}{N}\right)^{11/4} r^2
    \left(1-\frac{r_h^4}{r^4}\right) \log
   (r_h) (2 \log(N) N_f-6 N_f \log (r))^{2/3}(r-r_h)^2}{\log(N)^2 N_f r_h^2 \alpha _{\theta _2}^3 \left(9
   a^2+r_h^2\right)}\Biggr), \nonumber\\
& & g_{v r}(r)=\frac{1}{4} \left(\frac{3}{\pi }\right)^{2/3} N_f^{2/3} \sqrt{\frac{6 a^2+r^2}{9 a^2+r^2}} (2
   \log(N)-6 \log (r))^{2/3}\left(1+\frac{\beta}{2}\left({\cal{C}}_{zz}-2 {\cal{C}}_{\theta_1 z}+2 {\cal{C}}_{\theta_1 x}\right) \right), \nonumber\\
& & g_{x_1 x_1}(r)=g_{x_2 x_2}(r)=g_{x_3 x_3}(r)=\frac{3^{2/3} \sqrt{\frac{1}{N}} N_f^{2/3} r^2 (2 \log(N)-6 \log (r))^{2/3}}{8 \pi ^{7/6}
   \sqrt{g_s}}\nonumber\\
   & & -\beta \Biggl(\frac{b^8 \left(9 b^2+1\right)^4 \left(39 b^2-4\right) M \left(\frac{1}{N}\right)^{11/4} r^2 \Sigma _1 \left(6
   a^2+r_h^2\right) (r-r_h)^2 \log (r_h) (N_f (\log(N)-3 \log (r)))^{2/3}}{12
   \sqrt[3]{6} \pi ^{13/6} \left(3 b^2-1\right)^5 \left(6 b^2+1\right)^4 \sqrt{g_s} \log(N)^2
   N_f r_h^2 \alpha _{\theta _2}^3 \left(9 a^2+r_h^2\right)}\Biggr),\nonumber\\
   & &
\end{eqnarray}
where $b=\frac{1}{\sqrt{3}}+\epsilon$ and
\begin{eqnarray}
\label{Sigma_1-3-def}
& & \Sigma_1 \equiv 19683
   \sqrt{6} \alpha _{\theta _1}^6+6642 \alpha _{\theta _2}^2 \alpha _{\theta _1}^3-40 \sqrt{6} \alpha _{\theta _2}^4.
\end{eqnarray} 
As was shown in \cite{OR4}, the $C^{(1)}_{MNP}=0$-truncation of (\ref{ansaetze}), yields:
\begin{eqnarray}
\label{C1=0-truncation}
& & {\cal{C}}_{zz}-2 {\cal{C}}_{\theta_1 z} = 0,\nonumber\\
& & |{\cal{C}}_{\theta_1 x}|\ll 1.
\end{eqnarray}
Hence, $g_{vr}$ receives no ${\cal O}(\beta)$ correction.

To obtain the linearized equations of motion at ${\cal O}(\beta^0)$ and ${\cal O}(\beta)$, we will simplify ``${\rm EOM_{MN}^{\beta^0}}$'' and ``${\rm EOM_{MN}^{\beta}}$'' appearing in (\ref{EOMs-beta0-beta}) using the perturbed metric defined in (\ref{metric-perturbations}) in subsections \ref{beta0-LP-BV} and \ref{beta-LP-BV} respectively. For this purpose we will use the unperturbed metric given in (\ref{gvv-etc}). Linearzied equations of motion are the ${\cal O}(\eta)$ contribution of the ``${\rm EOM_{MN}^{\beta^0}}$'' and ${\cal O}(\eta)$ contribution of the ``${\rm  EOM_{MN}^{\beta}}$'' at ${\cal O}(\beta^0)$ and ${\cal O}(\beta)$ respectively.

\subsection{${\cal O}(\beta^0)$ Contributions}
\label{beta0-LP-BV}
For the metric perturbations defined in (\ref{metric-perturbations}), the  linearized equations of motion at ${\cal O}(\beta^0)$ are obtained by simplifying ``${\rm EOM}_{MN}^{\beta^0}$'' given in (\ref{EOMs-beta0-beta}) and retaining the terms of ${\cal O}(\eta)$. Therefore, using (\ref{EOMs-beta0-beta}), (\ref{metric-perturbations}) and (\ref{gvv-etc})\footnote{For ``${\rm EOM}_{MN}^{\beta^0}$'', we use ${\cal O}(\beta^0)$ contribution of the metric (\ref{gvv-etc}) whereas for ``${\rm EOM}_{MN}^{\beta}$'', we use ${\cal O}(\beta^0)$ as well as ${\cal O}(\beta)$ contribution of the metric (\ref{gvv-etc}) as per the requirement because ``${\rm EOM}_{MN}^{\beta}$'' involves $R^{\beta^0}, R^\beta$ etc. and in general metric (\ref{gvv-etc}) has the form: $g_{MN}=g_{MN}^{\beta^0}+\beta g_{MN}^{\beta}$.}, the  linearized equations of motion at ${\cal O}(\beta^0)$ are given in appendix \ref{EOMs-beta0}\footnote{Use is also made of:
\begin{eqnarray*}
\label{flux-beta0-subdominant}
& &  G_{rMNP}G_r^{\ \ MNP} = \left.\frac{\sqrt{\frac{2}{3}} {g_s}^{7/4} M N^{3/5}
   \sqrt[4]{g_s} \left(108 a^2
   {N_f} \log ^2(r)+72 a^2 {N_f} \log
   (r)+{N_f} r+4 {N_f} r \log (r)\right)}{27
   \pi  \alpha ^2 r^3 \alpha _{\theta _1}^2 \alpha
   _{\theta _2}^3 {b_1}(r)^2}\right|_{(\ref{C1=0-truncation})} \nonumber\\
   & & \hskip 0.9in \sim\frac{1}{r^2N^{2/5}}<\left.R_{rr}^{\beta^0}\right|_{(\ref{C1=0-truncation})}\sim\frac{1}{N^{1/3}r^2},
\end{eqnarray*}
where $b_1(r) = \frac{3 {gs}^{7/4} M  {N_f} \left(r^2-3
   a^2\right) \log ^2(r)}{\sqrt{2} \pi ^{5/4} r^2
   \alpha _{\theta _1} \alpha _{\theta _2}^2}$.}. We can write linearized equations of motion (\ref{pEOM-beta0}) into a single equation of motion using the gauge-invariant combination of the metric perturbation \cite{Kovtun:2005ev}. For our case, the gauge-invariant combination of scalar perturbations is \cite{Sil:2020jhr}:
{\footnotesize
\begin{eqnarray}
\label{Zs}
& &  {Z_s}(r)=-2 {H_{x_2x_2}}(r) \left(\frac{q^2 {g_{vv}}'(r)}{{g^{x_1x_1}}(r)
   {g_{x_1x_1}}'(r)}+w^2 {g_{x_1x_1}}(r)\right)+4 q w {g_{x_1x_1}}(r) {H_{vx_1}}(r)+2 w^2
   {g_{x_1x_1}}(r) {H_{x_1x_1}}(r)\nonumber\\
   & & \hskip 0.6in +2 q^2 {g_{vv}}(r) {H_{vv}}(r).
\end{eqnarray}
}
Up to NLO in $N$,
\begin{itemize}
\item
the coefficient of $H_{vv}(r)$ in $Z_s(r)$ is given by:
\begin{eqnarray}
\label{ZsHvv}
& &
\frac{2187 {g_s}^{3/2} M^2 \left(\frac{1}{N}\right)^{3/2} {N_f}^3 {r_h}^2 \left(36 w^2-529 q^2\right)
   (r-{r_h})^2 \log ^4(r)}{33856 \pi ^{9/2} w^2} \nonumber\\
 & &  +\frac{243 \sqrt{\frac{1}{N}} {N_f}^2 {r_h}^2 \left(529 q^2-36
   w^2\right) (r-{r_h})^2 \log ^2(r)}{4232 \pi ^{5/2} \sqrt{{g_s}} w^2};
\end{eqnarray}
\item the coefficient of $H_{vv}(r)$ in $Z_s'(r)$ is:
\begin{eqnarray}
\label{dZsHvv}
& &
-\frac{2187 \sqrt{\frac{1}{N}} {N_f}^2 {r_h}^2 (r-{r_h}) \log ^2(r)}{529 \pi ^{5/2} \sqrt{{g_s}}}+\frac{243
   \sqrt{3} {N_f}^2 {r_h} \left(3 w^2-2 q^2\right) \left(36 w^2-529 q^2\right) \log ^2(r)}{33856 \pi ^2 \iota 
   w^3}\nonumber\\
 & & +\frac{243 \sqrt{3} {N_f}^2 {r_h} \left(36 w^2-529 q^2\right) \log (r)}{33856 \pi ^2 \iota  w};
\end{eqnarray}
\item the coefficient of $H_{vv}'(r)$ in $Z_s'(r)$ is:
\begin{eqnarray}
\label{dZsHvv'}
\frac{24\ 2^{2/3} 3^{5/6} \iota  {N_f}^{2/3} q^2 {r_h} (r-{r_h})^2 \log ^2(r)}{\pi ^{5/3} {g_s} N w (-\log
   (r))^{4/3}}-\frac{243 \sqrt{\frac{1}{N}} {N_f}^2 {r_h}^2 \left(529 q^2+36 w^2\right) (r-{r_h})^2 \log
   ^2(r)}{4232 \pi ^{5/2} \sqrt{{g_s}} w^2};
\end{eqnarray}
\item the coefficient of $H_{vv}(r)$ in $Z_s''(r)$ is:
\begin{eqnarray}
\label{ddZsHvv}
& &
\frac{19683 \left(\frac{3}{2}\right)^{2/3} \sqrt{{g_s}} \sqrt{N} {N_f}^{10/3} \left(529 q^2-36 w^2\right)
   (r-{r_h})^2 (-\log (r))^{10/3} \left(-529 \left(\iota ^2+4\right) q^2-36 \iota ^2 w^2\right)}{143278592 \pi ^{17/6}
   \iota ^2 w^4}\nonumber\\
 & & -\frac{243 \sqrt{3} {N_f}^2 q^2 {r_h} \left(2 q^2-3 w^2\right) \log ^2(r)}{32 \pi ^2 \iota  w^3;
   (r-{r_h})}
\end{eqnarray}
\item the coefficient of $H_{vv}'(r)$ in $Z_s''(r)$ is:
\begin{eqnarray}
\label{ddZsHvv'}
& &
\frac{12\ 2^{2/3} \sqrt[3]{3} \sqrt{\frac{1}{N}} {N_f}^{2/3} \left(3 q^2 w^2-2 q^4\right) \log (r)}{\pi ^{7/6}
   \sqrt{{g_s}} w^2 \sqrt[3]{-\log (r)}}+\frac{243 \sqrt{3} {N_f}^{7/3} {r_h} \left(3 w^2-2 q^2\right) \left(529
   q^2+36 w^2\right) \log ^2(r)}{33856 \pi ^2 \sqrt[3]{{\iota Nf}} w^3}\nonumber\\
 & & +\frac{243 \sqrt{3} {N_f}^{7/3} {r_h}
   \left(529 q^2+36 w^2\right) \log (r)}{33856 \pi ^2 \sqrt[3]{{\iota Nf}} w}.
\end{eqnarray}
\end{itemize}
Writing the equation of motion for gauge-invariant variable $Z_s(r)$ as:
\begin{equation}
\label{Zs-EOM}
Z_s''(r) = l(r) Z_s'(r) + m(r) Z_s(r),
\end{equation}
where $l(r) = {\rm Coefficient\ of}\ H_{vv}'(r)\ {\rm in}\ Z_s''(r)/{\rm Coefficient\ of}\ H_{vv}'(r)\ {\rm in}\ Z_s'(r)$ and is given by:
{\footnotesize
\begin{eqnarray}
\label{l(r)}
& & l(r)=-\frac{i }{8 \sqrt{\frac{1}{N}} \sqrt[3]{{N_f}} {r_h} w (r-{r_h})^2 \left(-81 \sqrt{{g_s}} {N_f}^{4/3}
   {r_h} \left(529 q^2+36 w^2\right) (-\log (r))^{4/3}+33856 i 2^{2/3} (3 \pi )^{5/6} \sqrt{\frac{1}{N}} q^2 w\right)}\nonumber\\
   & & \times \Biggl(135424 i 2^{2/3} \sqrt[3]{3} \pi ^{4/3} \sqrt{{g_s}} \sqrt{\frac{1}{N}} \sqrt[3]{{N_f}} q^2 w
   \left(2 q^2-3 w^2\right)-81 \sqrt{3 \pi } {g_s} {N_f}^{5/3} {r_h} \left(-1515 q^2 w^2+1058 q^4-108 w^4\right)
   (-\log (r))^{4/3}\nonumber\\
   & &-81 \sqrt{3 \pi } {g_s} {N_f}^{5/3} {r_h} w^2 \left(529 q^2+36 w^2\right) \sqrt[3]{-\log
   (r)}\Biggr).
\end{eqnarray}
}
Using (\ref{l(r)}) and equating coefficients of $H_{vv}(r)$ in (\ref{Zs-EOM}), one obtains:
{\footnotesize
\begin{eqnarray}
\label{m(r)}
& & m(r)=-\frac{\pi ^{5/3} \sqrt{{g_s}} }{8464 \left(\frac{1}{N}\right)^{3/2} {r_h}^2 w^2 \left(36 w^2-529 q^2\right) (r-{r_h})^4 \log ^2(r)
   \left(8 \pi ^2 N-9 {g_s}^2 M^2 {N_f} \log ^2(r)\right)}\nonumber\\
   & & \times \Biggl[-81 \sqrt[3]{2} 3^{2/3} \sqrt{{g_s}} \sqrt{N} {N_f}^{4/3} \left(1587
   q^2-36 w^2\right) \left(36 w^2-529 q^2\right) (r-{r_h})^4 (-\log (r))^{10/3}\nonumber\\
   & &+\frac{1058 \pi ^{5/6} N \log (r)
   }{33856\
   2^{2/3} (3 \pi )^{5/6} q^2 w+81 i {N_f}^{4/3} {r_h} \sqrt{{g_s} N} \left(529 q^2+36 w^2\right) (-\log
   (r))^{4/3}} \Biggl(\sqrt{3 \pi } \sqrt{{g_s}} w^2 \left(36 w^2-529 q^2\right)\nonumber\\
   & &+\log (r) \left(\sqrt{3 \pi } \sqrt{{g_s}}
   \left(-1659 q^2 w^2+1058 q^4+108 w^4\right)+576 i \sqrt{\frac{1}{N}} {r_h} w^3 ({r_h}-r)\right)\Biggr)
   \Biggl(\sqrt[3]{3} w  \nonumber\\
  & &\left(135424\ 2^{2/3} \pi ^{5/6} \sqrt{\frac{1}{N}} q^2 \left(2 q^2-3 w^2\right)  +81 i \sqrt[6]{3}
   \sqrt{{g_s}} {N_f}^{4/3} {r_h} w \left(529 q^2+36 w^2\right) \sqrt[3]{-\log (r)}\right) \nonumber\\
  & &+81 i \sqrt{3}
   \sqrt{{g_s}} {N_f}^{4/3} {r_h} \left(-1515 q^2 w^2+1058 q^4-108 w^4\right) (-\log (r))^{4/3}\Biggr)\nonumber\\
   & &+8954912 \sqrt{3} \pi ^{5/6} \iota  q^2 {r_h} w \left(2 q^2-3 w^2\right) (r-{r_h}) \log
   ^2(r)\Biggr].
\end{eqnarray}
}
Now, 
\begin{equation}
\label{m(r)-Laurent-expansion}
m(r) = \sum_{n=-4}^\infty m_i(r_h; N, M, N_f, g_s)(r - r_h)^n,
\end{equation}
to ensure $r=r_h$ is a regular singular point of (\ref{Zs-EOM}), we need to set:
\begin{eqnarray}
\label{absent-IRSP}
& & m_{-4} = m_{-3} = 0.
\end{eqnarray}
One can show that $m_{-4} = 0$ implies:
\begin{eqnarray}
\label{vb-i}
& & \hskip -0.6in {\cal W} = \sqrt{\frac{2}{3}}{\cal Q}\left(1 -\frac{1 - 3|\log r_h| + \sqrt{3}\sqrt{\left(3|\log r_h| - 1\right)|\log r_h|}}{1 - 3|\log r_h|}\right) =  \sqrt{\frac{2}{3}}{\cal Q}\left(1 + \frac{1}{6|\log r_h|} + {\cal O}\left(\frac{1}{(\log r_h)^2}\right)\right),\nonumber\\
& & 
\end{eqnarray}
(where ${\cal Q}, {\cal W} \equiv \frac{q, w}{2 \pi T}$) i.e., the analog of the butterfly velocity $v_b$ in our setup, is given by:
\begin{eqnarray}
\label{vb}
v_b = \sqrt{\frac{2}{3}}\left(1 + \frac{1}{6|\log r_h|}\right).
\end{eqnarray}
One hence sees that the non-conformal butterfly velocity exceeds its conformal counterpart's value.\par
We can also verify the value of butterfly velocity using (\ref{Ageev-results}). The discussion given below is valid for AdS-BH spacetime. The expression of metric components relevant for this purpose are given as:
\begin{eqnarray}
\label{metric-tt-11}
& & \hskip -0.6in g_{tt}(r)=\frac{\left(\frac{1}{N}\right)^{3/2} r^2 g_2(r) (a_1(r) (N-B(r))-a_2(r))}{2 \sqrt{\pi } \sqrt{g_s}
   \sqrt[3]{a_1(r)}} \nonumber\\
   & & \hskip -0.6in +\beta \Biggl(\frac{b^8 \left(9 b^2+1\right)^3 \left(4374 b^6+1035 b^4+9 b^2-4\right) M r^2 \Sigma _1 \left(6 a^2+r_h^2\right) a_1(r)^{2/3}
   g_2(r) (r-r_h)^2 \log (r_h)}{54 \pi ^{3/2} \left(18 b^4-3 b^2-1\right)^5 \sqrt{g_s} \log(N)^2 N^{11/4} N_f
   r_h^2 \alpha _{\theta _2}^3 \left(9 a^2+r_h^2\right)} + {\cal O}((r-r_h)^2)\Biggr),\nonumber\\
\end{eqnarray}
where
\begin{eqnarray}
\label{ai-bi's}
& & 
g_2(r)=1-\frac{r_h^4}{r^4},\nonumber\\
& & a_1(r)= \frac{3 (2 \log(N) N_f-6 N_f \log (r))}{8 \pi } ,\nonumber\\
& & a_2(r)= \frac{12 a^2 g_s M^2 N_f ({c_1}+{c_2} \log (r_h))}{9 a^2+r^2},\nonumber\\
& & B(r)=\frac{3 g_s M^2 \log (r) (12 g_s N_f \log (r)-g_s \log(N) N_f)}{32 \pi ^2},
\end{eqnarray}
and one has worked out the ${\cal O}(\beta)$ corrections up to ${\cal O}((r-r_h)^2)$ as one is working in the IR. For the metric (\ref{metric-tt-11}) and (\ref{gvv-etc}), we can show that
\begin{eqnarray}
\label{d-gii}
& & g_{tt}'(r)=\frac{3^{2/3} \sqrt{\frac{1}{N}} \left(\log(N) N_f r^4+\log(N) N_f r_h^4-N_f r^4-3 N_f r^4 \log (r)-3
   N_f r_h^4 \log (r)+N_f r_h^4\right)}{2 \sqrt[3]{2} \pi ^{7/6} \sqrt{g_s} r^3 \sqrt[3]{N_f (\log(N)-3
   \log (r))}}\nonumber\\
   & & +\beta \Biggl(\frac{b^8 \left(9 b^2+1\right)^3 \left(4374 b^6+1035 b^4+9 b^2-4\right) \beta  M \left(\frac{1}{N}\right)^{11/4} r \Sigma _1 \left(6
   a^2+r_h^2\right) (r-r_h) \log (r_h)}{81 \sqrt[3]{3} \pi ^{7/6} \left(18 b^4-3 b^2-1\right)^5 \sqrt{g_s}
   \log(N)^2 N_f r_h^2 \alpha _{\theta _2}^3 \left(9 a^2+r_h^2\right) \sqrt[3]{2 \log(N) N_f-6 N_f \log
   (r)}} \nonumber\\
   & & \times \frac{9 \left(\frac{4 r_h^4 (r-r_h)}{r^4}+(4 r-2 r_h) \left(1-\frac{r_h^4}{r^4}\right)\right) (2 \log(N)
   N_f-6 N_f \log (r))}{8 \pi }-\frac{9 N_f (r-r_h) \left(1-\frac{r_h^4}{r^4}\right)}{2 \pi } \Biggr),\nonumber\\
   & & g_{x^1 x^1}'(r)=\frac{3^{2/3} \sqrt{\frac{1}{N}} r (\log(N) N_f-3 N_f \log (r))}{2 \sqrt[3]{2} \pi ^{7/6} \sqrt{g_s}
   \sqrt[3]{N_f (\log(N)-3 \log (r))}} -\beta \Biggl(\frac{b^8 \left(9 b^2+1\right)^4 \left(39 b^2-4\right) M \left(\frac{1}{N}\right)^{11/4} r (r-r_h)}{6 \sqrt[3]{6} \pi ^{13/6} \left(3
   b^2-1\right)^5 \left(6 b^2+1\right)^4 \sqrt{g_s} \log(N)^2 r_h^2}\nonumber\\
   & & \times \left(\frac{\Sigma _1 \left(6 a^2+r_h^2\right) \log (r_h) (2 \log(N) r-\log(N) r_h+3 r_h \log (r)-r-6 r \log
   (r)+r_h)}{\alpha _{\theta _2}^3 \left(9 a^2+r_h^2\right) \sqrt[3]{N_f (\log(N)-3 \log (r))}}\right)\Biggr).
\end{eqnarray}
We can see explicitily from (\ref{d-gii}) that when $r=r_h$, the ${\cal O}(\beta)$ contributions to the derivatives of the metric vanish. Now, we can compute the butterfly velocity using (\ref{Ageev-results}) for the metric (\ref{metric-tt-11}) (for our case, $d=4$) and the same is given as:
\begin{eqnarray}
\label{vb-ageev-formula}
v_b=\sqrt{\frac{2}{3}}.
\end{eqnarray}
Therefore, we see that there is no ${\cal O}(\beta)$ contribution to the butterfly velocity even from the use of (\ref{Ageev-results}). As expected, (\ref{Ageev-results}) being applicable for $AdS$ backgrounds, misses out the non-conformal correction worked out in (\ref{vb}). But, it is nice to see that both approaches yield the same conformal contribution. \par
 One can similarly show that $m_{-3}=0$ substituting (\ref{vb-i}) implies \footnote{Note, $r$ in our paper is in fact the dimensionless $\frac{r}{{\cal R}_{D5/\overline{D5}}},\ {\cal R}_{D5/\overline{D5}}$ being the $D5-\overline{D5}$-brane separation in the parent type IIB dual of \cite{metrics}.}:
\begin{eqnarray}
\label{lambdaL-i}
& & q = \frac{8464 i \sqrt{2} \log r_h ^3   \left(1-\frac{9 {g_s}^2 \log r_h ^2
   M^2 {N_f}}{8 \pi ^2 N}\right)}{505\sqrt{\pi g_s N}}r_h^2.
\end{eqnarray}
Replacing one of the two powers of the multiplicative $r_h^2$ in (\ref{lambdaL-i}) with $\sqrt{3}\pi^{3/2}\sqrt{g_s N} T$, one identifies the following expression for the analog of the Lyapunov exponent:
\begin{eqnarray}
\label{LambdaL-i}
& & \lambda_L = \frac{16928  \pi  {r_h}  \left(1-\frac{1}{6 \log ({r_h})}\right) |\log ^3({r_h})| \left(1-\frac{9 {g_s}^2 M^2
  {N_f} \log ^2({r_h})}{8 \pi ^2 N}\right)}{505}T. 
\end{eqnarray}
We should note, all radial distances $r$ are in fact the dimensionaless $\frac{r}{{\cal R}_{D5/\overline{D5}}}=\frac{r}{\sqrt{3}\left(\frac{1}{\sqrt{3}} + \epsilon\right)r_h}, \epsilon \sim r_h^2\left(\log r_h\right)^{9/2}N^{- 9/10 - \alpha} + \frac{g_sM^2}{N}\left(c_1 + c_2\log r_h\right), c_{1,2}<0$ \cite{OR4}, \cite{Bulk-Viscosity-McGill-IIT-Roorkee}. Writing $r_h = \tilde{\kappa}_{r_h} e^{-0.3 N^{1/3}}: \tilde{\kappa}_{r_h}\stackrel{<}{\sim}4$ ensuring $r_h(N=100, {\rm c/o\ Table}\ \tcb{2}) \sim 1 - \epsilon$, the chaos bound  $\lambda_L<2\pi T$ \cite{Maldacena:2015waa} is satisifed by (\ref{LambdaL-i}). 

The differential equation satisfied by $Z_s(r)$ involves:
\begin{eqnarray}
\label{l+m}
& & l(r) = \frac{\gamma_2}{r-{r_h}}+\gamma_3+\gamma_4 (r-{r_h}) + \gamma_5(r-r_h)^2 + {\cal O}((r-r_h)^3);\nonumber\\
& & m(r)=\frac{\gamma_m}{(r-{r_h})^2}+\frac{\gamma_m^{(2)}}{r-{r_h}}+\gamma_m^{(3)}
 + \gamma_m^{(4)}(r-r_h) + \gamma_m^{(5)}(r-r_h)^2 + {\cal O}((r-r_h)^3),
\end{eqnarray}
where
\begin{eqnarray}
\label{gammas}
& & \left.\gamma_m\right|_{(\ref{vb-i}), (\ref{lambdaL-i})} \equiv 1.5 \log r_h  {r_h} \left(\frac{307 (-\log r_h )^{8/3}
   {r_h}}{{g_s} N {N_f}^{4/3}}+0.7\right)\approx 0,\nonumber\\
& & \left.\gamma_{2}\right|_{(\ref{vb-i}), (\ref{lambdaL-i})}\equiv \frac{2116 \log ({r_h})}{505}-\frac{1212710002688
   \left(\frac{2}{3}\right)^{2/3} \sqrt[3]{\pi } {r_h} (-\log
   ({r_h}))^{8/3}}{1370800179 {g_s} N {N_f}^{4/3}},\nonumber\\
& & \left.\gamma_{3}\right|_{(\ref{vb-i}), (\ref{lambdaL-i})} \equiv \frac{606355001344 \left(\frac{2}{3}\right)^{2/3} \sqrt[3]{\pi }
   (-\log r_h )^{8/3}}{1370800179 {g_s} N
   {N_f}^{4/3}}-\frac{1058 \log r_h }{505 {r_h}},\nonumber\\
& & \left.\gamma_{4}\right|_{(\ref{vb-i}), (\ref{lambdaL-i})} \equiv \frac{2116 \log r_h }{1515 {r_h}^2}-\frac{2425420005376
   \left(\frac{2}{3}\right)^{2/3} \sqrt[3]{\pi }
   (-\log r_h )^{8/3}}{4112400537 {g_s} N {N_f}^{4/3}
   {r_h}},\nonumber
   \end{eqnarray}
   \begin{eqnarray}
& & \left.\gamma_5 \right|_{(\ref{vb-i}), (\ref{lambdaL-i})} \equiv -\frac{529 {\log r_h}}{505 {r_h}^3},\nonumber\\
& & \left.\gamma_m^{(2)}\right|_{(\ref{vb-i}), (\ref{lambdaL-i})} \equiv \frac{839523 {g_s}^2 \log r_h ^4 M^2 {N_f}}{51005 \pi
   ^2 N {r_h}}+\frac{2238728 \log r_h ^2}{153015
   {r_h}},\nonumber\\
& & \left.\gamma_m^{(3)}\right|_{(\ref{vb-i}), (\ref{lambdaL-i})} = \frac{3418281 \left(\frac{3}{2}\right)^{2/3} {g_s}^3
   (-{\log r_h})^{10/3} M^2 {N_f}^{7/3}}{541696 \pi
   ^{7/3} {r_h}^2}-\frac{379809
   \left(\frac{3}{2}\right)^{2/3} {g_s}
   (-{\log r_h})^{4/3} N {N_f}^{4/3}}{67712
   \sqrt[3]{\pi } {r_h}^2},\nonumber\\
& &  \left.\gamma_m^{(4)}\right|_{(\ref{vb-i}), (\ref{lambdaL-i})} = -\frac{126603 \left(\frac{3}{2}\right)^{2/3} {g_s}
   \sqrt[3]{-{\log r_h}} N {N_f}^{4/3}}{16928
   \sqrt[3]{\pi } {r_h}^3},\nonumber\\
& &  \left.\gamma_m^{(5)}\right|_{(\ref{vb-i}), (\ref{lambdaL-i})} = \frac{126603 \left(\frac{3}{2}\right)^{2/3} {g_s}
   \sqrt[3]{-{\log r_h}} N {N_f}^{4/3}}{33856
   \sqrt[3]{\pi } {r_h}^4}.
\end{eqnarray}
Therefore, differential equation of $Z_s(r)$ using (\ref{Zs-EOM}), (\ref{l+m}) and (\ref{gammas}) is obtained as:
\begin{eqnarray}
\label{EOM-Zs}
& & Z_s''(r) -\left(\frac{\gamma_2}{(r-r_h)} + \gamma_3 + \gamma_4(r-r_h)\right)Z_s'(r)
-\left(\frac{\gamma_m^{(2)}}{(r-r_h)} + \gamma_m^{(3)}\right)Z_s(r) = 0.
\end{eqnarray}
With $Z_s(r) = (r - r_h)^\alpha z_s(r)$, $\alpha = 0, 1+\gamma_2$. We choose $\alpha=0$. Comparing  (\ref{EOM-Zs}) with the canonical $(r - r_h)^2 Z_s''(r) + (r - r_h) P(r - r_h) Z_s'(r) + Q(r - r_h) Z_s(r) = 0$, 
\begin{eqnarray}
\label{P+Q}
& & P(r - r_h) = - \gamma_2 - \gamma_3(r - r_h) + ....\equiv \sum_{n=0}^\infty P_n (r - r_h)^n,\nonumber\\
& & Q(r - r_h) = - (r - r_h) \gamma_m^{(2)} - \gamma_m^{(3)} (r - r_h)^2 + ....\equiv \sum_{n=0}^\infty Q_n (r -r_h)^n.
\end{eqnarray} 
Writing
\begin{equation}
\label{zs} 
z_s(r) =\sum_{n=0}^\infty A_n(r - r_h)^n, 
\end{equation}
one obtains:
\begin{eqnarray}
\label{Zs_EOM_IR}
& &  \frac{-{A_0} {\gamma_m^{(2)}}-{A_1}
   \gamma_2}{r-{r_h}}+(-{A_0}
   {\gamma_m^{(3)}}-{A_1} (\gamma_3+{\gamma_m^{(2)}})-2
   {A_2} (\gamma_2-1))\nonumber\\
& & +(r-{r_h}) (-A_0\gamma_m^{(4)} -{A_1}
   (\gamma_4+{\gamma_m^{(3)}})-{A_2} (2
   \gamma_3+{\gamma_m^{(2)}}) - 3 A_3(-2+\gamma_2))\nonumber\\
& & +   (-{A_0}
   \gamma_m^{(5)}-{A_1} {\gamma_5}-{A_1}
   \gamma_m^{(4)}-{A_2} (2
   {\gamma_4}+\gamma_m^{(3)})-{A_3} (3
   {\gamma_3}+{\gamma_m^{(2)}}))(r-{r_h})^2  \nonumber\\
& &  +  (-{A_1} \gamma_m^{(5)}-{A_2} (2
   {\gamma_5}+\gamma_m^{(4)})-{A_3} (3
   {\gamma_4}+\gamma_m^{(3)}) - {A_4}(4\gamma_3 + \gamma_m^{(2)})) (r-{r_h})^3 \nonumber\\
& & + {\cal O}\left((r-{r_h})^4\right) = 0.\nonumber\\ 
\end{eqnarray}

Substituting (\ref{vb-i}) and (\ref{lambdaL-i}) into (\ref{Zs_EOM_IR}),
\begin{itemize}
\item {\bf n=1}
$C_{00} = -\gamma_m^{(2)}, C_{01} = -\gamma_2\neq0$

\item {\bf n=2}
$C_{10} = -\gamma_m^{(3)}, C_{11} = -\left(\gamma_3 + \gamma_m^{(2)}\right),
C_{12} = -2\left(-1 + \gamma_2\right)\neq0$

\item {\bf n=3}
$C_{20} = -\gamma_m^{(4)}, C_{21} = -\left(\gamma_4 + \gamma_m^{(4)}\right), C_{22} = -\left(2\gamma_3 + \gamma_m^{(2)}\right), C_{23} = -3(-2+\gamma_2)\neq0$

\item {\bf n=4}
$C_{30} = -\gamma_m^{(5)}, C_{31} = -\left(\gamma_m^{(4)} + \gamma_5\right),
C_{32} = -\left(2\gamma_4 + \gamma_m^{(3)}\right), C_{33} = -\left(3\gamma_3 + \gamma_m^{(2)}\right), C_{34}=0$.

Now, the $M^{(3)}$ matrix is defined as under:
\begin{eqnarray}
\label{C4x4}
& & M^{(3)} = \left(\begin{array}{cccc} C_{00} & C_{01} & 0 & 0\\ C_{10} & C_{11} & C_{12} & 0 \\
C_{20} & C_{21} & C_{22} & C_{23} \\
C_{30} & C_{31} & C_{32} & C_{33} 
 \end{array}\right),
\end{eqnarray}
where (for n=3) $C_{23} = -3(-2 + \gamma_2)$.  To find an appropriate $r_0/r_h$ \cite{Bulk-Viscosity-McGill-IIT-Roorkee} it would be easier to work with the type IIB side instead of its type IIA mirror as the mirror {\it a la} SYZ keeps the radial coordinate unchanged. To proceed then, let us define an {\it effective} number of three-brane charge as:
\begin{equation}
\label{Neff}
N_{\rm eff}(r) = \int_{\mathbb{M}_5} F_5^{\rm IIB} + \int_{\mathbb{M}_5} B_2^{\rm IIB} \wedge F_3^{\rm IIB}, 
\end{equation}
where $B_2^{\rm IIB}, F_3^{\rm IIB}$ and $F_5^{\rm IIB}$ are given in \cite{metrics}. The five-dimensional internal space $\mathbb{M}_5$,
with coordinates ($\theta_i, \phi_i, \psi$), is basically the base of the resolved warped-defomed conifold. As shown in \cite{Bulk-Viscosity-McGill-IIT-Roorkee}, 
{\footnotesize
\begin {eqnarray}
& & \hskip -0.8in N_{\rm eff}(r_0) = N + {3g_sM^2 \log~r\over 10 r^4} \bigg\{18\pi r (g_sN_f)^2 \log~N \sum_{k = 0}^1\left(18a^2 (-1)^k\log~r + r^2\right)\left({108a^2 \log~r\over 2k+1} + r\right)\nonumber \\
 &&  \hskip -0.8in + 5 \left(3 a^2 ({g_s}-1)+r^2\right) (3 {g_s} {N_f} \log ~r+2\pi ) (9 {g_s} {N_f} \log ~r+4 \pi )
 \left[9 a^2 {g_s} {N_f} \log\left({e^2\over r^3}\right) +4 \pi  r^2\right]\biggr\}, \nonumber\\
 &  &  \hskip -0.8in = N\left[1 + 6\pi\log~r\left(3 g_s N_f \log~r + 2\pi\right)\left(9 g_s N_f \log~r + 4\pi\right){g_sM^2\over N}  \right] +
 {\cal O}\left[{g_sM^2\over N}\left(g_sN_f\right)^2 \log~N\right]. 
 \end {eqnarray}
 }
Note that the assumption of small $r_0$ is crucial here as the same implies the dominance of $g_sN_f|\log~r_0|$ over other constant pieces. Solving for $r_0: N_{\rm eff}(r_0)=0$ yields
\begin{equation}
\label{rh-estimate}
r_0\sim r_h\sim e^{-\kappa_{r_h}(M, N_f, g_s)N^{\frac{1}{3}}}, \kappa_{r_h} = 
\frac{1}{3(6\pi)^{1/3}(g_sN_f)^{2/3}(g_sM^2)^{1/3}}.
\end{equation}
 One sees that for $\kappa_{r_h}(g_s=0.1, M=N_f=3)=0.3$, ${\rm det}M^{(3)}=0$ is satisfied.

\item {\bf n=5}
$C_{40} =0, C_{41} = -\gamma_m^{(5)},
C_{42} = -\left(2\gamma_5 + \gamma_m^{(4)}\right), C_{43} = -\left(3\gamma_4 + \gamma_m^{(3)}\right), C_{44}=-(4\gamma_3 + \gamma_m^{(2)}), C_{45}=0$.

Now, the $M^{(4)}$ matrix has the following structure:
\begin{eqnarray}
\label{C5x5}
& & M^{(4)} = \left(\begin{array}{ccccc} C_{00} & C_{01} & 0 & 0 & 0\\ C_{10} & C_{11} & C_{12} & 0 & 0\\
C_{20} & C_{21} & C_{22} & C_{23} & 0 \\
C_{30} & C_{31} & C_{32} & C_{33} & C_{34} \\
C_{40} & C_{41} & C_{42} & C_{43} & C_{44} \\
 \end{array}\right),
\end{eqnarray}
where (for n=5) $C_{23} = -3(-2 + \gamma_2), C_{34} = -4(-3 + \gamma_2)$. One can verify that in the intermediate-$N$ MQGP limit, for the same set of values of $(g_s, N, M, N_f)$ as for $n=4$, ${\rm det} M^{(4)}\left(g_s=0.1, N=100, M=N_f=3,\kappa_{r_h}=0.3\right)\neq0$.

\end{itemize}

\subsection{${\cal O}(R^4)$ Contributions}
\label{beta-LP-BV}
For the ${\cal O}(\beta)$ metric perturbations defined in (\ref{metric-perturbations}) as $g_{MN}^{\beta}$, linearized equations of motion at ${\cal O}(\beta)$ are obtained by simplifying ``${\rm EOM}_{MN}^{\beta}$'' given in (\ref{EOMs-beta0-beta}) by using $\left|{\cal{C}}_{zz}-2 {\cal{C}}_{\theta_1 z}+2 {\cal{C}}_{\theta_1 x}\right| = 2\left|{\cal{C}}_{\theta_1 x}\right|\ll1$ for a consistent truncation $C^{(1)}_{MNP}=0$ \cite{OR4}. The same are given in (\ref{EOMs-metric-fluctuations}).

Given the decoupling of $S^1_t\times_w\mathbb{R}^3\times\mathbb{R}_{>0}$-metric and the unwarped $S^1_{\cal M}\times_w T^{\rm NE}$-metric, in (\ref{EOMs-beta0-beta}),
\begin{eqnarray}
\label{NC-sufficient-i}
\left(g^{MN}\frac{\delta J_0}{\delta g^{MN}}\right)^{\beta^0} = \left(g^{\mu\nu}\frac{\delta J_0}{\delta g^{\mu\nu}}\right)^{\beta^0}
+ \left(g^{mn}\frac{\delta J_0}{\delta g^{mn}}\right)^{\beta^0},
\end{eqnarray}
where $\mu, \nu = v, x^{1, 2, 3}, r$ and $m, n = \theta_{1,2}, x/\phi_1, y/\phi_2, z/\psi, x^{10}$. Further, for instance, 
\begin{eqnarray}
\label{NC-sufficient-ii}
& & \left.\left(g_{x^1x^1}g^{\mu\nu}\frac{\delta J_0}{\delta g^{\mu\nu}}\right)^{\beta^0}\right|_{(\ref{alpha_theta_12})}
\sim \frac{10^3 M \left(\frac{1}{N}\right)^{19/10} r^2\Sigma_1}{{g_s}^{5/4} {\log N}^2 {N_f} \alpha_{\theta _2}^5}\stackrel{{\rm Table}\ \tcb{2}}{\longrightarrow}{\cal O}(10^{-1})r_h^2\Sigma_1;\nonumber\\
& &  \left.\left(g_{x^1x^1}g^{mn}\frac{\delta J_0}{\delta g^{mn}}\right)^{\beta^0}\right|_{(\ref{alpha_theta_12})}
\sim 10^4\frac{r^2 \Sigma_1}{\sqrt{{g_s}} N^{3/2}  N_f^2 \alpha _{\theta_1}^4 \alpha_{\theta _2}^2}\stackrel{{\rm Table}\ \tcb{2}}{\longrightarrow}{\cal O}(10^{-2})r_h^2\Sigma_1;\nonumber\\
& & \left.\left(\frac{\delta J_0}{\delta g^{x^1x^1}}\right)^{\beta^0}\right|_{(\ref{alpha_theta_12})} \sim 10^7\frac{M \left(\frac{1}{N}\right)^{9/4}
   {r_h}^2 \Sigma_1 \log ({r_h})}{{N_f} \alpha _{\theta _2}^3
   \log ^2(N)}\stackrel{{\rm Table}\ \tcb{2}}{\longrightarrow}{\cal O}(10^2)r_h^2\Sigma_1
\end{eqnarray}
In the intermediate-$N$ MQGP limit (\ref{MQGP_limit}) restricted to (\ref{Ouyang-definition})
\begin{eqnarray}
\label{NC-sufficient-iii}
& & \hskip -0.8in  \left.\left(g_{\mu\nu}g^{\tilde{m}\tilde{n}}\frac{\delta J_0}{\delta g^{\tilde{m}\tilde{n}}}\right)^{\beta^0}\right|_{(\ref{MQGP_limit})\cap (\ref{Ouyang-definition}) } <  \left.\left(g_{\mu\nu}g^{\tilde{\mu}\tilde{\nu}}\frac{\delta J_0}{\delta g^{\tilde{\mu}\tilde{\nu}}}\right)^{\beta^0}\right|_{(\ref{MQGP_limit})\cap (\ref{Ouyang-definition}) } < \left.\left(\frac{\delta J_0}{\delta g^{\mu\nu}}\right)^{\beta^0}\right|_{(\ref{MQGP_limit})\cap (\ref{Ouyang-definition}) }.
\end{eqnarray}
To estimate the contribution of the ${\cal O}(R^4)$ terms to the equations of motion of the metric fluctuations $h^\beta_{\mu\nu}$ from (\ref{EOMs-beta0-beta}), one will hence consider only the linear fluctuations in $g_{\mu\nu}g^{\tilde{\mu}\tilde{\nu}}\frac{\delta J_0}{\delta g^{\tilde{\mu}\tilde{\nu}}} +
\frac{\delta J_0}{\delta g^{\mu\nu}}$. 

Substituting (\ref{metric-perturbations}) into ${\bf {EOM_{MN}^{\beta}}}$ in (\ref{EOMs-beta0-beta}), and looking at the ${\cal O}(\eta^\beta)$ term in \\  $G_{MPQR}G_{N\tilde{P}\tilde{Q}\tilde{R}}g^{P\tilde{P}, \beta}g^{Q\tilde{Q},\beta^0}g^{R\tilde{R},\beta^0}$, one obtains:
\begin{eqnarray}
\label{flux-beta-perturb-i}
& & \left(G_{MPQR}G_{N\tilde{P}\tilde{Q}\tilde{R}}g^{P\tilde{P}, \beta}g^{Q\tilde{Q},\beta^0}g^{R\tilde{R},\beta^0}\right)^{\eta} \nonumber\\
& &  = G_{rPQR}G_{rSTU}\left(h^{PS, \beta} g^{QT,\beta^0}g^{RU, \beta^0} +g^{PS, \beta}h^{QT, \beta^0} g^{RU, \beta^0} + g^{PS, \beta} g^{QT, \beta^0} h^{RU, \beta^0} \right).
\end{eqnarray}
This will be relevant for $M=N=r$ in (\ref{EOMs-beta0-beta}) when looking at metric perturbations only along the non-compact directions $v, r, x^{1, 2, 3}$.

One can show that:
\begin{itemize}
\item
\begin{eqnarray}
\label{flux-beta-perturb-ii}
& & \hskip -0.8in G_{rPQR}G_{rSTU}h^{PS, \beta} g^{QT,\beta^0}g^{RU, \beta^0} \sim
- \frac{g_s^{1/4} N^{3/10} N_f^{2/3}M(h^{\theta_1\theta_1, \beta} + h^{\theta_1\theta_2, \beta})}{r^3}\left(2\log N - \log\left(r^6 + 9 a^2 r^4\right)^{2/3}\right)\nonumber\\
& & \hskip -0.8in  \times \left(N_f r + 72 a^2 N_f \log r + 4 N_f r \log r + 108 a^2 N_f(\log r)^2\right)=0, 
\end{eqnarray}
as we have only considered metric perturbations along the non-compact directions;
\item
\begin{eqnarray}
\label{flux-beta-perturb-iii}
& & \hskip -0.8in G_{rPQR}G_{rSTU}g^{PS, \beta}h^{QT, \beta^0} g^{RU, \beta^0} \nonumber\\
& & \hskip -0.8in  \sim  -\frac{g_s^4 M^2 a_1(r)^{2/3}\left({\cal C}_{zz} - 2 {\cal C}_{\theta_1z}\right)h^{\theta_2\theta_2, \beta^0}N^{13/10}}{b_1(r)^2}\left(N_f r + 72 a^2 N_f \log r + 4 N_f r \log r + 108 a^2 N_f(\log r)^2\right)=0,\nonumber\\
& & 
\end{eqnarray}
for a pair of reasons: ${\cal C}_{zz} - 2 {\cal C}_{\theta_1z} = 0$ if $C^{(\beta)}_{MNP}=0$ \cite{OR4}, and we have only considered metric perturbations along the non-compact directions. In (\ref{flux-beta-perturb-iii}),  $a_1(r) = \left(\frac{3}{8\pi}\right)\left(2\log N - \log\left(r^6 + 9 a^2 r^4\right)\right)$.

\item
\begin{eqnarray}
\label{flux-beta-perturb-iv}
& & \hskip -0.8in G_{rPQR}G_{rSTU}g^{PS, \beta} g^{QT, \beta^0} h^{RU, \beta^0} \nonumber\\
& & \hskip -0.8in  \sim 
\frac{g_s^{7/2}M^2N_f^4\left({\cal C}_{zz} - 2 {\cal C}_{\theta_1z}\right) N h^{xx, \beta^0}}{a_1(r)^{4/3}b_1(r)^2}\left(N_f r + 72 a^2 N_f \log r + 4 N_f r \log r + 108 a^2 N_f(\log r)^2\right)^2=0,\nonumber\\
& & 
\end{eqnarray}
for the same pair of reasons as (\ref{flux-beta-perturb-iii}).

As one will see in (\ref{EOMs-beta}), $p_{37}....p_{42}$ figure in ${\bf EOM}_{rr}^\beta$ as coefficients of $h_{x_{1,2}x_{1,2}}^{\beta,\ (n)}, n=0, 1, 2$ (corresponding respectively to $h_{x_{1,2}x_{1,2}}$, its first order and second order derivatives) and $h_{rr}^\beta$, which are like $\frac{\sqrt{g_s N}}{N_f^{2/3}|\log r_h|^{2/3}r_h^{2/3/4}}$. Hence, the flux contributions may be disregarded as compared to other terms on the LHS of (\ref{EOMs-beta}).

\end{itemize}
 In the IR, the ${\cal O}(\beta)$ equations of motion (\ref{EOMs-beta}) in the $h_{r\mu}=0(\mu=v, x^{1, 2, 3})$-gauge, are simplified to read:
\begin{eqnarray*}
& & \hskip -0.4in {\rm EOM}_{vv}^\beta: -\frac{1760512 \sqrt[3]{2} \sqrt[6]{\pi }
   \sqrt{\frac{1}{N}} {r_h}^7 \log
   ^5({r_h}) \left(3 \sqrt{2}
   {h_{vx^1}^\beta}'(r)+2 \sqrt{3}
   \left({h_{x^1x^1}^\beta}'(r)+2
   {h_{x^2x^2}^\beta}'(r)\right)\right)}{49995\
   3^{5/6} \sqrt{{g_s}}
   {N_f}^{2/3}} \nonumber\\
 & & \hskip -0.4in -\frac{1664
   \sqrt[3]{\frac{2}{3}} \pi ^{8/3}
   {h_{vv}^{\ \beta^0}}(r)}{19683 {N_f}^{8/3}}=0;\nonumber\\
& & \hskip -0.4in {\rm EOM}_{vx_1}^\beta: \frac{128 i \sqrt[3]{2} \pi ^{13/6}
   \sqrt{\frac{1}{N}} }{9939915\ 3^{5/6}
   \sqrt{{g_s}} {N_f}^{8/3} {r_h}^5
   |\log r_h|^{8/3}}\Biggl(974145
   {h_{vv}^{\ \beta^0}}(r) +406272
   {h_{x^2x^2}^{\ \beta^0}}(r) \log
   ^3({r_h})\nonumber\\
   & &\hskip -0.4in -5 \left(125745
   {h_{x^2x^2}^{\ \beta^0}}'(r)+102616
   {h_{x^3x^3}^{\ \beta^0}}'(r)+406272 {h_{x^3x^3}^{\ \beta^0}}(r)
   \log ^3({r_h})\right)\Biggr)=0;\nonumber\\
& & \hskip -0.4in {\rm EOM}_{vr}^{\ \beta^0}: -\frac{3848 \sqrt[3]{2} \pi ^{19/6}
   \sqrt{{g_s}} \sqrt{N}
   {h_{vv}^{\ \beta^0}}(r)}{19683\ 3^{5/6}
   {N_f}^{8/3} {r_h}^8 |\log r_h|^{8/3}}-\frac{16928 \sqrt[3]{2} \pi
   ^{2/3} |\log r_h|^{7/3} \left(35
   \sqrt{6} {h_{vx^1}^\beta}(r)+48
   {h_{x^1x^1}^\beta}(r)\right)}{16665\ 3^{5/6}
   {N_f}^{2/3} {r_h}}=0;\nonumber
   \end{eqnarray*}
   \begin{eqnarray*}
& & \hskip -0.4in {\rm EOM}_{x^1x^1}^\beta:\nonumber\\
& & \hskip -0.4in -\frac{16928 \sqrt[3]{2} \sqrt[6]{\pi } \sqrt{\frac{1}{N}} {r_h} \log ^2({r_h}) }{25247475\ 3^{5/6} \sqrt{{g_s}} {N_f}^{2/3}} 
 \Biggl[4 {r_h} \Biggl(505
   \sqrt[3]{-\log ({r_h})} \left(3 \sqrt{2} {h_{vx^1}^\beta}'(r)+2 \sqrt{3} \left({h_{x^1x^1}^\beta}'(r)+13
   {h_{x^2x^2}^\beta}'(r)\right)\right)\nonumber\\
& & \hskip -0.4in   -50784 \sqrt{3} {h_{x^2x^2}^\beta}(r) \log ^4({r_h})\Biggr)-13635 \sqrt{2}
   {h_{vx^1}^\beta}(r) \sqrt[3]{-\log ({r_h})}-31310 \sqrt{3} {h_{x^1x^1}^\beta}(r) \sqrt[3]{-\log
   ({r_h})}\Biggr] \nonumber\\ 
& & \hskip -0.4in = \frac{118400 \sqrt[3]{\frac{2}{3}} \pi ^{8/3} {h_{vv}^{\beta^0}}(r)}{2187  {N_f}^{8/3} {r_h}^6 (-\log (r))^{8/3}};\nonumber\\   
& & \hskip -0.4in {\rm EOM}_{x^{2,3}x^{2,3}}^\beta:\nonumber\\
& & \hskip -0.4in -\frac{16928 \sqrt[3]{2} \sqrt[6]{\pi } \sqrt{\frac{1}{N}} {r_h} (-\log ({r_h}))^{7/3} }{5049495\ 3^{5/6} \sqrt{{g_s}} {N_f}^{2/3}}\nonumber\\
& & \hskip -0.4in \times \Biggl[2 {r_h}
   \left(3939 \sqrt{2} {h_{vx^1}^{\beta}}'(r)+202 \sqrt{3} \left(13 {h_{x^1x^1}^{\beta}}'(r)+15{h_{x^2x^2}^\beta}'(r)\right)+76176
   \sqrt{3} {h_{x^2x^2}^\beta}(r) \log ^3({r_h})\right)\nonumber\\
& & \hskip -0.4in-2727 \sqrt{2} {h_{vx^1}^\beta}(r)+404 \sqrt{3}
{h_{x^1x^1}^\beta}(r)\Biggr] = \frac{4736 \sqrt[3]{\frac{2}{3}} \pi ^{8/3} {h_{vv}^{\beta^0}}(r)}{19683
   {N_f}^{8/3} {r_h}^7 (-\log ({r_h}))^{8/3}};\nonumber\\
\end{eqnarray*}
\begin{eqnarray}
\label{EOMs-IR-simp}
& & \hskip -0.4in {\rm EOM}_{x_1r}^\beta: \frac{8 \sqrt[3]{2} \pi ^{2/3}
   {h_{vx^1}^\beta}(r)}{9\ 3^{5/6} {N_f}^{2/3}
   {r_h}^2 |\log r_h|^{2/3}}-\frac{256 i
   \sqrt[3]{\frac{2}{3}} \pi ^{8/3} \left(11
   {h_{vv}^{\ \beta^0}}(r)-75 {h_{x^2x^2}^{\ \beta^0}}'(r)-67
   {h_{x^3x^3}^{\ \beta^0}}'(r)\right)}{6561
   {N_f}^{8/3} {r_h}^7 |\log r_h|^{8/3}}=0;\nonumber\\
& & \hskip -0.4in {\rm EOM}_{rr}^\beta: -\frac{26 \sqrt[3]{\frac{2}{3}} \pi ^{7/6}
   \sqrt{{g_s}} \sqrt{N}
   ({h_{x^1x^1}^\beta}(r)+2 {h_{x^2x^2}^\beta}(r))}{9
   {N_f}^{2/3} {r_h}^4 |\log r_h|^{2/3}}-\frac{832
   \sqrt[3]{\frac{2}{3}} \pi ^{8/3} r
   {h_{vv}^{\ \beta^0}}(r)}{59049 {N_f}^{8/3}
   {r_h}^7 |\log r_h|^{8/3}}=0.         
\end{eqnarray}
One sees that (\ref{EOMs-IR-simp}) can be solved consistently by either of the following: 
\begin{itemize}
\item
\begin{eqnarray}
\label{hMNbeta-i}
& & h^\beta_{x^1x^1, x^{2/3}x^{2/3}, vv, vx^1}(r)=0;\nonumber\\
& & h^{\beta^0}_{vv}(r) = h^{\beta^0}_{x^2x^2}(r) = h^{\beta^0}_{x^3x^3}(r) = 0;
\end{eqnarray}
$h^{\beta^0}_{x^1x^1,vx^1}(r)$ are determined from (\ref{pEOM-beta0}).
One hence sees that $Z_s$ receives no ${\cal O}(R^4)$ corrections in the IR implying a non-renormalization of $v_b$ and $\lambda_L$ up to ${\cal O}(R^4)$.

\item {\bf A more general solution}:

Using (\ref{rh-estimate}), we see that in the IR $r^x\left(-\log r\right)^y \sim e^{-\kappa_{r_h} x N^{1/3}}\kappa_{r_h}^y N^{y/3}, x>0, y\in\mathbb{R}$. Hence, in the intermediate-$N$ limit (\ref{MQGP_limit}), $r^x\left(-\log r\right)^y$ may be dropped largely due to the $N$-dependent exponential suppression. In the IR, assuming $h_{vv}^{\beta^0}(r)\sim h_{x^2x^2}^{\beta^0}\ '(r)$, one therefore sees that (\ref{EOMs-IR-simp}) can be consistently solved in this approximation by:
\begin{eqnarray}
\label{hMNbeta-ii}
& & h^\beta_{x^1x^1, x^{2/3}x^{2/3}, vv, vx^1}(r)=0;\nonumber\\
& & 974145
   {h_{vv}^{\ \beta^0}}(r) +406272
   {h_{x^2x^2}^{\ \beta^0}}(r) \log
   ^3({r_h})\nonumber\\
& &  -5 \left(125745
   {h_{x^2x^2}^{\ \beta^0}}'(r)+102616
   {h_{x^3x^3}^{\ \beta^0}}'(r)+406272 {h_{x^3x^3}^{\ \beta^0}}(r)
   \log ^3({r_h})\right) = 0;\nonumber\\
& & 11
   {h_{vv}^{\ \beta^0}}(r)-75 {h_{x^2x^2}^{\ \beta^0}}'(r)-67
   {h_{x^3x^3}^{\ \beta^0}}'(r).
\end{eqnarray}
Making an ansatz,
\begin{equation}
\label{hvvbeta0-ansatz}
h^{\beta^0}_{vv}(r) = {\cal C}_{ h^{\beta^0}_{vv}}\ r^{\mu+7}\left(\log r\right)^\nu, \mu>0, \nu, {\cal C}_{ h^{\beta^0}_{vv}}\in\mathbb{R},
\end{equation}
and assuming $h_{x^2x^2}^{\ \beta^0}(r) = h_{x^3x^3}^{\ \beta^0}(r)$ the second equation of (\ref{hMNbeta-ii}) is solved to yield:
\begin{equation}
\label{hx2x2beta0-solution-i}
h_{x^2x^2}^{\ \beta^0}(r\sim r_h) = e^{-\frac{1625088 r \log ^3\left(r_h\right)}{1141805}} \left(\int _{r_h}^r\frac{1929
   e^{\frac{1625088 \xi \log ^3\left(r_h\right)}{1141805}} \xi ^{X+7} \log ^\nu(\xi )
   {\cal C}_{ h^{\beta^0}_{vv}}}{2261}d\xi +c_1\right).
\end{equation}
Setting $c_1=0$, expanding the exponentials inside and outside the integral, and using:
\begin{equation}
\label{integral-1}
\int_{r_h}^r r^{7 + \mu + n}(\log r)^y dr = \left.\frac{\log ^\nu(r) ((-z-1) \log (r))^{-y} \Gamma (\nu+1,(-\mu - n - 8) \log (r))}{8 + \mu + n}\right|_{r_h}^r,\ n\in\mathbb{Z}^+\cup\left\{0\right\}
\end{equation}
which, given that $|\log r|\sim N^{1/3}>1$ and assuming $r = r_h(1 + \delta), 0<\delta\ll1$ for $r\in$IR, yields:
\begin{equation}
\label{integral-2}
\int_{r_h}^{r = r_h(1+\delta)} r^{7 + \mu + n}(\log r)^\nu dr\sim \epsilon r_h^{8 + \mu + n}(\log r_h)^\nu < r_h^{7 + \mu}(\log r)^y.
\end{equation}
From (\ref{hx2x2beta0-solution-i}) - (\ref{integral-2}), one sees that:
\begin{equation}
\label{dhx2x2beta0-IR}
h_{x^2x^2}^{\ \beta^0}\ '(r\in{\rm IR}) \approx 0.9 {\cal C}_{ h^{\beta^0}_{vv}}r_h^{7+\mu}.
\end{equation}
From (\ref{hMNbeta-ii}), one sees that ${h_{x^2x^2}^{\ \beta^0}}'(r\sim r_h)
= \frac{11}{142}h_{vv}^{\beta^0}(r\sim r_h)$. Thus, upon comparison with (\ref{dhx2x2beta0-IR}) yields:
\begin{equation}
\label{Cvvbeta0}
0.9{\cal C}_{ h^{\beta^0}_{vv}} =  \frac{11}{142}.
\end{equation}

\end{itemize}

\section{Conclusion}
\label{conclusion}

In this paper, we have addressed the question: ``Is thermal QCD chaotic above the deconfinement temperature?'' by looking at the phenomenon of pole-skipping in the solution to the equation of motion of a gauge-invariant combination $Z_s$ of scalar modes of metric perturbations in a top-down finite-but-large-$N$ thermal ${\cal M}$-theory dual at intermediate coupling. The following summarizes the important Physics takeaways.

\begin{itemize}
\item
\begin{itemize}
\item
The horizon $r=r_h$ unlike all bottom-up computations available in the literature, turns out to be an irregular singular point of the second-order homogenous differential equation corresponding to the equation of motion for $Z_s(r)$. The coefficient of $Z_s(r)$ near $r=r_h$, instead of having a pole of order two, turns out to have a pole of order four; in fact the residue of the term of pole order three is also generically non-trivial. However, at specific values of the frequency and momentum, residues of both pole orders, vanish. We then use these values to read off the analogs of the ``Lyapunov exponent'' $\lambda_L$ and ``butterfly velocity'' $v_b$.

\item
The ``Lyapunov exponent'' $\lambda_L$ satisfies the Maldacena-Shenker-Stanford (MSS) bound.

\item
The ``butterfly velocity'' $v_b $ obtained by requiring the horizon to be a regular singular point, matches, up to LO in $N$ the conformal expression obtained from general arguments of \cite{Ageev:2021poy} that is valid for $AdS$ backgrounds; the non-conformal corrections yield positive shifts to the conformal value. Requiring $\mathbb{R}^3$-isotropy yields non-renormalization up to ${\cal O}(R^4)$ of the gauge-invariant combination of scalar modes of metric perturbations, implying a non-renormalization of the ``butterfuly velocity''. Interestingly, assuming the expression for $v_b$ for $AdS$ backgrounds as obtained in \cite{Ageev:2021poy} to also be applicable for our framework also yields a vanishing correction up to ${\cal O}(R^4)$ of $v_b$.

\end{itemize}  

\item
Truncating the incoming mode solution of $Z_s$ as a power series around $r_h$,  yields a ``missing pole'', i.e., ``$C_{n, n+1}=0,\ {\rm det}\ M^{(n)}=0'', n\in\mathbb{Z}^+$ is satisfied for {\it a single} $n\geq3$ depending on the values of  $g_s, M$ (number of fractional $D3$ branes), $N$ (number of color $D3$-branes) and flavor $D7$-branes $N_f$ in the parent type IIB set \cite{metrics}. Further, as an example, for the QCD(EW-scale)-inspired $N=100, M=N_f=3, g_s=0.1$, one finds a missing pole at $n=3$. For integral $n>3$, truncating $Z_s$ at ${\cal O}((r-r_h)^n)$, yields $C_{n, n+1}=0$ at order $n,\ \forall n\geq3$. 

\item
Generically, there is a one-parameter family of ${\cal O}(R^4)$ corrections to  $Z_s$. However, assuming preservation of isotropy in $\mathbb{R}^3$ even with the inclusion of ${\cal O}(R^4)$ corrections, $Z_s$ receives no ${\cal O}(R^4)$ corrections. Hence, (the aforementioned analog of) the Lyapunov exponent, just like the analog of butterfuly velocity, is unrenormalized up to ${\cal O}(R^4)$ in ${\cal M}$ theory.

\end{itemize}

\section*{Acknowledgements}
 GY thanks the Infosys Foundation for the partial support at CMI. SSK is supported by a Junior
Research Fellowship (JRF) from the Ministry of Human Resource and Development (MHRD), Govt. of India. AM is partly supported by a Core Research Grant number SER-1829-PHY from the Science and Engineering Research Board, Govt. of India. One of us (AM) would like to thank J. Maldacena, M. Mezei and K. Sil for very useful clarifications. One of us (SSK) would like to thank IIT Roorkee for high-end computational facilities. We thank the anonymous referee for the many useful comments that improved the paper's presentation. 

\appendix
\section{Linearized Equations of Motion at ${\cal O}(\beta^0)$ and  ${\cal O}(\beta)$ }
\label{EOMs}
\setcounter{equation}{0}
\seceqaa
\subsection{${\cal O}(\beta^0)$ Equations of Motion}
\label{EOMs-beta0}
\begin{eqnarray}
\label{pEOM-beta0}
& & {\rm {\bf EOM_{vv}^{\beta^0}}}:\nonumber\\
   & &
h_{vv}(r)y_1(r)+12 \iota  q y_{3}(r) h_{v x_1}'(r)+h_{v x_1}(r) y_{2}(r)+12 \iota  w y_{3}(r) h_{x_1 x_1}'(r)+h_{x_1 x_1}(r) y_{4}(r)\nonumber\\
   & &+24 \iota  w y_{3}(r)
   h_{x_2 x_2}'(r)+h_{x_2 x_2}(r) y_{5}(r)=0, \nonumber\\
   & & {\rm {\bf EOM_{v x_1}^{\beta^0}}}:\nonumber\\
   & & \iota  w y_{22}(r) h_{v x_1}'(r)-\Biggl[-\iota  q y_{22}(r) h_{vv}'(r)+h_{x_2 x_2}(r) y_{23}(r) -\frac{1}{y_{1}(r)}\Biggl(y_{21}(r) \Biggl(12 \iota  q y_{3}(r) h_{vx1}'(r)\nonumber\\
   & &+h_{v x_1}(r) y_{2}(r)+12 \iota  w y_{3}(r)
   h_{x_1 x_1}'(r)+h_{x_1 x_1}(r) y_{4}(r)+24 \iota  w y_{3}(r) h_{x_2 x_2}'(r)+h_{x_2 x_2}(r) y_{5}(r)\Biggr)\Biggr)\Biggr]=0,
   \nonumber\\
   & & {\rm {\bf EOM_{v r}^{\beta^0}}}:\nonumber\\
   & &
   h_{vv}(r) y_{6}(r)+\iota  q y_{7}(r) h_{v x_1}'(r)+h_{v x_1}(r) y_{8}(r)+\iota  w y_{7}(r)h_{x_1 x_1}'(r)+h_{x_1 x_1}(r) y_{9}(r)\nonumber\\
   & &+2 \iota  w y_{7}(r)
   h_{x_2 x_2}'(r)+h_{x_2 x_2}(r) y_{10}(r)=0
    \nonumber\\
   & & {\rm {\bf EOM_{x_1 x_1}^{\beta^0}}}:\nonumber\\
   & & h_{x_1 x_1}(r)y_{13}(r)+6 \iota  q y_{12}(r) h_{v x_1}'(r)+h_{v x_1}(r) y_{11}(r)+6 \iota  w y_{12}(r) h_{x_1 x_1}'(r)+6 \iota  w y_{12}(r) h_{x_2 x_2}'(r)
 \nonumber\\
   & &   +h_{x_2 x_2}(r)
   y_{14}(r)=0,
   \nonumber\\
   & & {\rm  {\bf EOM_{x_1 r}^{\beta^0}}}:\nonumber\\
   & & h_{rr}(r) s_{14}(r)+s_{16}(r) h_{v x_1}''(r)+s_{15}(r) h_{v x_1}'(r)+h_{v x_1}(r) s_{11}(r)+h_{x_1 x_1}(r) s_{12}(r)+s_{15}(r) h_{x_2 x_2}'(r)\nonumber\\
   & &-2 h_{x_2 x_2}(r)
   s_{12}(r)+s_{15}(r) h_{x_3 x_3}'(r)=0, 
   \nonumber\\
   & & {\rm {\bf EOM_{x_2 x_2}^{\beta^0}}}={\rm {\bf EOM_{x_3 x_3}^{\beta^0}}}: \nonumber\\
   & & h_{x_2 x_2}(r)y_{19}(r)+h_{vv}(r) y_{15}(r)+6 \iota  q y_{17}(r) h_{v x_1}'(r)+h_{v x_1}(r) y_{16}(r)+6 \iota  w y_{17}(r) h_{x_1 x_1}'(r) \nonumber\\
   & &+h_{x_1 x_1}(r) y_{18}(r)+6 \iota 
   w y_{17}(r) h_{x_2 x_2}'(r)=0, 
   \nonumber\\
   & & {\rm {\bf EOM_{rr}^{\beta^0}}}: \nonumber\\
   & & w^2 s_{7}(r)h_{rr}(r)-s_{7}(r) h_{vv}''(r)+h_{vv}(r) s_{1}(r)+s_{9}(r) h_{v x_1}'(r)+h_{v x_1}(r) s_{2}(r)+s_{10}(r) h_{x_1 x_1}'(r)\nonumber\\
   & &+h_{x_1 x_1}(r) s_{4}(r)+2
   s_{10}(r) h_{x_2 x_2}'(r)+h_{x_2 x_2}(r) s_{6}(r),
\end{eqnarray}
where $y_i(r)$ and $s_i(r)$ are defined in appendix \ref{yi-si} and $h_{MN} \equiv h_{MN}^{\beta^0}$ in (\ref{pEOM-beta0}).

\subsection{${\cal O}(\beta)$ Equations of Motion}
\label{EOMs-beta-ap}
{\footnotesize
\begin{eqnarray}
\label{EOMs-metric-fluctuations}
& & \hskip -0.3in {\bf EOM_{vv}^{\beta}}:\nonumber\\
  & &  {h_{vv}^{\beta}}(r) {p_1}(r)+{p_3}(r) {h_{v x_1}^{\beta}}^{'}(r)+{h_{v x1}^{\beta}}(r) p_{_2}(r)+p_{_5}(r)
   {h{x_1 x_1}^{\beta}}'(r)+{h_{x_1 x_1}^{\beta}}(r) p_{4}(r)+p_{7}(r) {h_{x_2 x_2}^{\beta}}'(r)+{h{_{x_2 x_2}}^{\beta}}(r)
   p_{6}(r)=\nonumber\\
  & &- \frac{1}{2} \Biggl(v_{19}(r)(r) {h_{x_1 x_1}^{\beta}}''(r)+v_{16}(r) {h_{rr}^{\beta}}'(r)+{h_{rr}^{\beta}}(r)
   v_{8}(r)+v_{11}(r) {h_{v r}^{\beta}}'(r)+{h_{v r}^{\beta}}(r) v_{3}(r)+v_{17}(r)
   {h_{vv}^{\beta}}''(r)+v_{9}(r) {h_{vv}^{\beta}}'(r)\nonumber\\
   & &+{h_{vv}^{\beta}}(r) v_{1}(r) +v_{18}(r)
   {h_{v x_1}^{\beta}}''(r)+v_{10}(r) {h_{v x_1}^{\beta}}'(r)+{h_{v x_1}^{\beta}}(r) v_{2}(r)+v_{13}(r)
   {h_{x_1 r}^{\beta}}'(r)+{h_{x_1 r}^{\beta}}(r) v_{5}(r)+v_{12}(r) {h_{x_1 x_1}^{\beta}}'(r)\nonumber\\
   & &+{h_{x_1 x_1}^{\beta}}(r)
   v_{4}(r)+v_{20}(r) {h_{x_2 x_2}^{\beta}}''(r)+v_{14}(r) {h_{x_2 x_2}^{\beta}}'(r)+{h_{x_2 x_2}^{\beta}}(r)
   v_{6}(r)+v_{21}(r) {h_{x_3 x_3}^{\beta}}''(r)+v_{15}(r) {h_{x_3 x_3}^{\beta}}'(r)+{h_{x_3 x_3}^{\beta}}(r)
   v_{7}(r)\Biggr), \nonumber\\
  & & \hskip -0.3in {\bf EOM_{v x_1}^{\beta}}:\nonumber\\
  & &  p_{9}(r) {h_{vv}^{\beta}}'(r)+{h_{vv}^{\beta}}(r) p_{8}(r)+p_{10}(r)
   {h_{v x_1}^{\beta}}'(r)+{h_{x_2 x_2}^{\beta}}(r) p_{11}(r)\nonumber\\
  & & = -\frac{1}{2} \Biggl({h_{rr}^{\beta}}(r) v_{29}(r)+v_{32}(r) {h_{v r}^{\beta}}'(r)+{h_{v r}^{\beta}}(r) v_{24}(r)+v_{30}(r)
   {h_{vv}^{\beta}}'(r)+{h_{vv}^{\beta}}(r) v_{22}(r)+v_{36}(r) {h_{v x_1}^{\beta}}''(r)+v_{31}(r)
   {h_{v x_1}^{\beta}}'(r)\nonumber\\
   & &+{h_{v x_1}^{\beta}}(r) v_{23}(r)+v_{33}(r) {h_{x_1 r}^{\beta}}'(r)+{h_{x_1 r}^{\beta}}(r)
   v_{26}(r)+{h_{x_1 x_1}^{\beta}}(r) v_{25}(r)+v_{34}(r) {h_{x_2 x_2}^{\beta}}'(r)+{h_{x_2 x_2}^{\beta}}(r)
   v_{27}(r)+v_{35}(r) {h_{x_3 x_3}^{\beta}}'(r)\nonumber\\
   & &+{h_{x_3 x_3}^{\beta}}(r) v_{28}(r)\Biggr),  \nonumber\\
  & & \hskip -0.3in {\bf EOM_{v r}^{\beta}}:\nonumber\\
  & &   p_{13}(r) {h_{v x_1}^{\beta}}'(r)+{h_{v x_1}^{\beta}}(r) p_{12}(r)+p_{15}(r)
   {h_{x_1 x_1}^{\beta}}'(r)+{h_{x_1 x_1}^{\beta}}(r) p_{14}(r)+p_{17}(r) {h_{x_2 x_2}^{\beta}}'(r)+{h_{x_2 x_2}^{\beta}}(r)
   p_{16}(r) \nonumber\\
  & & = -\frac{1}{2} \Biggl(v_{52}(r) {h_{rr}^{\beta}}'(r)+{h_{rr}^{\beta}}(r) v_{44}(r)+v_{47}(r)
   {h_{v r}^{\beta}}'(r)+{h_{v r}^{\beta}}(r) v_{39}(r)+v_{53}(r) {h_{vv}^{\beta}}''(r)+v_{45}(r)
   {h_{vv}^{\beta}}'(r)+{h_{vv}^{\beta}}(r) v_{37}(r)\nonumber\\
   & &+v_{54}(r) {h_{v x_1}^{\beta}}''(r)+v_{46}(r)
   {h_{v x_1}^{\beta}}'(r)+{h_{v x_1}^{\beta}}(r) v_{38}(r)+v_{49}(r) {hx1r}'(r)+{h_{x_1 r}^{\beta}}(r)
   v_{41}(r)+v_{55}(r) {h_{x_1 x_1}^{\beta}}''(r)+v_{48}(r) {h_{x_1 x_1}^{\beta}}'(r)\nonumber\\
   & &+{h_{x_1 x_1}^{\beta}}(r)
   v_{40}(r)+v_{56}(r) {h_{x_2 x_2}^{\beta}}''(r)+v_{50}(r) {h_{x_2 x_2}^{\beta}}'(r)+{h_{x_2 x_2}^{\beta}}(r)
   v_{42}(r)+v_{57}(r) {h_{x_3 x_3}^{\beta}}''(r)+v_{51}(r) {h_{x_3 x_3}^{\beta}}'(r)\nonumber\\
   & &+{h_{x_3 x_3}^{\beta}}(r)
   v_{43}(r)\Biggr)-{h_{rr}^{\beta}}(r) z_{1}(r), \nonumber\\
  & & \hskip -0.3in {\bf EOM_{x_1 x_1}^{\beta}}: \nonumber\\
  & &  p_{19}(r) {h_{v x_1}^{\beta}}'(r)+{h_{v x_1}^{\beta}}(r) p_{18}(r)+p_{21}(r)
   {h_{x_1 x_1}^{\beta}}'(r)+{h_{x_1 x_1}^{\beta}}(r) p_{20}(r)+p_{23}(r) {h_{x_2 x_2}^{\beta}}'(r)+{h_{x_2 x_2}^{\beta}}(r)
   p_{22}(r) \nonumber\\
  & & = -\frac{1}{2} \Biggl(v_{73}(r) {h_{rr}^{\beta}}'(r)+{h_{rr}^{\beta}}(r) v_{65}(r)+v_{68}(r)
   {h_{v r}^{\beta}}'(r)+{h_{v r}^{\beta}}(r) v_{60}(r)+v_{74}(r) {h_{vv}^{\beta}}''(r)+v_{66}(r)
   {h_{vv}^{\beta}}'(r)+{h_{vv}^{\beta}}(r) v_{58}(r)\nonumber\\
   & &+v_{75}(r) {h_{v x_1}^{\beta}}''(r)+v_{67}(r)
   {h_{v x_1}^{\beta}}'(r)+{h_{v x_1}^{\beta}}(r) v_{59}(r)+v_{70}(r) {h_{x_1 r}^{\beta}}'(r)+{h_{x_1 r}^{\beta}}(r)
   v_{62}(r)+v_{76}(r) {h_{x_1 x_1}^{\beta}}''(r)+v_{69}(r) {h_{x_1 x_1}^{\beta}}'(r)\nonumber\\
   & &+{h_{x_1 x_1}^{\beta}}(r)
   v_{61}(r)+v_{77}(r) {h_{x_2 x_2}^{\beta}}''(r)+v_{71}(r) {h_{x_2 x_2}^{\beta}}'(r)+{h_{x_2 x_2}^{\beta}}(r)
   v_{63}(r)+v_{78}(r) {h_{x_3 x_3}^{\beta}}''(r)+v_{72}(r) {h_{x_3 x_3}^{\beta}}'(r)+{h_{x_3 x_3}^{\beta}}(r)
   v_{64}(r)\Biggr)\nonumber\\
   & &-{h_{rr}^{\beta}}(r) z_{2}(r), \nonumber\\
   & &\hskip -0.3in {\bf EOM_{x_1 r}^{\beta}}: \nonumber\\
  & &   {h_{rr}^{\beta}}(r) p_{30}(r)+p_{26}(r) {h_{v x_1}^{\beta}}''(r)+p_{25}(r)
   {h_{v x_1}^{\beta}}'(r)+{h_{v x_1}^{\beta}}(r) p_{24}(r)+{h_{x_1 x_1}^{\beta}}(r) p_{27}(r)+p_{29}(r)
   {h_{x_2 x_2}^{\beta}}'(r)+{h_{x_2 x_2}^{\beta}}(r) p_{28}(r)\nonumber\\
  & & = -\frac{1}{2} \Biggl({h_{rr}^{\beta}}(r) v_{86}(r)+v_{89}(r) {h_{v r}^{\beta}}'(r)+{h_{v r}^{\beta}}(r) v_{81}(r)+v_{87}(r)
   {h_{vv}^{\beta}}'(r)+{h_{vv}^{\beta}}(r) v_{79}(r)+v_{93}(r) {h_{v x_1}^{\beta}}''(r)+v_{88}(r)
   {h_{v x_1}^{\beta}}'(r)\nonumber\\
   & &+{h_{v x_1}^{\beta}}(r) v_{80}(r)+v_{90}(r) {h_{x_1 r}^{\beta}}'(r)+{h_{x_1 r}^{\beta}}(r)
   v_{83}(r)+{h_{x_1 x_1}^{\beta}}(r) v_{82}(r)+v_{91}(r) {h_{x_2 x_2}^{\beta}}'(r)+{h_{x_2 x_2}^{\beta}}(r)
   v_{84}(r)+v_{92}(r) {h_{x_3 x_3}^{\beta}}'(r)\nonumber\\
   & &+{h_{x_3 x_3}^{\beta}}(r) v_{85}(r)\Biggr), \nonumber
   \end{eqnarray}   
   }
   {\footnotesize
   \begin{eqnarray}
   \label{EOMs-beta}
  & & \hskip -0.3in {\bf EOM_{x_2 x_2}^{\beta}}:\nonumber\\
  & & \hskip -0.3in  p_{32}(r) {h_{v x_1}^{\beta}}'(r)+{h_{v x_1}^{\beta}}(r) p_{31}(r)+p_{34}(r)
   {h_{x_1 x_1}^{\beta}}'(r)+{h_{x_1 x_1}^{\beta}}(r) p_{33}(r)+p_{36}(r) {h_{x_2 x_2}^{\beta}}'(r)+{h_{x_2 x_2}^{\beta}}(r)
   p_{35}(r)\nonumber\\
  & & \hskip -0.3in = -\frac{1}{2} \Biggl(v_{109}(r) {h_{rr}^{\beta}}'(r)+{h_{rr}^{\beta}}(r) v_{101}(r)+v_{104}(r)
   {h_{v r}^{\beta}}'(r)+{h_{v r}^{\beta}}(r) v_{96}(r)+v_{110}(r) {h_{vv}^{\beta}}''(r)+v_{102}(r)
   {h_{vv}^{\beta}}'(r)+{h_{vv}^{\beta}}(r) v_{94}(r)\nonumber\\
   & &+v_{103}(r) {h_{v x_1}^{\beta}}'(r)+{h_{v x_1}^{\beta}}(r)
   v_{95}(r)+v_{106}(r) {h_{x_1 r}^{\beta}}'(r)+{h_{x_1 r}^{\beta}}(r) v_{98}(r)+v_{111}(r)
   {h_{x_1 x_1}^{\beta}}''(r)+v_{105}(r) {h_{x_1 x_1}^{\beta}}'(r)+{h_{x_1 x_1}^{\beta}}(r) v_{97}(r)\nonumber\\
   & &+v_{112}(r)
   {h_{x_2 x_2}^{\beta}}''(r)+v_{107}(r) {h_{x_2 x_2}^{\beta}}'(r)+{h_{x_2 x_2}^{\beta}}(r) v_{99}(r)+v_{113}(r)
   {h_{x_3 x_3}^{\beta}}''(r)+v_{108}(r) {h_{x_3 x_3}^{\beta}}'(r)+{h_{x_3 x_3}^{\beta}}(r) v_{100}(r)\Biggr)-{h_{vv}^{\beta}}(r) z_{3}(r)\nonumber\\
  & & \hskip -0.3in -{h_{v x_1}^{\beta}}(r) z_{4}(r)-{h_{x_1 x_1}^{\beta}}(r)
   z_{5}(r)-{h_{x_2 x_2}^{\beta}}(r) (z_{6}(r)+z_{7}(r)),  \nonumber\\  
   & & \hskip -0.3in {\bf EOM_{x_3 x_3}^{\beta}}:\nonumber\\
  & & \hskip -0.3in p_{32}(r) {h_{v x_1}^{\beta}}'(r)+{h_{v x_1}^{\beta}}(r) p_{31}(r)+p_{34}(r)
   {h_{x_1 x_1}^{\beta}}'(r)+{h_{x_1 x_1}^{\beta}}(r) p_{33}(r)+p_{36}(r) {h_{x_2 x_2}^{\beta}}'(r)+{h_{x_2 x_2}^{\beta}}(r)
   p_{35}(r)\nonumber\\
  & &\hskip -0.3in =-\frac{1}{2} \Biggl(v_{129}(r) {h_{rr}^{\beta}}'(r)+{h_{rr}^{\beta}}(r) v_{121}(r)+v_{124}(r)
   {h_{v r}^{\beta}}'(r)+{h_{v r}^{\beta}}(r) v_{116}(r)+v_{130}(r) {h_{vv}^{\beta}}''(r)+v_{122}(r)
   {h_{vv}^{\beta}}'(r)+{h_{vv}^{\beta}}(r) v_{114}(r)\nonumber\\
   & &+v_{123}(r) {h_{v x_1}^{\beta}}'(r)+{h_{v x_1}^{\beta}}(r)
   v_{115}(r)+v_{126}(r) {h_{x_1 r}^{\beta}}'(r)+{h_{x_1 r}^{\beta}}(r) v_{118}(r)+v_{131}(r)
   {h_{x_1 x_1}^{\beta}}''(r)+v_{125}(r) {h_{x_1 x_1}^{\beta}}'(r)+{h_{x_1 x_1}^{\beta}}(r) v_{117}(r)\nonumber\\
   & &+v_{132}(r)
   {h_{x_2 x_2}^{\beta}}''(r)+v_{127}(r) {h_{x_2 x_2}^{\beta}}'(r)+{h_{x_2 x_2}^{\beta}}(r) v_{119}(r)+v_{133}(r)
   {h_{x_3 x_3}^{\beta}}''(r)+v_{128}(r) {h_{x_3 x_3}^{\beta}}'(r)+{h_{x_3 x_3}^{\beta}}(r) v_{120}(r)\Biggr)-{h_{vv}^{\beta}}(r) {z_8}(r)\nonumber\\
  & & -{h_{v x_1}^{\beta}}(r) z_{9}(r)-{h_{x_1 x_1}^{\beta}}(r)
   z_{10}(r)-{h_{x_2 x_2}^{\beta}}(r) (z_{11}(r)+z_{12}(r)),\nonumber\\
  & &  \hskip -0.3in {\bf EOM_{rr}^\beta}:\nonumber\\
  & & \hskip -0.3in {h_{rr}^{\beta}}(r) p_{43}(r)+p_{39}(r) {h_{x_1 x_1}^{\beta}}''(r)+p_{38}(r)
   {h_{x_1 x_1}^{\beta}}'(r)+{h_{x_1 x_1}^{\beta}}(r) p_{37}(r)+p_{42}(r) {h_{x_2 x_2}^{\beta}}''(r)+p_{41}(r)
   {h_{x_2 x_2}^{\beta}}'(r)+{h_{x_2 x_2}^{\beta}}(r) p_{40}(r)\nonumber\\
  & & \hskip -0.3in = -\frac{1}{2} \Biggl(v_{149}(r) {h_{rr}^{\beta}}'(r)+{h_{rr}^{\beta}}(r) v_{141}(r)+v_{144}(r)
   {h_{v r}^{\beta}}'(r)+h_{vr}^{\beta}(r) v_{136}(r)+v_{150}(r) {h_{vv}^{\beta}}''(r)+v_{142}(r)
   {h_{vv}^{\beta}}'(r)+{h_{vv}^{\beta}}(r) v_{134}(r)\nonumber\\
   & &+v_{151}(r) {h_{v x_1}^{\beta}}''(r)+v_{143}(r)
   {h_{v x_1}^{\beta}}'(r)+{h_{v x_1}^{\beta}}(r) v_{135}(r)+v_{146}(r) {h_{x_1 r}^{\beta}}'(r)+{h_{x_1 r}^{\beta}}(r)
   v_{138}(r)+v_{152}(r) {h_{x_1 x_1}^{\beta}}''(r)+v_{145}(r) {h_{x_1 x_1}^{\beta}}'(r)\nonumber\\
   & &+{h_{x_1 x_1}^{\beta}}(r)
   v_{137}(r)+v_{153}(r) {h_{x_2 x_2}^{\beta}}''(r)+v_{147}(r) {h_{x_2 x_2}^{\beta}}'(r)+{h_{x_2 x_2}^{\beta}}(r)
   v_{139}(r)+v_{154}(r) {h_{x_3 x_3}^{\beta}}''(r)+v_{148}(r) {h_{x_3 x_3}^{\beta}}'(r)\nonumber\\
  & &+{h_{x_3 x_3}^{\beta}}(r)
   v_{140}(r)\Biggr)  -{h_{vv}^{\beta}}(r) z_{_{13}}(r)-{h_{v x_1}^{\beta}}(r) z_{14}(r)-{h_{x_1 x_1}^{\beta}}(r)
   z_{15}(r)-{h_{x_2 x_2}^{\beta}}(r) (z_{16}(r)+z_{17}(r)),
\end{eqnarray}
}
where $p_i(r), \ z_i(r)$ and $v_i(r)$ are defined in appendix \ref{pi-zi-vi}.

\section{$y_i(r)$s, $s_i(r)$s}
\label{yi-si}
\setcounter{equation}{0}
\seceqbb
The $y_i(r)$s appearing in equation (\ref{pEOM-beta0}) are given as:
\begin{eqnarray}
\label{yi's}
& & y_1(r)= \frac{6 \sqrt[3]{2} 3^{2/3} \sqrt[6]{\pi } \sqrt{g_s} \sqrt{N} \sqrt[3]{{N_f}} q^2 \sqrt[3]{-\log (r)}}{r_h^2} , \ \ \ \ \ \ y_2(r)= \frac{12 \sqrt[3]{2} 3^{2/3} \sqrt[6]{\pi } \sqrt{g_s} \sqrt{N} \sqrt[3]{N_f} q w \sqrt[3]{-\log (r)}}{r_h^2},\nonumber\\
& & y_3(r)=\frac{4 \sqrt[3]{2} \sqrt[6]{3} \sqrt[3]{N_f} (r-r_h) \sqrt[3]{-\log (r)}}{\sqrt[3]{\pi } r_h}, \ \ \ \ \ \ \ \ \ \ \ \ \ \ \  y_4(r)=\frac{6 \sqrt[3]{2} 3^{2/3} \sqrt[6]{\pi } \sqrt{g_s} \sqrt{N} \sqrt[3]{N_f} w^2 \sqrt[3]{-\log (r)}}{r_h^2},\nonumber\\
& & y_5(r)=\frac{12 \sqrt[3]{2} 3^{2/3} \sqrt[6]{\pi } \sqrt{g_s} \sqrt{N} \sqrt[3]{N_f} w^2 \sqrt[3]{-\log (r)}}{r_h^2}, \ \ \ \ \ \ y_6(r)=\frac{81\ 3^{5/6} \sqrt{g_s} \sqrt{N} N_f^{2/3} r_h^2 (-\log (r))^{2/3}}{4232 \sqrt[3]{2} \sqrt[6]{\pi }},\nonumber\\
& & y_7(r)=\frac{2 \sqrt[3]{\frac{2}{3}} \pi ^{2/3} r_h^2}{3 N_f^{2/3} (-\log (r))^{2/3}}, \ \ \ \ \ \ \ \ \ \ \ \ \ \ \ \ \ \ \ \ \ \ \ \ \ \ \ \ \ \ \ \ \   y_8(r)=\frac{2 \sqrt[3]{\frac{2}{3}} \pi ^{2/3} \iota  q r_h}{N_f^{2/3} (-\log (r))^{2/3}},\nonumber\\
& & y_9(r)=\frac{9\ 3^{5/6} \sqrt{g_s} \sqrt{N} N_f^{2/3} r_h^2 (-\log (r))^{2/3}}{32 \sqrt[3]{2} \sqrt[6]{\pi }}, \ \ \ \ \ \ \ \ \ \ \ \ \ \  y_{10}(r)=\frac{9\ 3^{5/6} \sqrt{g_s} \sqrt{N} N_f^{2/3} r_h^2 (-\log (r))^{2/3}}{16 \sqrt[3]{2} \sqrt[6]{\pi }},\nonumber\\
& & y_{11}(r)=\frac{12 \sqrt[3]{2} \sqrt[6]{3} \iota  \sqrt[3]{N_f} q r_h \sqrt[3]{-\log (r)}}{\sqrt[3]{\pi }}, \ \ \ \ \ \  \ \ \ \ \ \ \ \ \ \ \ \ \ \ y_{12}(r)=\frac{\sqrt[3]{2} \sqrt[6]{3} \sqrt[3]{N_f} r_h^2 (r-r_h) \sqrt[3]{-\log (r)}}{\sqrt[3]{\pi }},\nonumber\\
& & y_{13}(r)=\frac{243 \sqrt[3]{\frac{3}{2}} \sqrt{g_s} \sqrt{N} N_f^{5/3} r_h^2 (-\log (r))^{5/3}}{64 \pi ^{7/6}}, \ \ \ \ \ \ \ \ \ \ \ \ y_{14}(r)=\frac{243 \sqrt[3]{\frac{3}{2}} \sqrt{g_s} \sqrt{N} N_f^{5/3} r_h^2 (-\log (r))^{5/3}}{32 \pi ^{7/6}},\nonumber\\
& & y_{15}(r)=\frac{2187 \sqrt[3]{\frac{3}{2}} \sqrt{g_s} \sqrt{N} N_f^{5/3} r_h^2 (-\log (r))^{5/3}}{4232 \pi ^{7/6}}, \ \ \ \ \ \ \ \ \ \ y_{16}(r)=24 \sqrt[6]{3} \sqrt[3]{\frac{2}{\pi }} \iota  \sqrt[3]{N_f} q r_h (r-r_h) \sqrt[3]{-\log (r)}, \nonumber\\
 & & y_{17}(r)=\frac{\sqrt[3]{2} \sqrt[6]{3} \sqrt[3]{N_f} r_h^2 (r-r_h) \sqrt[3]{-\log (r)}}{\sqrt[3]{\pi }}, \ \ \ \ \ \ \ \ \ \ \ \ \ \  y_{18}(r)=\frac{243 \sqrt[3]{\frac{3}{2}} \sqrt{g_s} \sqrt{N} N_f^{5/3} r_h^2 (-\log (r))^{5/3}}{32 \pi ^{7/6}}, \nonumber\\
& & y_{19}(r)=\frac{243 \sqrt[3]{\frac{3}{2}} \sqrt{g_s} \sqrt{N} N_f^{5/3} r_h^2 (-\log (r))^{5/3}}{16 \pi ^{7/6}}\, \ \ \ \ \ \ \ \ \ \ \ \ y_{vv}(r)=\frac{2187 \sqrt[3]{\frac{3}{2}} \sqrt{g_s} \sqrt{N} N_f^{5/3} r_h^2 (-\log (r))^{5/3}}{8464 \pi ^{7/6}},\nonumber\\
& & y_{21}(r)=\frac{27 \iota  N_f q r_h \log (r)}{4 \pi }, \ \ \ \ \ \ \ \ \ \ \ \ \ \ \ \ \ \ \ \ \ \ \ \ \ \ \ \ \ \ \ \ \ \ \ \ y_{22}(r)=\frac{9 N_f r_h^2 \log (r)}{4 \pi },\nonumber\\
& & y_{23}(r)=\frac{9}{2} \sqrt{\frac{3}{\pi }} \sqrt{g_s} \sqrt{N} N_f q w (r-r_h) \log (r).
\end{eqnarray}
The $s_i(r)$s appearing in equation (\ref{pEOM-beta0}) are given as:
\begin{eqnarray}
\label{si's}
& & s_1(r)= \frac{2187 N_f^2 r_h \log ^2(r)}{64 \pi ^2}, \ \ \ \ \ \ \ \ \ \  \ \ \ \ \ \ \ \  \ \ \ \ \ \ \ \  \ \ \ \ \ \  s_2(r)= \frac{2187 \sqrt{3} \sqrt{g_s} \iota  \sqrt{N} N_f^2 q \log ^2(r)}{16 \pi ^{3/2}},\nonumber\\
& & s_3(r)=-\frac{2187 \sqrt{3} \sqrt{g_s} \iota ^2 \sqrt{N} N_f^2 q^2 r_h \log ^2(r)}{32 \pi ^{3/2}}, \ \ \ \ \ \ \ \ \ \ s_4(r)=-\frac{2187 \sqrt{3} \sqrt{g_s} \iota  \sqrt{N} N_f^2 w \log ^2(r)}{32 \pi ^{3/2}},\nonumber\\
& & s_5(r)=-\frac{2187 \sqrt{3} \sqrt{g_s} \iota ^2 \sqrt{N} N_f^2 q r_h w \log ^2(r)}{32 \pi ^{3/2}}, \ \ \ \ \ \ \ \ \ \ s_6(r)=\frac{2187 \sqrt{3} \sqrt{g_s} \iota  \sqrt{N} N_f^2 w \log ^2(r)}{16 \pi ^{3/2}},\nonumber\\
& & s_7(r)=-\frac{2187 N_f^2 r_h^3 \log ^2(r)}{32 \pi ^2}, \ \ \ \ \ \ \ \ \ \  \ \ \ \ \ \ \ \  \ \ \ \ \ \ \ \  \ \ \ \ \  s_8(r)=-\frac{26973 N_f^2 r_h^2 \log ^2(r)}{128 \pi ^2},\nonumber\\
& & s_9(r)=\frac{2187 \sqrt{3} \sqrt{g_s} \iota  \sqrt{N} N_f^2 q r_h \log ^2(r)}{32 \pi ^{3/2}}, \ \ \ \ \ \ \ \ \ \  \ \ \ \ \ \ \  s_{10}(r)=-\frac{2187 \sqrt{3} \sqrt{g_s} \iota  \sqrt{N} N_f^2 r_h w \log ^2(r)}{32 \pi ^{3/2}},\nonumber\\
& & s_{11}(r)=-\frac{243 N_f^2 r_h \log ^2(r)}{8 \pi ^2}, \ \ \ \ \ \ \ \ \ \  \ \ \ \ \ \ \ \  \ \ \ \ \ \ \ \  \ \ \ \ \ \  s_{12}(r)=-\frac{729 \sqrt{3} \sqrt{g_s} \iota  \sqrt{N} N_f^2 q \log ^2(r)}{16 \pi ^{3/2}},\nonumber\\
& & s_{14}(r)=\frac{729 N_f^2 q r_h^3 w \log ^2(r)}{32 \pi ^2}, \ \ \ \ \ \ \ \ \ \  \ \ \ \ \ \ \ \  \ \ \ \ \ \ \ \  \ \ \ \ \  s_{15}(r)=-\frac{729 \sqrt{3} \sqrt{g_s} \iota  \sqrt{N} N_f^2 q r_h \log ^2(r)}{32 \pi ^{3/2}},\nonumber\\
& & s_{16}(r)=\frac{729 N_f^2 r_h^3 \log ^2(r)}{32 \pi ^2}.
\end{eqnarray}

\section{$p_i(r)$s, $z_i(r)$s, $v_i(r)$s}
\label{pi-zi-vi}
\setcounter{equation}{0}
\seceqcc
The $p_i(r)$s appearing in equation (\ref{EOMs-metric-fluctuations}) are given as:
\begin{eqnarray}
\label{pi's}
& & p_1(r)= \frac{2 \sqrt[3]{\frac{2}{3}} \pi ^{2/3} \left(\sqrt{\pi } \sqrt{g_s} \sqrt{N} q^2+i \sqrt{3} r_h
   w\right)}{3 |\log r|^{2/3} N_f^{2/3} r_h^2}  , \ \ \ \ \ \ \ \ \ \  \ \ \  p_2(r)= \frac{4 \sqrt[3]{\frac{2}{3}} \pi ^{2/3} q \left(3 \sqrt{\pi } \sqrt{g_s} \sqrt{N} w-2 i \sqrt{3}
   r_h\right)}{9 |\log r|^{2/3} N_f^{2/3} r_h^2}, \nonumber\\
& & p_3(r)= \frac{208 i \sqrt[3]{2} \pi ^{2/3} q}{33\ 3^{5/6} |\log r|^{2/3} N_f^{2/3} r_h}, \ \ \ \ \ \ \ \ \ \  \ \ \ \ \ \ \  \ \ \ \ \ \ \  \  p_4(r)=\frac{2 \sqrt[3]{\frac{2}{3}} \pi ^{2/3} w \left(3 \sqrt{\pi } \sqrt{g_s} \sqrt{N} w-2 i \sqrt{3}
   r_h\right)}{9 |\log r|^{2/3} N_f^{2/3} r_h^2},  \nonumber\\
& & p_5(r)=  \frac{208 i \sqrt[3]{2} \pi ^{2/3} w}{33\ 3^{5/6} |\log r|^{2/3} N_f^{2/3} r_h}  , \ \ \ \ \ \ \ \ \ \  \ \ \ \ \ \ \  \ \ \ \ \ \ \  \ \ \ p_6(r)= \frac{4 \sqrt[3]{2} \pi ^{2/3} w \left(\sqrt{3 \pi } \sqrt{g_s} \sqrt{N} w-2 i r_h\right)}{3\
   3^{5/6} |\log r|^{2/3} N_f^{2/3} r_h^2},\nonumber\\
& & p_7(r)=\frac{416 i \sqrt[3]{2} \pi ^{2/3} w}{33\ 3^{5/6} |\log r|^{2/3} N_f^{2/3} r_h}
  , \ \ \ \ \ \ \ \ \ \  \ \ \ \ \ \ \  \ \ \ \ \ \ \  \ \ \ p_8(r)=-\frac{2 i \sqrt[3]{2} \pi ^{2/3} q}{3^{5/6} |\log r|^{2/3} N_f^{2/3} r_h},\nonumber\\
& & p_9(r)= \frac{2 i \sqrt[3]{2} \pi ^{2/3} q}{3\ 3^{5/6} |\log r|^{2/3} N_f^{2/3}}, \ \ \ \ \ \ \ \ \ \  \ \ \ \ \ \ \  \ \ \ \ \ \ \  \ \ \ \ \ \ \  \ \  p_{10}(r)=\frac{2 i \sqrt[3]{2} \pi ^{2/3} w}{3\ 3^{5/6} |\log r|^{2/3} N_f^{2/3}}, \nonumber\\
& & p_{11}(r)= -\frac{4 \sqrt[3]{\frac{2}{3}} \pi ^{7/6} \sqrt{g_s} \sqrt{N} q w}{3 |\log r|^{2/3} N_f^{2/3}
   r_h^2}, \ \ \ \ \ \ \ \ \ \  \ \ \ \ \ \ \  \ \ \ \ \ \ \  \ \ \ \ \ \ \  \ p_{12}(r)=\frac{70 i \sqrt[3]{\frac{2}{3}} \pi ^{7/6} \sqrt{g_s} \log(r) \sqrt{N} q}{33 |\log r|^{5/3}
   N_f^{2/3} r_h^3},  \nonumber
\end{eqnarray}
\begin{eqnarray}
& & p_{13}(r)= -\frac{10 i \sqrt[3]{\frac{2}{3}} \pi ^{7/6} \sqrt{g_s} \sqrt{N} q}{11 |\log r|^{2/3}
   N_f^{2/3} r_h^2}, \ \ \ \ \ \ \ \ \ \  \ \ \ \ \ \ \  \ \ \ \ \ \ \  \ \ \ \ \ p_{14}(r)=\frac{16 i \sqrt[3]{\frac{2}{3}} \pi ^{7/6} \sqrt{g_s} \log(r) \sqrt{N} w}{11 |\log r|^{5/3}
   N_f^{2/3} r_h^3}, \nonumber\\
& & p_{15}(r)= -\frac{10 i \sqrt[3]{\frac{2}{3}} \pi ^{7/6} \sqrt{g_s} \sqrt{N} w}{11 |\log r|^{2/3}
   N_f^{2/3} r_h^2}, \ \ \ \ \ \ \ \ \ \  \ \ \ p_{16}(r)= -\frac{4 \sqrt[3]{\frac{2}{3}} \pi ^{7/6} \sqrt{g_s} \sqrt{N} \left(13 \sqrt{3 \pi } \sqrt{g_s}
   \sqrt{N} q^2+24 i r_h w\right)}{33 |\log r|^{2/3} N_f^{2/3} r_h^4},  \nonumber\\
& & p_{17}(r)=-\frac{20 i \sqrt[3]{\frac{2}{3}} \pi ^{7/6} \sqrt{g_s} \sqrt{N} w}{11 |\log r|^{2/3}
   N_f^{2/3} r_h^2},  \ \ \ \ \ \ \ \ \ \ \ \ \ \ \ \ \ \ \ \ \ \ \ \ \ \ \ \ \ \ p_{18}(r)= \frac{2 i \sqrt[3]{2} \sqrt[6]{3} \pi ^{2/3} q}{11 |\log r|^{2/3} N_f^{2/3} r_h},   \nonumber\\
   & &  p_{19}(r)=-\frac{8 i \sqrt[3]{2} \pi ^{2/3} q}{33\ 3^{5/6} |\log r|^{2/3} N_f^{2/3}}, \ \ \ \ \ \ \ \ \ \  \ \ \ \ \ \ \  \ \ \ \ \ \ \  \ \  p_{20}(r)=\frac{62 i \sqrt[3]{2} \pi ^{2/3} w}{33\ 3^{5/6} |\log r|^{2/3} N_f^{2/3} r_h},\nonumber\\
& &  p_{21}(r)= -\frac{8 i \sqrt[3]{2} \pi ^{2/3} w}{33\ 3^{5/6} |\log r|^{2/3} N_f^{2/3}}, \ \ \ \ \ \ \ \ \ \   p_{22}(r)=-\frac{8 \sqrt[3]{2} \pi ^{2/3} \left(\sqrt{3 \pi } \sqrt{g_s} |\log r|^{2/3} \sqrt{N} q^2+i
   r_h w\right)}{33\ 3^{5/6} |\log r|^{2/3} N_f^{2/3} r_h^2}, \nonumber\\
& & p_{23}(r)=  -\frac{104 i \sqrt[3]{2} \pi ^{2/3} w}{33\ 3^{5/6} |\log r|^{2/3} N_f^{2/3}}, \ \ \ \ \ \ \ \ \ \  \ \ \ \ \ \ \  \ \ \ \ \ \ \  \ \ \  p_{24}(r)=-\frac{8 \sqrt[3]{2} \pi ^{2/3}}{9\ 3^{5/6} |\log r|^{2/3} N_f^{2/3} r_h^2}, \nonumber\\
& & p_{25}(r)=-\frac{61 \left(\frac{\pi }{2}\right)^{2/3}}{9\ 3^{5/6} |\log r|^{2/3} N_f^{2/3} r_h}, \ \ \ \ \ \ \ \ \ \  \ \ \ \ \ \ \  \ \ \ \ \ \ \  \ \ p_{26}(r)= \frac{2 \sqrt[3]{2} \pi ^{2/3}}{3\ 3^{5/6} |\log r|^{2/3} N_f^{2/3}},\nonumber\\
& & p_{27}(r)=\frac{4 i \sqrt[3]{\frac{2}{3}} \pi ^{7/6} \sqrt{g_s} \log(r) \sqrt{N} q}{3 |\log r|^{5/3}
   N_f^{2/3} r_h^3}, \ \ \ \ \ \ \ \ \ \  \ \ \ \ \ \ \  \ \ \ \ \ \ \  \ \ \ \  p_{28}(r)= -\frac{8 i \sqrt[3]{\frac{2}{3}} \pi ^{7/6} \sqrt{g_s} \sqrt{N} q}{3 \log(r) N_f^{2/3}
   r_h^3}, \nonumber\\
& & p_{29}(r)= -\frac{4 i \sqrt[3]{\frac{2}{3}} \pi ^{7/6} \sqrt{g_s} \sqrt{N} q}{3 |\log r|^{2/3} N_f^{2/3}
   r_h^2}, \ \ \ \ \ \ \ \ \ \  \ \ \ \ \ \ \  \ \ \ \ \ \ \  \ \ \ \ \ \ \  p_{30}(r)= \frac{2 \sqrt[3]{2} \pi ^{2/3} q w}{3\ 3^{5/6} |\log r|^{2/3} N_f^{4/3}},\nonumber\\
& & p_{31}(r)=\frac{2 i \sqrt[3]{2} \sqrt[6]{3} \pi ^{2/3} q}{11 |\log r|^{2/3} N_f^{2/3} r_h}, \ \ \ \ \ \ \ \ \ \  \ \ \ \ \ \ \  \ \ \ \ \ \ \  \ \ \ \ \ \ \ \ \ p_{32}(r)= -\frac{52 i \sqrt[3]{2} \pi ^{2/3} q}{33\ 3^{5/6} |\log r|^{2/3} N_f^{2/3}},\nonumber\\
& & p_{33}(r)=-\frac{4 i \sqrt[3]{2} \pi ^{2/3} w}{33\ 3^{5/6} |\log r|^{2/3} N_f^{2/3} r_h} , \ \ \ \ \ \ \ \ \ \  \ \ \ \ \ \ \  \ \ \ \ \ \ \  \ \  p_{34}(r)=-\frac{52 i \sqrt[3]{2} \pi ^{2/3} w}{33\ 3^{5/6} |\log r|^{2/3} N_f^{2/3}},  \nonumber\\
& & p_{35}(r)= \frac{2 i \sqrt[3]{\frac{2}{3}} \pi ^{2/3} \left(29 \sqrt{3} r_h w+45 i \sqrt{\pi } \sqrt{g_s}
   \sqrt{N} q^2\right)}{99 |\log r|^{2/3} N_f^{2/3} r_h^2}, \ \ \ \ \ \ \ \ \   p_{36}(r)= -\frac{20 i \sqrt[3]{2} \pi ^{2/3} w}{11\ 3^{5/6} |\log r|^{2/3} N_f^{2/3}},\nonumber\\
   & &  p_{37}(r)= -\frac{26 \sqrt[3]{\frac{2}{3}} \pi ^{7/6} \sqrt{g_s} \sqrt{N}}{9 |\log r|^{2/3} N_f^{2/3}
   r_h^4}, \ \ \ \ \ \ \ \ \ \  \ \ \ \ \ \ \  \ \ \ \ \ \ \  \ \ \ \ \ \ \ \ \ p_{38}(r)=\frac{49 \pi ^{7/6} \sqrt{g_s} \sqrt{N}}{9\ 2^{2/3} \sqrt[3]{3} |\log r|^{2/3} N_f^{2/3}
   r_h^3}, \nonumber\\
   & & p_{39}(r)=-\frac{2 \sqrt[3]{\frac{2}{3}} \pi ^{7/6} \sqrt{g_s} \sqrt{N}}{3 |\log r|^{2/3} N_f^{2/3}
   r_h^2}, \ \ \ \ \ \ \ \ \ \  \ \ \ \ \ \ \  \ \ \ \ \ \ \  \ \ \ \ \ \ \ \ \ p_{40}(r)=-\frac{52 \sqrt[3]{\frac{2}{3}} \pi ^{7/6} \sqrt{g_s} \sqrt{N}}{9 |\log r|^{2/3} N_f^{2/3}
   r_h^4}, \nonumber
   \end{eqnarray}
\begin{eqnarray}
   & &  p_{41}(r)= \frac{49 \sqrt[3]{\frac{2}{3}} \pi ^{7/6} \sqrt{g_s} \sqrt{N}}{9 |\log r|^{2/3} N_f^{2/3}
   r_h^3}, \ \ \ \ \ \ \ \ \ \  \ \ \ \ \ \ \  \ \ \ \ \ \ \  \ \ \ \ \ \ \ \ \ \ p_{42}(r)= -\frac{4 \sqrt[3]{\frac{2}{3}} \pi ^{7/6} \sqrt{g_s} \sqrt{N}}{3 |\log r|^{2/3} N_f^{2/3}
   r_h^2}, \nonumber\\
   & & p_{43}(r)=  \frac{2 \sqrt[3]{\frac{2}{3}} \pi ^{7/6} \sqrt{g_s} \sqrt{N} q^2}{3 |\log r|^{2/3} N_f^{2/3}
   r_h^2}.
\end{eqnarray}

The $z_i(r)$s appearing in equation (\ref{EOMs-metric-fluctuations}) are given as:
\begin{eqnarray}
\label{zi's}
& & z_1(r)= \frac{4096 \sqrt[3]{\frac{2}{3}} \pi ^{2/3} (r-{r_h})^2}{19683 {g_s}^2 |\log r|^{8/3} N^2
   N_f^{8/3} r_h^2}, \ \ \ \ \ \ \ \ \ \  \ \ \ \ \ \ \  \ \ \ \     z_2(r)=\frac{2351104 \sqrt[3]{\frac{2}{3 \pi }} r_h^3 (r-r_h)}{177147 {g_s}^3 |\log r|^{8/3} N^3
   N_f^{8/3}},  \nonumber\\
& & z_3(r)= \frac{8192 \sqrt[3]{\frac{2}{3}} \pi ^{5/3} (r-r_h)^2}{19683 g_s|\log r|^{8/3} N
   N_f^{8/3} r_h^4}, \ \ \ \ \ \ \ \ \ \  \ \ \ \ \ \ \  \ \ \ \ \ \ \     z_4(r)=\frac{16384 \sqrt[3]{\frac{2}{3}} \pi ^{5/3} w (r-r_h)^2}{19683 g_s|\log r|^{8/3} N
   N_f^{8/3} r_h^4}, \nonumber\\
& & z_5(r)= \frac{8192 \sqrt[3]{\frac{2}{3}} \pi ^{5/3} w^2 (r-r_h)^2}{19683 g_s|\log r|^{8/3} N
   N_f^{8/3} r_h^4}, \ \ \ \ \ \ \ \ \ \  \ \ \ \ \ \ \  \ \ \ \ \ \ \     z_6(r)=-\frac{8192 \sqrt[3]{\frac{2}{3}} \pi ^{5/3} w^2 (r-r_h)}{6561 g_s|\log r|^{8/3} N
   N_f^{8/3} r_h^3}, \nonumber\\
& & z_7(r)=-\frac{1343488 \sqrt[3]{\frac{2}{3}} \pi ^{5/3} (r-r_h)^3}{6561 g_s|\log r|^{8/3} N
   N_f^{8/3} r_h^5}, \ \ \ \ \ \ \ \ \ \  \ \ \ \ \ \ \  \ \ \ \ \ \    z_8(r)=-\frac{8192 \sqrt[3]{\frac{2}{3}} \pi ^{5/3} (r-r_h)^2}{19683 g_s|\log r|^{8/3} N
   N_f^{8/3} r_h^4}, \nonumber\\
& & z_9(r)= -\frac{16384 \sqrt[3]{\frac{2}{3}} \pi ^{5/3} w (r-r_h)^2}{19683 g_s|\log r|^{8/3} N
   N_f^{8/3} r_h^4}, \ \ \ \ \ \ \ \ \ \  \ \ \ \ \ \ \  \ \ \ \ \   z_{10}(r)=-\frac{8192 \sqrt[3]{\frac{2}{3}} \pi ^{5/3} w^2 (r-r_h)^2}{19683 g_s|\log r|^{8/3} N
   N_f^{8/3} r_h^4},\nonumber\\
& & z_{11}(r)=-\frac{8192 \sqrt[3]{\frac{2}{3}} \pi ^{5/3} w^2 (r-r_h)^2}{19683 g_s|\log r|^{8/3} N
   N_f^{8/3} r_h^4} , \ \ \ \ \ \ \ \ \ \  \ \ \ \ \ \ \  \ \ \ \    z_{12}(r)=-\frac{16384 \sqrt[3]{\frac{2}{3}} \pi ^{5/3} w^2 (r-r_h)}{6561 g_s|\log r|^{8/3} N
   N_f^{8/3} r_h^3}, \nonumber\\
& & z_{13}(r)= -\frac{832 \sqrt[3]{\frac{2}{3}} \pi ^{8/3} (r-r_h)}{59049 |\log r|^{8/3} N_f^{8/3}
   r_h^7}, \ \ \ \ \ \ \ \ \ \  \ \ \ \ \ \ \  \ \ \ \ \ \ \  \ \ \   z_{14}(r)=-\frac{1664 \sqrt[3]{\frac{2}{3}} \pi ^{8/3} w (r-r_h)}{59049 |\log r|^{8/3} N_f^{8/3}
   r_h^7}, \nonumber\\
& & z_{15}(r)=z_{16}(r)= -\frac{832 \sqrt[3]{\frac{2}{3}} \pi ^{8/3} w^2 (r-r_h)}{59049 |\log r|^{8/3} N_f^{8/3}
   r_h^7}.
\end{eqnarray}
The $v_i(r)$s appearing in equation (\ref{EOMs-metric-fluctuations}) are given as:
\begin{eqnarray}
\label{v_{i}'s}
& & v_{1}(r)=-\frac{3328 \sqrt[3]{\frac{2}{3}} \pi ^{8/3} }{19683 |\log r|^{8/3} N_f^{8/3} {r_h}^6} , \ \ \ \ \ \ \ \ \ \  \ \ \ \ \ \ \  \ \ \ \ \ \ \  v_{2}(r)=-\frac{6656 \sqrt[3]{\frac{2}{3}} \pi ^{8/3}  w}{19683 |\log r|^{8/3} N_f^{8/3} r_h^6}, \nonumber\\
& & v_{3}(r)=\frac{665600 \sqrt[3]{2} \pi ^{13/6} {g1} \sqrt{\frac{1}{N}}}{6561\ 3^{5/6} \sqrt{g_s} |\log r|^{8/3} N_f^{8/3} r_h^4}, \ \ \ \ \ \ \ \ \ \  \ \ \ \ \ \ \  \ \  v_{4}(r)=-\frac{3328 \sqrt[3]{\frac{2}{3}} \pi ^{8/3}  w^2}{19683 |\log r|^{8/3} N_f^{8/3} r_h^6}, \nonumber
\end{eqnarray}
\begin{eqnarray}
& & v_{5}(r)=\frac{133120 \sqrt[3]{2} \pi ^{13/6}  \sqrt{\frac{1}{N}} w}{6561\ 3^{5/6} \sqrt{g_s} |\log r|^{8/3} N_f^{8/3} r_h^4}, \ \ \ \ \ \ \ \ \ \  \ \ \ \ \ \ \  \ \ \  v_{6}(r)=-\frac{3328 \sqrt[3]{\frac{2}{3}} \pi ^{8/3}  w^2}{19683 |\log r|^{8/3} N_f^{8/3} r_h^6}, \nonumber\\
& & v_{7}(r)=-\frac{3328 \sqrt[3]{\frac{2}{3}} \pi ^{8/3}  w^2}{19683 |\log r|^{8/3} N_f^{8/3} r_h^6} , \ \ \ \ \ \ \ \ \ \  \ \ \ \ \ \ \  \ \ \ \ \ \ \  \ \ \ v_{8}(r)=\frac{4096 \sqrt[3]{\frac{2}{3}} \pi ^{5/3}  w^2}{2187 g_s |\log r|^{8/3} N N_f^{8/3} r_h^2}, \nonumber\\
& & v_{9}(r)=\frac{40960 \sqrt[3]{\frac{2}{3}} \pi ^{5/3} }{6561 g_s |\log r|^{8/3} N N_f^{8/3} r_h^3} , \ \ \ \ \ \ \ \ \ \  \ \ \ \ \ \ \  \ \ \ \ \ \ \  \ \ v_{10}(r)=\frac{133120 i \sqrt[3]{2} \pi ^{13/6}  \sqrt{\frac{1}{N}}}{6561\ 3^{5/6} \sqrt{g_s} |\log r|^{8/3} N_f^{8/3} r_h^4}, \nonumber\\
& & v_{11}(r)=-\frac{8192 i \sqrt[3]{\frac{2}{3}} \pi ^{5/3}  w}{2187 g_s |\log r|^{8/3} N N_f^{8/3} r_h^2}, \ \ \ \ \ \ \ \ \ \  \ \ \ \ \ \ \  \ \ \ \ \ \  v_{12}(r)=\frac{26624 i \sqrt[3]{2} \pi ^{13/6}  \sqrt{\frac{1}{N}} w}{6561\ 3^{5/6} \sqrt{g_s} |\log r|^{8/3} N_f^{8/3} r_h^4}, \nonumber\\
& & v_{13}(r)=-\frac{1361920 i \sqrt[3]{\frac{2}{3}} \pi ^{5/3}  (r-r_h)}{59049 g_s |\log r|^{8/3} N N_f^{8/3} r_h^3}, \ \ \ \ \ \ \ \ \ \  \ \ \ \ \ \ \  \ \ \ \ \ \ v_{14}(r)=\frac{26624 i \sqrt[3]{2} \pi ^{13/6}  \sqrt{\frac{1}{N}} w}{6561\ 3^{5/6} \sqrt{g_s} |\log r|^{8/3} N_f^{8/3} r_h^4}, \nonumber\\
& & v_{15}(r)=\frac{26624 i \sqrt[3]{2} \pi ^{13/6}  \sqrt{\frac{1}{N}} w}{6561\ 3^{5/6} \sqrt{g_s} |\log r|^{8/3} N_f^{8/3} r_h^4}, \ \ \ \ \ \ \ \ \ \  \ \ \ \ \ \ \  \  v_{16}(r)=-\frac{106496 \sqrt[3]{\frac{2}{3}} \pi ^{2/3}  r_h}{19683 g_s^2 |\log r|^{8/3} N^2 N_f^{8/3}}, \nonumber \\
& & v_{17}(r)=-\frac{4096 \sqrt[3]{\frac{2}{3}} \pi ^{5/3} }{2187 g_s |\log r|^{8/3} N N_f^{8/3} r_h^2} , \ \ \ \ \ \ \ \ \ \  \ \ \ \ \ \ \  \ \ \ \  v_{18}(r)=\frac{697984 \sqrt[3]{\frac{2}{3}} \pi ^{2/3}  r_h (r-r_h)}{177147 g_s^2 |\log r|^{8/3} N^2 N_f^{8/3}}, \nonumber \\
& & v_{19}(r)=\frac{320512 \sqrt[3]{\frac{2}{3}} \pi ^{5/3}  (r-r_h)}{59049 g_s |\log r|^{8/3} N N_f^{8/3} r_h^3}, \ \ \ \ \ \ \ \ \ \  \ \ \ \ \ \ \  \ \ \ \ \  v_{20}(r)=\frac{320512 \sqrt[3]{\frac{2}{3}} \pi ^{5/3}  (r-r_h)}{59049 g_s |\log r|^{8/3} N N_f^{8/3} r_h^3}, \nonumber\\
& & v_{21}(r)=\frac{320512 \sqrt[3]{\frac{2}{3}} \pi ^{5/3}  (r-r_h)}{59049 g_s |\log r|^{8/3} N N_f^{8/3} r_h^3}, \ \ \ \ \ \ \ \ \ \  \ \ \ \ \ \ \  \ \ \ \ \ v_{22}(r)=-\frac{164608 i \sqrt[3]{2} \pi ^{13/6}  \sqrt{\frac{1}{N}}}{6561\ 3^{5/6} \sqrt{g_s} |\log r|^{8/3} N_f^{8/3} r_h^5}, \nonumber \\
& & v_{23}(r)=\frac{5632 i \sqrt[3]{2} \pi ^{13/6}  \sqrt{\frac{1}{N}} w}{2187\ 3^{5/6} \sqrt{g_s} |\log r|^{8/3} N_f^{8/3} r_h^5}, \ \ \ \ \ \ \ \ \ \  \ \ \ \ \ \ \  \ v_{24}(r)=\frac{14080 \sqrt[3]{2} \pi ^{13/6}  \sqrt{\frac{1}{N}} w}{2187\ 3^{5/6} \sqrt{g_s} |\log r|^{8/3} N_f^{8/3} r_h^4}, \nonumber \\
& & v_{25}(r)=-\frac{34445824 i \sqrt[3]{2} \pi ^{7/6}  \left(\frac{1}{N}\right)^{3/2} q (r-r_h)}{59049\ 3^{5/6} g_s^{3/2} |\log r|^{8/3}N_f^{8/3} r_h^2}, \ \ \ \ \ \ \ \ \ \  \ \ \ \   v_{26}(r)=\frac{2816 \sqrt[3]{2} \pi ^{13/6}  \sqrt{\frac{1}{N}} w^2}{2187\ 3^{5/6} \sqrt{g_s} |\log r|^{8/3} N_f^{8/3} r_h^4}, \nonumber \\
& & v_{27}(r)=-\frac{1024 \sqrt[3]{\frac{2}{3}} \pi ^{8/3}  w (2 r-97 r_h)}{6561 |\log r|^{8/3} N_f^{8/3} r_h^7}, \ \ \ \ \ \ \ \ \ \  \ \ \ \ \ \ \  \ \ \ \ \ \ \ \   v_{28}(r)=\frac{1024 \sqrt[3]{\frac{2}{3}} \pi ^{8/3}  w (10 r+97 r_h)}{6561 |\log r|^{8/3} N_f^{8/3} r_h^7}, \nonumber\\
& & v_{29}(r)=\frac{3160832 \sqrt[3]{\frac{2}{3}} \pi ^{5/3}  w (r-r_h)}{59049 g_s |\log r|^{8/3} N N_f^{8/3} r_h^3}, \ \ \ \ \ \ \ \ \ \  \ \ \ \ \ \ \  \  v_{30}(r)=-\frac{14080 i \sqrt[3]{2} \pi ^{13/6}  \sqrt{\frac{1}{N}}}{2187\ 3^{5/6} \sqrt{g_s} |\log r|^{8/3} N_f^{8/3} r_h^4}, \nonumber \\
& & v_{31}(r)=-\frac{2816 i \sqrt[3]{2} \pi ^{13/6}  \sqrt{\frac{1}{N}} w}{2187\ 3^{5/6} \sqrt{g_s} |\log r|^{8/3} N_f^{8/3} r_h^4}, \ \ \ \ \ \ \ \ \ \  \ \ \ v_{32}(r)=-\frac{3160832 i \sqrt[3]{\frac{2}{3}} \pi ^{5/3}  (r-r_h)}{59049 g_s |\log r|^{8/3} N N_f^{8/3} r_h^3}, \nonumber
\end{eqnarray}
\begin{eqnarray}
& & v_{33}(r)=\frac{875264 i \sqrt[3]{\frac{2}{3}} \pi ^{5/3}  w (r-r_h)}{59049 g_s |\log r|^{8/3} N N_f^{8/3} r_h^3}, \ \ \ \ \ \ \ \ \ \  \ \ \ \ \ \ \  \ \ \ v_{34}(r)=\frac{106240 i \sqrt[3]{2} \pi ^{13/6}  \sqrt{\frac{1}{N}} (r-r_h)}{6561\ 3^{5/6} \sqrt{g_s} |\log r|^{8/3} N_f^{8/3} r_h^5}, \nonumber\\
& & v_{35}(r)=\frac{260096 i \sqrt[3]{2} \pi ^{13/6}  \sqrt{\frac{1}{N}} (r-r_h)}{19683\ 3^{5/6} \sqrt{g_s} |\log r|^{8/3} N_f^{8/3} r_h^5}, \ \ \ \ \ \ \ \ \ \  \ \ \ \  v_{36}(r)=-\frac{256 \sqrt[3]{\frac{2}{3}} \pi ^{5/3}  (-3468 r+3468 r_h+49)}{59049 g_s |\log r|^{8/3} N N_f^{8/3} r_h^3}, \nonumber\\
& & v_{37}(r)=-\frac{7696 \sqrt[3]{2} \pi ^{19/6}  \sqrt{g_s} \sqrt{N}}{19683\ 3^{5/6} |\log r|^{8/3} N_f^{8/3} r_h^8}, \ \ \ \ \ \ \ \ \ \  \ \ \ \ \ \ \  \  v_{38}(r)=-\frac{15392 \sqrt[3]{2} \pi ^{19/6}  \sqrt{g_s} \sqrt{N} w}{19683\ 3^{5/6} |\log r|^{8/3} N_f^{8/3} r_h^8}, \nonumber \\
& & v_{39}(r)=\frac{3200 \sqrt[3]{\frac{2}{3}} \pi ^{8/3}  (864 r-383 r_h)}{19683 |\log r|^{8/3} N_f^{8/3} r_h^7}, \ \ \ \ \ \ \ \ \ \  \ \ \ \ \ \ \  \ \ \ \ \ \ \  v_{40}(r)=-\frac{7696 \sqrt[3]{2} \pi ^{19/6}  \sqrt{g_s} \sqrt{N} w^2}{19683\ 3^{5/6} |\log r|^{8/3} N_f^{8/3} r_h^8}, \nonumber\\
& & v_{41}(r)=\frac{640 \sqrt[3]{\frac{2}{3}} \pi ^{8/3}  w (864 r-383 r_h)}{19683 |\log r|^{8/3} N_f^{8/3} r_h^7}, \ \ \ \ \ \ \ \ \ \  \ \ \ \ \ \ \  \ \ \ \ \ \  v_{42}(r)=-\frac{7696 \sqrt[3]{2} \pi ^{19/6}  \sqrt{g_s} \sqrt{N} w^2}{19683\ 3^{5/6} |\log r|^{8/3} N_f^{8/3} r_h^8}, \nonumber\\
& & v_{43}(r)=-\frac{7696 \sqrt[3]{2} \pi ^{19/6}  \sqrt{g_s} \sqrt{N} w^2}{19683\ 3^{5/6} |\log r|^{8/3} N_f^{8/3} r_h^8}, \nonumber\\
& & v_{44}(r)=\frac{256 \sqrt[3]{2} \pi ^{13/6}  \sqrt{\frac{1}{N}} \left(2336 r^2-4672 r r_h+r_h^2 \left(37 w^2+2336\right)\right)}{2187\ 3^{5/6}
\sqrt{g_s} |\log r|^{8/3} N_f^{8/3} r_h^6}, \nonumber\\
& & v_{45}(r)=\frac{64 \sqrt[3]{2} \pi ^{13/6}  \sqrt{\frac{1}{N}} \left(-128 r^2+256 r r_h+205 r_h^2\right)}{2187\ 3^{5/6} \sqrt{g_s}
   |\log r|^{8/3} N_f^{8/3} r_h^7}, \ \ \ \ \ \ \ \ \ \  \ v_{46}(r)=\frac{640 i \sqrt[3]{\frac{2}{3}} \pi ^{8/3}  (864 r-383 r_h)}{19683 |\log r|^{8/3} N_f^{8/3} r_h^7}, \nonumber \\
& & v_{47}(r)=-\frac{18944 i \sqrt[3]{2} \pi ^{13/6}  \sqrt{\frac{1}{N}} w}{2187\ 3^{5/6} \sqrt{g_s} |\log r|^{8/3} N_f^{8/3} r_h^4}, \nonumber\\
& & v_{48}(r)=\frac{128 i \sqrt[3]{\frac{2}{3}} \pi ^{8/3}  w \left(864 r^2-1728 r r_h+r_h (864 r_h+481)\right)}{19683 |\log r|^{8/3}
   N_f^{8/3} r_h^7}, \nonumber \\
& & v_{49}(r)=\frac{128 i \sqrt[3]{2} \pi ^{13/6}  \sqrt{\frac{1}{N}} (107136 r-114869 r_h) (r-r_h)}{59049\ 3^{5/6} \sqrt{g_s} |\log r|^{8/3}
   N_f^{8/3} r_h^6}, \ \ \ \ \ \  v_{50}(r)=\frac{128 i \sqrt[3]{\frac{2}{3}} \pi ^{8/3}  w (864 r-383 r_h)}{19683 |\log r|^{8/3} N_f^{8/3} r_h^7}, \nonumber\\
& & v_{51}(r)=\frac{128 i \sqrt[3]{\frac{2}{3}} \pi ^{8/3}  w (864 r-383 r_h)}{19683 |\log r|^{8/3} N_f^{8/3} r_h^7}, \ \ \ \ \ \ \ \ \ \  \ \ \ \ \ \ \  \ \ \ \ \ \ v_{52}(r)=\frac{512 \sqrt[3]{2} \pi ^{7/6}  \left(\frac{1}{N}\right)^{3/2} (438 r-919 r_h)}{19683\ 3^{5/6} g_s^{3/2} |\log r|^{8/3} N_f^{8/3}
   r_h^2}, \nonumber \\
& & v_{53}(r)=-\frac{9472 \sqrt[3]{2} \pi ^{13/6}  \sqrt{\frac{1}{N}}}{2187\ 3^{5/6} \sqrt{g_s} |\log r|^{8/3} N_f^{8/3} r_h^4} , \nonumber\\
& &  v_{54}(r)=-\frac{73472 \sqrt[3]{2} \sqrt[6]{\pi }  \left(\frac{1}{N}\right)^{5/2} r_h^2 (35 r-33 r_h) (r-r_h)}{177147\ 3^{5/6} g_s^{5/2}
   |\log r|^{8/3} N_f^{8/3}}, \nonumber 
   \end{eqnarray}
\begin{eqnarray}
& & v_{55}(r)=-\frac{64 \sqrt[3]{2} \pi ^{13/6}  \sqrt{\frac{1}{N}} (48384 r-51529 r_h) (r-r_h)}{59049\ 3^{5/6} \sqrt{g_s} |\log r|^{8/3}
   N_f^{8/3} r_h^6} , \nonumber\\
   & & v_{56}(r)=-\frac{64 \sqrt[3]{2} \pi ^{13/6}  \sqrt{\frac{1}{N}} (27648 r-30793 r_h) (r-r_h)}{59049\ 3^{5/6} \sqrt{g_s} |\log r|^{8/3}
   N_f^{8/3} r_h^6}, \nonumber \\
& & v_{57}(r)=-\frac{64 \sqrt[3]{2} \pi ^{13/6}  \sqrt{\frac{1}{N}} (6912 r-10057 r_h) (r-r_h)}{59049\ 3^{5/6} \sqrt{g_s} |\log r|^{8/3}
   N_f^{8/3} r_h^6}, \nonumber \\
& &v_{58}(r)=-\frac{236800 \sqrt[3]{\frac{2}{3}} \pi ^{8/3} }{2187 |\log r|^{8/3} N_f^{8/3} r_h^6} , \ \ \ \ \ \ \ \ \ \  \ \ \ \ \ \ \  \ \ \ \ \ \ v_{59}(r)=-\frac{94720 \sqrt[3]{\frac{2}{3}} \pi ^{8/3}  w}{2187 |\log r|^{8/3} N_f^{8/3} r_h^6}, \nonumber \\
& &v_{60}(r)=-\frac{947200 \sqrt[3]{2} \pi ^{13/6}  \sqrt{\frac{1}{N}} (r-r_h)}{2187\ 3^{5/6} \sqrt{g_s} |\log r|^{8/3} N_f^{8/3} r_h^5} , \ \ \ \ \ \ \ \ \ \  \ \ \ v_{61}(r)=-\frac{9472 \sqrt[3]{\frac{2}{3}} \pi ^{8/3}  w^2}{2187 |\log r|^{8/3} N_f^{8/3} r_h^6}, \nonumber \\
& &v_{62}(r)=-\frac{189440 \sqrt[3]{2} \pi ^{13/6}  \sqrt{\frac{1}{N}} w (r-r_h)}{2187\ 3^{5/6} \sqrt{g_s} |\log r|^{8/3} N_f^{8/3} r_h^5}  , \ \ \ \ \ \ \ \ \ \  \ \ \  v_{63}(r)=-\frac{9472 \sqrt[3]{\frac{2}{3}} \pi ^{8/3}  w^2 (r-r_h)}{19683 |\log r|^{8/3} N_f^{8/3} r_h^7}, \nonumber \\
& &v_{64}(r)=-\frac{9472 \sqrt[3]{\frac{2}{3}} \pi ^{8/3}  w^2 (r-r_h)}{19683 |\log r|^{8/3} N_f^{8/3} r_h^7} , \ \ \ \ \ \ \ \ \ \  \ \ \ \ \ \ \  \ \ \ \ \ \ v_{65}(r)=\frac{151552 i \sqrt[3]{2} \pi ^{7/6}  \left(\frac{1}{N}\right)^{3/2} w (r-r_h)^2}{6561\ 3^{5/6} g_s^{3/2} |\log r|^{8/3} N_f^{8/3}
   r_h^3}, \nonumber \\
& &v_{66}(r)=\frac{512 \sqrt[3]{\frac{2}{3}} \pi ^{2/3}  (2801 r+511 r_h)}{177147 g_s^2 |\log r|^{8/3} N^2 N_f^{8/3}}, \ \ \ \ \ \ \ \ \ \  \ \ \ \ \ \ \  \ \ \  v_{67}(r)=-\frac{189440 i \sqrt[3]{2} \pi ^{13/6}  \sqrt{\frac{1}{N}} (r-r_h)}{2187\ 3^{5/6} \sqrt{g_s} |\log r|^{8/3} N_f^{8/3} r_h^5} \nonumber \\
& &v_{68}(r)=\frac{52736 i \sqrt[3]{\frac{2}{3}} \pi ^{2/3}  r_h w (r-r_h)}{19683 g_s^2 |\log r|^{8/3} N^2 N_f^{8/3}} , \ \ \ \ \ \ \ \ \ \  \ \ \ \ \ \ \  \ \ \ \ \  v_{69}(r)=-\frac{37888 i \sqrt[3]{2} \pi ^{13/6}  \sqrt{\frac{1}{N}} w (r-r_h)}{2187\ 3^{5/6} \sqrt{g_s} |\log r|^{8/3} N_f^{8/3}r_h^5}, \nonumber \\
& &v_{70}(r)=\frac{26559488 i \sqrt[3]{\frac{2}{3}} \pi ^{2/3}  r_h (r-r_h)}{177147 g_s^2 |\log r|^{8/3} N^2 N_f^{8/3}}, \ \ \ \ \ \ \ \ \ \  \ \ \ \ \ \ \  \ \ \ v_{71}(r)=-\frac{37888 i \sqrt[3]{2} \pi ^{13/6}  \sqrt{\frac{1}{N}} w (r-r_h)^2}{19683\ 3^{5/6} \sqrt{g_s} |\log r|^{8/3} N_f^{8/3}
   r_h^6}, \nonumber \\
& &v_{72}(r)=-\frac{37888 i \sqrt[3]{2} \pi ^{13/6}  \sqrt{\frac{1}{N}} w (r-r_h)^2}{19683\ 3^{5/6} \sqrt{g_s} |\log r|^{8/3} N_f^{8/3}
   r_h^6}, \ \ \ \ \ \ \ \ \ \ v_{73}(r)=-\frac{2718464 \sqrt[3]{\frac{2}{3}} \pi ^{2/3}  (r-r_h)}{177147 g_s^2 |\log r|^{8/3} N^2 N_f^{8/3}}, \nonumber \\
& &v_{74}(r)=\frac{26368 \sqrt[3]{\frac{2}{3}} \pi ^{2/3}  r_h (r-r_h)}{19683 g_s^2 |\log r|^{8/3} N^2 N_f^{8/3}} , \ \ \ \ \ \ \ \ \ \  \ \ \ \ \ \ \  \ \ v_{75}(r)=\frac{274432\ 2^{2/3}  r_h (287 r-296 r_h) (r-r_h)^2}{4782969 \sqrt[6]{3} \pi ^{2/3} g_s^4 |\log r|^{10/3} N^4
   N_f^{10/3}}, \nonumber \\
& & v_{76}(r)=-\frac{611840 \sqrt[3]{\frac{2}{3}} \pi ^{2/3}  r_h (r-r_h)}{19683 g_s^2 |\log r|^{8/3} N^2 N_f^{8/3}}, \ \ \ \ \ \ \ \ \ \  \ \ \ \ \ \ \ v_{77}(r)=\frac{32768 \sqrt[3]{\frac{2}{3}} \pi ^{2/3}  (47 r-92 r_h) (r-r_h)^2}{59049 g_s^2 |\log r|^{8/3} N^2 N_f^{8/3} r_h}, \nonumber 
\end{eqnarray}
\begin{eqnarray}
& & v_{78}(r)=\frac{32768 \sqrt[3]{\frac{2}{3}} \pi ^{2/3}  (47 r-92 r_h) (r-r_h)^2}{59049 g_s^2 |\log r|^{8/3} N^2 N_f^{8/3} r_h}, \ \ \ \ \ \ \ \ \ \  \ \ v_{79}(r)=-\frac{5632 i \sqrt[3]{\frac{2}{3}} \pi ^{8/3} }{6561 |\log r|^{8/3} N_f^{8/3} r_h^7}, \nonumber \\
& & v_{80}(r)=\frac{2048 i \sqrt[3]{\frac{2}{3}} \pi ^{8/3}  w}{729 |\log r|^{8/3} N_f^{8/3} r_h^7} , \ \ \ \ \ \ \ \ \ \  \ \ \ \ \ \ \  \ \ \ \ \ \ \ \ \ \ \  v_{81}(r)=\frac{5120 \sqrt[3]{\frac{2}{3}} \pi ^{8/3}  w}{729 |\log r|^{8/3} N_f^{8/3} r_h^6}, \nonumber \\
& & v_{82}(r)=-\frac{1024 i \sqrt[3]{\frac{2}{3}} \pi ^{5/3}  q \left(96 r^2+2063 r r_h-2315 r_h^2\right)}{59049 g_s |\log r|^{8/3} N
   N_f^{8/3} r_h^5}, \nonumber \\
& & v_{83}(r)=\frac{1024 \sqrt[3]{\frac{2}{3}} \pi ^{8/3}  w^2}{729 |\log r|^{8/3} N_f^{8/3} r_h^6} , \ \ \ \ \ \ \ \ \ \  \ \ \ \ \ \ \  \ \ \ \ \ \ \ \ \ \ \ \ v_{84}(r)=\frac{24320 \sqrt[3]{2} \pi ^{19/6}  \sqrt{g_s} \sqrt{N} w}{2187\ 3^{5/6} |\log r|^{8/3} N_f^{8/3} r_h^8}, \nonumber \\
& & v_{85}(r)=\frac{27392 \sqrt[3]{2} \pi ^{19/6}  \sqrt{g_s} \sqrt{N} w}{2187\ 3^{5/6} |\log r|^{8/3} N_f^{8/3} r_h^8} , \ \ \ \ \ \ \ \ \ \  \ \ \ \ \ \ \  \ \ \ \ \ v_{86}(r)=\frac{88064 \sqrt[3]{2} \pi ^{13/6}  \sqrt{\frac{1}{N}} w (r-r_h)}{2187\ 3^{5/6} \sqrt{g_s} |\log r|^{8/3} N_f^{8/3} r_h^5}, \nonumber \\
& & v_{87}(r)=-\frac{5120 i \sqrt[3]{\frac{2}{3}} \pi ^{8/3} }{729 |\log r|^{8/3} N_f^{8/3} r_h^6} , \ \ \ \ \ \ \ \ \ \  \ \ \ \ \ \ \  \ \ \ \ \ v_{88}(r)=-\frac{1024 i \sqrt[3]{\frac{2}{3}} \pi ^{8/3}  w}{729 |\log r|^{8/3} N_f^{8/3} r_h^6}, \nonumber \\
& & v_{89}(r)=-\frac{88064 i \sqrt[3]{2} \pi ^{13/6}  \sqrt{\frac{1}{N}} (r-r_h)}{2187\ 3^{5/6} \sqrt{g_s} |\log r|^{8/3} N_f^{8/3} r_h^5}, \ \ \ \ \ \ \ \ \ \  v_{90}(r)=\frac{8192 i \sqrt[3]{2} \pi ^{13/6}  \sqrt{\frac{1}{N}} w (r-r_h)}{729\ 3^{5/6} \sqrt{g_s} |\log r|^{8/3} N_f^{8/3} r_h^5}, \nonumber \\
& & v_{91}(r)=\frac{12800 i \sqrt[3]{\frac{2}{3}} \pi ^{8/3}  (r-r_h)}{2187 |\log r|^{8/3} N_f^{8/3} r_h^7}, \ \ \ \ \ \ \ \ \ \  \ \ \ \ \ \ \  \ \ \ \ \ \ \ \ v_{92}(r)=\frac{34304 i \sqrt[3]{\frac{2}{3}} \pi ^{8/3}  (r-r_h)}{6561 |\log r|^{8/3} N_f^{8/3} r_h^7}, \nonumber \\
& & v_{93}(r)=\frac{8192 \sqrt[3]{2} \pi ^{13/6}  \sqrt{\frac{1}{N}} (r-r_h)}{729\ 3^{5/6} \sqrt{g_s} |\log r|^{8/3} N_f^{8/3} r_h^5}, \ \ \ \ \ \ \ \ \ \  \ \ \ \ \ \ v_{94}(r)=\frac{9472 \sqrt[3]{\frac{2}{3}} \pi ^{8/3}  (r-r_h)}{6561 \log(r)^8 N_f^{8/3} r_h^7}, \nonumber \\
& & v_{95}(r)=\frac{18944 \sqrt[3]{\frac{2}{3}} \pi ^{8/3}  w (r-r_h)}{19683 |\log r|^{8/3} N_f^{8/3} r_h^7}, \ \ \ \ \ \ \ \ \ \  \ \ \ \ \ \ \  \ \ \ \ \ v_{96}(r)=\frac{350464 \sqrt[3]{2} \pi ^{13/6}  \sqrt{\frac{1}{N}} (r-r_h)^2}{19683\ 3^{5/6} \sqrt{g_s} |\log r|^{8/3} N_f^{8/3} r_h^6}, \nonumber \\
& & v_{97}(r)=\frac{9472 \sqrt[3]{\frac{2}{3}} \pi ^{8/3}  w^2 (r-r_h)}{19683 |\log r|^{8/3} N_f^{8/3} r_h^7}, \ \ \ \ \ \ \ \ \ \  \ \ \ \ \ \ \  \ \ \ \ \ v_{98}(r)=\frac{179968 \sqrt[3]{2} \pi ^{13/6}  \sqrt{\frac{1}{N}} w (r-r_h)^2}{19683\ 3^{5/6} \sqrt{g_s} |\log r|^{8/3} N_f^{8/3} r_h^6}, \nonumber \\
& & v_{99}(r)=-\frac{9472 \sqrt[3]{\frac{2}{3}} \pi ^{8/3}  w^2}{2187 |\log r|^{8/3} N_f^{8/3} r_h^6} , \ \ \ \ \ \ \ \ \ \  \ \ \ \ \ \ \  \ \ \ \ \ v_{100}(r)=\frac{9472 \sqrt[3]{\frac{2}{3}} \pi ^{8/3}  w^2 (r-r_h)}{19683 |\log r|^{8/3} N_f^{8/3} r_h^7}, \nonumber \\
& & v_{101}(r)=-\frac{833536 \sqrt[3]{\frac{2}{3}} \pi ^{5/3}  (r-r_h)^2}{6561 g_s |\log r|^{8/3} N N_f^{8/3} r_h^4}, \ \ \ \ \ \ \ \ \ \  \ \ \ \ \ \ \  \  v_{102}(r)=\frac{75776 \sqrt[3]{\frac{2}{3}} \pi ^{5/3}  (r-r_h)^2}{59049 g_s |\log r|^{8/3} N N_f^{8/3} r_h^5}, \nonumber 
\end{eqnarray}
\begin{eqnarray}
& & v_{103}(r)=\frac{179968 i \sqrt[3]{2} \pi ^{13/6}  \sqrt{\frac{1}{N}} (r-r_h)^2}{19683\ 3^{5/6} \sqrt{g_s} |\log r|^{8/3} N_f^{8/3} r_h^6}, \ \ \ \ \ \ \ \ \ \  \ \ \ \   v_{104}(r)=-\frac{1245184 \sqrt[3]{2} \pi ^{7/6}  \left(\frac{1}{N}\right)^{3/2} (r-r_h)^2}{19683\ 3^{5/6} g_s^{3/2} |\log r|^{8/3} N_f^{8/3}
   r_h^3}, \nonumber\\ 
& & v_{105}(r)=\frac{104192 i \sqrt[3]{2} \pi ^{13/6}  \sqrt{\frac{1}{N}} w (r-r_h)^2}{19683\ 3^{5/6} \sqrt{g_s} |\log r|^{8/3} N_f^{8/3}
   r_h^6}, \ \ \ \ \ \ \ \ \ \  \ \ \  \    v_{106}(r)=-\frac{757760 i \sqrt[3]{\frac{2}{3}} \pi ^{5/3}  (r-r_h)^2}{6561 g_s |\log r|^{8/3} N N_f^{8/3} r_h^4}, \nonumber \\
& & v_{107}(r)=-\frac{37888 i \sqrt[3]{2} \pi ^{13/6}  \sqrt{\frac{1}{N}} w (r-r_h)}{2187\ 3^{5/6} \sqrt{g_s} |\log r|^{8/3} N_f^{8/3} r_h^5}, \ \ \ \ \ \ \ \ \ \  \ \ \ \    v_{108}(r)=\frac{37888 i \sqrt[3]{2} \pi ^{13/6}  \sqrt{\frac{1}{N}} w (r-r_h)^2}{19683\ 3^{5/6} \sqrt{g_s} |\log r|^{8/3} N_f^{8/3}
   r_h^6}, \nonumber \\
& & v_{109}(r)=-\frac{100352 i \sqrt[3]{2} \pi ^{7/6}  \left(\frac{1}{N}\right)^{3/2} w}{6561\ 3^{5/6} g_s^{3/2} |\log r|^{8/3} N_f^{8/3} r_h}, \ \ \ \ \ \ \ \ \ \  \ \ \ \ \ v_{110}(r)=\frac{26368 \sqrt[3]{\frac{2}{3}} \pi ^{2/3}  r_h (r-r_h)}{19683 g_s^2 |\log r|^{8/3} N^2 N_f^{8/3}}, \nonumber \\
& & v_{111}(r)=\frac{530432 \sqrt[3]{\frac{2}{3}} \pi ^{5/3}  (r-r_h)^2}{19683 g_s |\log r|^{8/3} N N_f^{8/3} r_h^4}, \ \ \ \ \ \ \ \ \ \  \ \ \ \ \ \ \  \ \ \ \ \ v_{112}(r)=\frac{75776 \sqrt[3]{\frac{2}{3}} \pi ^{5/3}  (r-r_h)^2}{19683 g_s |\log r|^{8/3} N N_f^{8/3} r_h^4}, \nonumber \\
& & v_{113}(r)=\frac{32768 \sqrt[3]{\frac{2}{3}} \pi ^{2/3}  (47 r-92 r_h) (r-r_h)^2}{59049 g_s^2 |\log r|^{8/3} N^2 N_f^{8/3} r_h}, \ \ \ \ \ \ \ \ \ \  \ \ \ \ \ \ v_{114}(r)=-\frac{9472 \sqrt[3]{\frac{2}{3}} \pi ^{8/3}  (r-r_h)}{19683 |\log r|^{8/3} N_f^{8/3} r_h^7},\nonumber\\
& & v_{115}(r)=-\frac{18944 \sqrt[3]{\frac{2}{3}} \pi ^{8/3}  w (r-r_h)}{19683 |\log r|^{8/3} N_f^{8/3} r_h^7}, \ \ \ \ \ \ \ \ \ \ \ \ \ \ \ \ \ \ \  v_{116}(r)=\frac{37888 \sqrt[3]{2} \pi ^{13/6}  \sqrt{\frac{1}{N}} (r-r_h)^2}{2187\ 3^{5/6} \sqrt{g_s} |\log r|^{8/3} N_f^{8/3} r_h^6}, \nonumber\\
& & v_{117}(r)=-\frac{9472 \sqrt[3]{\frac{2}{3}} \pi ^{8/3}  w^2 (r-r_h)}{19683 |\log r|^{8/3} N_f^{8/3} r_h^7}, \ \ \ \ \ \ \ \ \ \ \ \ \ \ \ \ \ \ \ v_{118}(r)=\frac{37888 \sqrt[3]{2} \pi ^{13/6}  \sqrt{\frac{1}{N}} w (r-r_h)^2}{6561\ 3^{5/6} \sqrt{g_s} |\log r|^{8/3} N_f^{8/3} r_h^6}, \nonumber \\
& & v_{119}(r)=-\frac{9472 \sqrt[3]{\frac{2}{3}} \pi ^{8/3}  w^2 (r-r_h)}{19683 |\log r|^{8/3} N_f^{8/3} r_h^7}, \ \ \ \ \ \ \ \ \ \ \ \ \ \ \ \ \ \ \ v_{120}(r)=-\frac{9472 \sqrt[3]{\frac{2}{3}} \pi ^{8/3}  w^2}{2187 |\log r|^{8/3} N_f^{8/3} r_h^6}, \nonumber \\
& & v_{121}(r)=-\frac{3031040 \sqrt[3]{\frac{2}{3}} \pi ^{5/3}  (r-r_h)^2}{19683 g_s |\log r|^{8/3} N N_f^{8/3} r_h^4}, \ \ \ \ \ \ \ \ \ \ \ \ \ \  v_{122}(r)=\frac{151552 \sqrt[3]{\frac{2}{3}} \pi ^{5/3}  (r-r_h)^2}{59049 g_s |\log r|^{8/3} N N_f^{8/3} r_h^5}, \nonumber \\
& & v_{123}(r)=\frac{37888 i \sqrt[3]{2} \pi ^{13/6}  \sqrt{\frac{1}{N}} (r-r_h)^2}{6561\ 3^{5/6} \sqrt{g_s} |\log r|^{8/3} N_f^{8/3} r_h^6}, \ \ \ \ \ \ \ \ \ \ \ \  v_{124}(r)=-\frac{2490368 \sqrt[3]{2} \pi ^{7/6}  \left(\frac{1}{N}\right)^{3/2} (r-r_h)^2}{19683\ 3^{5/6} g_s^{3/2} |\log r|^{8/3} N_f^{8/3}
   r_h^3}, \nonumber \\
& & v_{125}(r)=\frac{9472 i \sqrt[3]{2} \pi ^{13/6}  \sqrt{\frac{1}{N}} w (r-r_h)^2}{6561\ 3^{5/6} \sqrt{g_s} |\log r|^{8/3} N_f^{8/3} r_h^6}, \ \ \ \ \ \ \ \ \ \ \ \ \ v_{126}(r)=-\frac{2424832 i \sqrt[3]{\frac{2}{3}} \pi ^{5/3}  (r-r_h)^2}{19683 g_s |\log r|^{8/3} N N_f^{8/3} r_h^4}, \nonumber \\
& & v_{127}(r)=\frac{9472 i \sqrt[3]{2} \pi ^{13/6}  \sqrt{\frac{1}{N}} w (r-r_h)^2}{6561\ 3^{5/6} \sqrt{g_s} |\log r|^{8/3} N_f^{8/3} r_h^6}, \ \ \ \ \ \ \ \ \ \ \ \ \  v_{128}(r)=-\frac{37888 i \sqrt[3]{2} \pi ^{13/6}  \sqrt{\frac{1}{N}} w (r-r_h)}{2187\ 3^{5/6} \sqrt{g_s} |\log r|^{8/3} N_f^{8/3} r_h^5}, \nonumber 
\end{eqnarray}
\begin{eqnarray}
& & v_{129}(r)=-\frac{2718464 \sqrt[3]{\frac{2}{3}} \pi ^{2/3}  (r-r_h)}{177147 g_s^2 |\log r|^{8/3} N^2 N_f^{8/3}},  \ \ \ \ \ \ \ \ \ \ \ \ \ \ \ v_{130}(r)=\frac{26368 \sqrt[3]{\frac{2}{3}} \pi ^{2/3}  r_h (r-r_h)}{19683 g_s^2 |\log r|^{8/3} N^2 N_f^{8/3}}, \nonumber \\
& & v_{131}(r)=\frac{530432 \sqrt[3]{\frac{2}{3}} \pi ^{5/3}  (r-r_h)^2}{19683 g_s |\log r|^{8/3} N N_f^{8/3} r_h^4},  \ \ \ \ \ \ \ \ \ \ \ \ \ \ \ v_{132}(r)=\frac{530432 \sqrt[3]{\frac{2}{3}} \pi ^{5/3}  (r-r_h)^2}{19683 g_s |\log r|^{8/3} N N_f^{8/3} r_h^4}, \nonumber \\
& & v_{133}(r)=\frac{151552 \sqrt[3]{\frac{2}{3}} \pi ^{5/3}  (r-r_h)^2}{19683 g_s |\log r|^{8/3} N N_f^{8/3} r_h^4},  \ \ \ \ \ \ \ \ \ \ \ \ \ \ \ v_{134}(r)=\frac{1147648 \sqrt[3]{\frac{2}{3}} \pi ^{8/3} }{6561 |\log r|^{8/3} N_f^{8/3} r_h^6}, \nonumber \\
& & v_{135}(r)=\frac{56576 \sqrt[3]{\frac{2}{3}} \pi ^{8/3}  w}{729 |\log r|^{8/3} N_f^{8/3} r_h^6} ,  \ \ \ \ \ \ \ \ \ \ \ \ \ \ \ \ \ \ \ \ \ \ \  v_{136}(r)=-\frac{1024 \sqrt[3]{2} \pi ^{13/6}  \sqrt{\frac{1}{N}} (1429 r+2017 r_h)}{6561\ 3^{5/6} \sqrt{g_s} |\log r|^{8/3} N_f^{8/3}
   r_h^5}, \nonumber \\
& & v_{137}(r)=\frac{66112 \sqrt[3]{\frac{2}{3}} \pi ^{8/3}  w^2}{6561 |\log r|^{8/3} N_f^{8/3} r_h^6} ,  \ \ \ \ \ \ \ \ \ \ \ \ \ \ \ \ \ \ \ \ \ \ v_{138}(r)=-\frac{1024 \sqrt[3]{2} \pi ^{13/6}  \sqrt{\frac{1}{N}} w (209 r+657 r_h)}{6561\ 3^{5/6} \sqrt{g_s} |\log r|^{8/3} N_f^{8/3}
   r_h^5}, \nonumber \\
& & v_{139}(r)=\frac{64 \sqrt[3]{\frac{2}{3}} \pi ^{8/3}  \left(1033 w^2+4368\right)}{6561 |\log r|^{8/3} N_f^{8/3} r_h^6},  \ \ \ \ \ \ \ \ \ \ \ \ \ \ \ \ \ \ v_{140}(r)=\frac{64 \sqrt[3]{\frac{2}{3}} \pi ^{8/3}  \left(1033 w^2+4368\right)}{6561 |\log r|^{8/3} N_f^{8/3} r_h^6}, \nonumber \\
& & v_{141}(r)=\frac{256 \sqrt[3]{\frac{2}{3}} \pi ^{5/3}  \left(3840 r^2-6480 r r_h-24 r_h^2 \left(87 w^2-110\right)-455\right)}{19683 g_s
   |\log r|^{8/3} N N_f^{8/3} r_h^4},\nonumber\\
   & & v_{142}(r)=-\frac{512 \sqrt[3]{\frac{2}{3}} \pi ^{5/3}  \left(16 r^2-438 r r_h+1307 r_h^2\right)}{6561 g_s |\log r|^{8/3} N N_f^{8/3}
   r_h^5}, \ \ \ \ \ \  v_{143}(r)=-\frac{1024 i \sqrt[3]{2} \pi ^{13/6}  \sqrt{\frac{1}{N}} (209 r+657 r_h)}{6561\ 3^{5/6} \sqrt{g_s} |\log r|^{8/3} N_f^{8/3}
   r_h^5}, \nonumber \\
& & v_{144}(r)=\frac{1024 i \sqrt[3]{\frac{2}{3}} \pi ^{5/3}  w \left(-64 r^2+119 r r_h+989 r_h^2\right)}{19683 g_s |\log r|^{8/3} N N_f^{8/3}
   r_h^4},  \ \ \ \ \ v_{145}(r)=\frac{35840 i \sqrt[3]{2} \pi ^{13/6}  \sqrt{\frac{1}{N}} w (r-11 r_h)}{6561\ 3^{5/6} \sqrt{g_s} |\log r|^{8/3} N_f^{8/3}
   r_h^5}, \nonumber \\
& & v_{146}(r)=\frac{327680 i \sqrt[3]{\frac{2}{3}} \pi ^{5/3}  (r-2 r_h) (r-r_h)}{6561 g_s |\log r|^{8/3} N N_f^{8/3} r_h^4},  \ \ \ \ \ \ \ \ \ \ \ \ \ \ \ \ \ v_{147}(r)=\frac{35840 i \sqrt[3]{2} \pi ^{13/6}  \sqrt{\frac{1}{N}} w (r-11 r_h)}{6561\ 3^{5/6} \sqrt{g_s} |\log r|^{8/3} N_f^{8/3}
   r_h^5}, \nonumber \\
& & v_{148}(r)=\frac{35840 i \sqrt[3]{2} \pi ^{13/6}  \sqrt{\frac{1}{N}} w (r-11 r_h)}{6561\ 3^{5/6} \sqrt{g_s} |\log r|^{8/3} N_f^{8/3}
   r_h^5},  \ \ \ \ \ \ \ \ \ \ \ \ \ \ \ \ \ \  v_{149}(r)=-\frac{13312 \sqrt[3]{\frac{2}{3}} \pi ^{5/3} }{6561 g_s |\log r|^{8/3} N N_f^{8/3} r_h^3}, \nonumber \\
& & v_{150}(r)=\frac{512 \sqrt[3]{\frac{2}{3}} \pi ^{5/3}  \left(-64 r^2+119 r r_h+989 r_h^2\right)}{19683 g_s |\log r|^{8/3} N N_f^{8/3}
   r_h^4},  \ \ \ \ \ \ \ \ \ \ \ \ v_{151}(r)=-\frac{73472 \sqrt[3]{\frac{2}{3}} \pi ^{2/3}  r_h (r-r_h)}{59049 g_s^2 |\log r|^{8/3} N^2 N_f^{8/3}}, \nonumber \\
& & v_{152}(r)=-\frac{512 \sqrt[3]{\frac{2}{3}} \pi ^{5/3}  (480 r-1033 r_h) (r-r_h)}{19683 g_s |\log r|^{8/3} N N_f^{8/3} r_h^4},  \ \ \ \ \ \ \ \ \ v_{153}(r)=-\frac{512 \sqrt[3]{\frac{2}{3}} \pi ^{5/3}  (384 r-1345 r_h) (r-r_h)}{19683 g_s |\log r|^{8/3} N N_f^{8/3} r_h^4}, \nonumber \\
& & v_{154}(r)=-\frac{512 \sqrt[3]{\frac{2}{3}} \pi ^{5/3}  (288 r-841 r_h) (r-r_h)}{19683 g_s |\log r|^{8/3} N N_f^{8/3} r_h^4}. \nonumber \\
\end{eqnarray}

\end{document}